\documentclass[12pt]{article}
\usepackage{jheppub}
\pdfoutput=1
\usepackage{epstopdf}
\usepackage{mathrsfs}
\usepackage{psfrag}

\usepackage{epsf}
\usepackage{graphicx,epsfig}
\usepackage{amssymb}
\usepackage{epstopdf}
\usepackage{float}
\usepackage[table]{xcolor}

\usepackage{amsmath,bbm,array,amsfonts,graphicx,wrapfig,lscape,float,slashbox,multirow,longtable,rotating,subfigure,makecell}
\usepackage{url}

\newcommand{\be}{\begin{equation}}
\newcommand{\ee}{\end{equation}}
\newcommand{\beq}{\begin{equation}}
\newcommand{\beql}[1]{\begin{equation}\label{#1}}
\newcommand{\eeq}{\end{equation}}
\newcommand{\ba}{\begin{array}}
\newcommand{\ea}{\end{array}}
\newcommand{\bea}{\begin{eqnarray}}
\newcommand{\beal}[1]{\begin{eqnarray}\label{#1}}
\newcommand{\eea}{\end{eqnarray}}
\newcommand{\ben}{\begin{enumerate}}
\newcommand{\een}{\end{enumerate}}
\newcommand{\bean}{\begin{eqnarray*}}
\newcommand{\eean}{\end{eqnarray*}}
\newcommand{\eref}[1]{(\ref{#1})}
\newcommand{\sref}[1]{\S\ref{#1}}

\newcommand{\fref}[1]{Figure \ref{#1}}
\newcommand{\btab}[1]{\begin{tabular}{#1}}
\newcommand{\etab}{\end{tabular}}

\newcommand{\comment}[1]{}

\newcommand{\qed}{\nobreak \ifvmode \relax \else
      \ifdim\lastskip<1.5em \hskip-\lastskip
      \hskip1.5em plus0em minus0.5em \fi \nobreak
      \vrule height0.75em width0.5em depth0.25em\fi}

\newcolumntype{C}[1]{>{\centering\arraybackslash}m{#1}}

\newcommand{\pl}{Pl\"ucker }

\newcommand{\A}{\mathcal{A}}

\newcolumntype{C}[1]{>{\centering\arraybackslash}m{#1}}
\newcommand{\N}{\mathcal{N}}
\newcommand{\tl}{\widetilde{\lambda}}

\newcommand{\MHVb}{\overline{\text{MHV}}}

\newcommand{\comments}[1]{}

\graphicspath{{./images/}}

\title{Non-Planar On-Shell Diagrams}

\author{Sebasti\'an Franco,$^{1,2}$}
\author{Daniele Galloni,$^3$}
\author{Brenda Penante,$^4$}
\author{Congkao Wen$^5$\\ [-.35cm]}

\affiliation{$^1$ Physics Department, The City College of the CUNY \\
160 Convent Avenue, New York, NY 10031, USA \\
$^2$ The Graduate School and University Center, The City University of New York  \\
365 Fifth Avenue, New York NY 10016, USA \\ 
$^3$ Institute for Particle Physics Phenomenology, Department of Physics \\
Durham University, Durham DH1 3LE, United Kingdom  \\ 
$^4$ Centre for Research in String Theory, School of Physics and Astronomy \\
Queen Mary University of London, Mile End Road, London E1 4NS, United Kingdom  \\ 
$^5$ Dipartimento di Fisica, Universit\`a di Roma ``Tor Vergata" \& I.N.F.N. Sezione di Roma ``Tor Vergata", Via della Ricerca Scientifica, 00133 Roma, Italy \\[-.25cm]
}

\emailAdd{sfranco@ccny.cuny.edu, daniele.galloni@durham.ac.uk, b.penante@qmul.ac.uk, Congkao.Wen@roma2.infn.it}

\abstract{We initiate a systematic study of non-planar on-shell diagrams in $\mathcal{N}=4$ SYM and develop powerful technology for doing so. We introduce canonical variables generalizing face variables, which make the $d\log$ form of the on-shell form explicit. We make significant progress towards a general classification of arbitrary on-shell diagrams by means of two classes of combinatorial objects: generalized matching and matroid polytopes. We propose a boundary measurement that connects general on-shell diagrams to the Grassmannian. Our proposal exhibits two important and non-trivial properties: positivity in the planar case and it matches the combinatorial description of the diagrams in terms of generalized matroid polytopes. Interestingly, non-planar diagrams exhibit novel phenomena, such as the emergence of constraints on \pl coordinates beyond \pl relations when deleting edges, which are neatly captured by the generalized matching and matroid polytopes. This behavior is tied to the existence of a new type of poles in the on-shell form at which combinations of \pl coordinates vanish. Finally, we introduce a prescription, applicable beyond the MHV case, for writing the on-shell form as a function of minors directly from the graph.}

\preprint{
\begin{flushright}IPPP/15/03\end{flushright} \vspace{-0.4cm}
\begin{flushright}DCPT/15/06\end{flushright} \vspace{-0.4cm}
\begin{flushright}QMUL-PH-15-01\end{flushright}\vspace{-0.4cm}
\begin{flushright}ROM2F/2015/01\end{flushright}
}

\begin{document}

\maketitle

\section{Introduction}

In recent years, there has been tremendous progress in our understanding of scattering amplitudes. This has been particularly impressive for planar $\mathcal{N}=4$ super Yang-Mills (SYM), see e.g.\ \cite{Elvang:2013cua,Henn:2014yza} for recent reviews. Extremely powerful tools have been developed and impressive results have been obtained to high loop order \cite{Goncharov:2010jf,Golden:2013xva,Golden:2014xqa,Golden:2014pua,Drummond:2014ffa,Dixon:2011pw,Dixon:2011nj,Dixon:2014voa,Dixon:2014iba, Basso:2013vsa,Basso:2013aha,Basso:2014koa}. These advances are closely related to the discovery of hidden symmetries and dualities in the theory \cite{Drummond:2008vq,Drummond:2009fd,Alday:2007hr,Drummond:2007aua,Brandhuber:2007yx}. Furthermore, new mathematical and geometric structures have been uncovered, most notably a dual formulation for the scattering amplitudes in this theory was developed: a Grassmannian formulation \cite{ArkaniHamed:2009dn,ArkaniHamed:2009vw,Kaplan:2009mh,Mason:2009qx,ArkaniHamed:2009dg}, on-shell diagrams \cite{ArkaniHamed:2012nw} and the geometrization of scattering amplitudes in terms of the amplituhedron \cite{Arkani-Hamed:2013jha,Arkani-Hamed:2013kca,Franco:2014csa,Lam:2014jda,Bai:2014cna,Arkani-Hamed:2014dca}.

At this point, there are very clear directions in which this program can be extended: considering quantum field theories in other dimensions or reduced supersymmetry and going beyond the planar limit of $\mathcal{N}=4$ SYM. This article is devoted to the latter, more concretely to non-planar on-shell diagrams. Although there has been important progress in the study of non-planar amplitudes in $\mathcal{N}=4$ SYM \cite{Bern:1997nh,Bern:2007hh,Bern:2010tq,Carrasco:2011mn,Bern:2012uc}, they are far less understood than amplitudes in the planar sector. Recently, building on the observation based on on-shell diagrams that the loop integrand in planar amplitudes has only logarithmic singularities and no poles at infinity, it has been conjectured that non-planar amplitudes share the same property \cite{Arkani-Hamed:2014via}. Further evidence supporting this conjecture was provided in \cite{Bern:2014kca}.

On-shell diagrams are extremely useful for studying scattering amplitudes. In particular, in the planar limit, the all-loop integrand in $\mathcal{N}=4$ SYM can be expressed in terms of on-shell diagrams. Currently there is no well-defined notion of loop integrands for the amplitudes beyond the planar limit, however non-planar on-shell diagrams are still certainly worth studying since, to say the least, they provide a description for the leading singularities of loop amplitudes.\footnote{More ambitiously, one could envision that a Grassmannian formulation of non-planar $\mathcal{N}=4$ SYM exists and, if so, it can perhaps be phrased in terms of non-planar on-shell diagrams.} This direction recently began to be explored in \cite{Arkani-Hamed:2014bca},\footnote{See also \cite{Du:2014jwa,Chen:2014ara} for relevant work.} primarily in the case of MHV leading singularities. In this paper, we initiate a systematic study of general non-planar on-shell diagrams in $\mathcal{N}=4$ SYM and develop powerful technology for doing so. We further explore their physical applications. General on-shell diagrams are constructed by gluing elementary MHV and $\overline{\rm{MHV}}$ three-point on-shell amplitudes together. Since every three-point amplitude also carries a color factor, so does the on-shell diagram built from them. While important, this color factor will be omitted from now on in our discussions.

We begin this article with a brief review of planar on-shell diagrams and of some basic bipartite technology in \sref{sec:review-planar}. Before studying non-planar on-shell diagrams in full generality, we discuss in \sref{section:non-adjacentBCFW} a concrete scenario in which non-planar on-shell diagrams appear and are relevant: the computation of tree-level amplitudes using non-adjacent BCFW shifts. In the following sections, we introduce powerful technology for a systematic understanding of the general non-planar case. Below we list some of the main concepts we will present.

\sref{sec:generalized-faces} introduces canonical variables for non-planar graphs generalizing face variables, which have proved extremely useful in the planar case. Among other things, these variables allow a straightforward determination of the degrees of freedom in a graph and automatically make the $d\log$ structure of the on-shell form manifest. We also discuss a systematic procedure for determining such canonical variables, based on embedding on-shell diagrams into bordered Riemann surfaces. Physical results are, of course, independent of the choice of such an embedding. 

In \sref{section_polytopes}, we present two combinatorial objects, the generalized matching and matroid polytopes, which provide a general characterization of non-planar on-shell graphs and are extremely useful in extending the notion of graph equivalence and reductions to the non-planar case. As for planar graphs, these two concepts can be exploited to reduce the infinite plethora of possible on-shell diagrams to a finite number of important ones. The canonical variables of \sref{sec:generalized-faces}, also give rise to a straightforward procedure for constructing these polytopes.

On-shell diagrams are mapped into the Grassmannian via the boundary measurement. In \sref{sec:general-boundary-measurement}, we propose a boundary measurement for completely general on-shell diagrams. So far, the boundary measurement was only known for graphs admitting a genus-zero embedding. Needless to say, the boundary measurement is an essential ingredient for developing a comprehensive theory of non-planar on-shell diagrams. A crucial ingredient in our construction is a delicate choice of signs, which achieves two important goals. First, the signs are necessary for positivity in the case of planar graphs and its generalization for non-planar ones. Second, our sign prescription beautifully leads to the combinatorial description based on generalized matroid polytopes.

While going from an on-shell diagram to the corresponding on-shell form in terms of face variables is straightforward, it is however much more challenging to directly obtain its expression in terms of minors. In \sref{sec:computeform}, we generalize the prescription introduced in \cite{Arkani-Hamed:2014bca} beyond the MHV case, which allows us to directly write the on-shell form of reduced diagrams as a function of minors starting from the graph. We compare the results of this proposal with those obtained using the boundary measurement, finding perfect agreement. An interesting new feature of non-planar on-shell diagrams we uncover is the possibility of a new kind of pole in the on-shell form, not given by the vanishing of the \pl coordinates. 

In \sref{sec:reducibility}, we present a comprehensive discussion of equivalences and reductions of non-planar graphs. We provide a systematic approach for beginning to address these issues based on generalized matching and matroid polytopes. Interestingly, non-planar graphs can exhibit new phenomena, such as non-unique reductions and the appearance of new constraints between \pl coordinates that are beyond \pl relations, to which we will refer as non-\pl constraints for short. The latter is directly tied to the emergence of the new type of poles found in \sref{sec:computeform}. Throughout the article, we collect several explicit examples illustrating our ideas.

\bigskip

\section{Planar On-Shell Diagrams and Bipartite Technology} 
\label{sec:review-planar}

In this section we quickly review some basic concepts regarding on-shell diagrams and tools for studying bipartite graphs.

\bigskip

\subsection{On-Shell Diagrams and On-Shell Forms}

N$^{k-2}$MHV leading singularities with $n$ external states in planar $\N=4$ SYM are given by contour integrals over the Grassmannian $Gr_{k,n}$ \cite{ArkaniHamed:2009dn}. $Gr_{k,n}$ is the space of $k$-dimensional planes in $\mathbb{C}^n$ passing through the origin, so points in it can be represented by $k\times n$ matrices $C$ modulo $\text{GL}(k)$. We thus have
\begin{equation}
\label{eq:planarG}
\mathcal{L}_{k,n}=\int\limits_{\Gamma_{k,n}} \frac{d^{k\times n} C}{\text{Vol(GL}(k))}\, \frac{\prod\limits_{\alpha=1}^k \delta^{4|4}\left(C_{\alpha a}\mathcal{W}^a\right)}{(1\cdots k)(2\cdots k+1)\cdots(n\cdots k-1)}\, ,
\end{equation}
where $\Gamma_{k,n}$ stands for the contour, namely a prescription for which particular combination of $k\times k$ consecutive minors of the matrix $C$ must be set to zero in order to compute the residues, and $\mathcal{W}^a$ encode the kinematical data in terms of supertwistors. Here and in what follows, $(i_1,\ldots,i_k)$ denotes the minor corresponding to columns $i_1,\ldots,i_k$.

The emergence of the Grassmannian in the context of scattering amplitudes was fully understood with the introduction of the {\it on-shell diagram} formalism \cite{ArkaniHamed:2012nw}, which is valid beyond leading singularities. In this section, we briefly review the main properties of planar on-shell diagrams, with the aim to introduce the basic concepts that will be generalized in coming sections to the non-planar case. For a detailed presentation, we refer the reader to the original work \cite{ArkaniHamed:2012nw}.

On-shell diagrams are graphs constructed by connecting vertices which represent three-point amplitudes along edges that represent on-shell momenta.\footnote{As we explain below, the valency of nodes can be increased by some simple operations.} There are two types of (non-vanishing) three-point amplitudes,  $A^{\text{MHV}}_3$ and $A^{\MHVb}_3$, which are represented by black and white vertices, respectively. Nodes are glued together via the integration over the on-shell phase space of the (super) momentum associated to the edge shared by two vertices.\footnote{In this article, following a standard approach in the combinatorics literature, we chose to include external nodes at the endpoints of legs of on-shell diagrams. We would like to emphasize that we are dealing with ordinary on-shell diagrams and that such external nodes have no physical significance. They can become useful bookkeeping devices when performing certain transformations of the diagram.}

In the Grassmannian formulation, $A^{\text{MHV}}_3$ is given by an integral over $Gr_{2,3}$ while $A^{\overline{\text{MHV}}}_3$ corresponds to an integral over $Gr_{1,3}$. As vertices are glued together, they give rise to a larger Grassmannian $Gr_{k,n}$. For a trivalent on-shell diagram with $n_B$ internal black nodes, $n_W$ internal white nodes and $n_I$ internal edges, the value of $k$ is given by
\begin{equation}
k\,=\,2 n_B + n_W - n_I .
\label{k_trivalent}
\end{equation}

The number of degrees of freedom $d$ of a general on-shell diagram is obtained by starting from the edge weights and subtracting the GL($1$) gauge redundancy associated to every internal node. This means that for a diagram with $E$ edges and $N$ internal nodes, we have 
\beq
d=E-N.
\eeq
The previous expression is completely general. For a planar on-shell diagram with $F$ faces, this is equal to $d=F-1$. This means that all edge weights can be expressed in terms of $F-1$ independent ones. Another useful parametrization of an on-shell diagram is in terms of face variables $f_i,\,i=1,\dots,F$, which are subject to the constraint $\prod_{i=1}^{F}f_i=1$. They are given by the product of all edge weights around a face (closed or open) and, for concreteness, they can be taken to be oriented clockwise. Face variables constitute a GL($1$) invariant way of parametrizing the degrees of freedom of the graph. In \sref{sec:generalized-faces}, we will generalize them to non-planar diagrams and discuss how the counting of degrees of freedom is modified. 

Generalizing \eqref{eq:planarG}, every on-shell diagram, either planar or non-planar, is associated to a differential form

\beq
\left(\prod_{\text{int. nodes }v} \dfrac{1}{\text{Vol}(\text{GL}(1)_v)}\right) 
\left(\prod_{\text{edges } X_e} \dfrac{dX_e}{X_e}\right) \prod\limits_{\alpha=1}^k \delta^{4|4}\left(C_{\alpha a}\mathcal{W}^a\right) ,
\label{on-shell_form}
\eeq

\noindent where the first product is taken over all internal nodes.  
We will refer to the form excluding the $\delta$-functions as the {\it on-shell form} $\Omega$ corresponding to a given on-shell diagram. The full on-shell form associated to a $d$-dimensional planar on-shell diagram in terms of edge or face variables is of the ``$d\log$'' form \cite{ArkaniHamed:2012nw}
\beq
\label{eq:edges}
\Omega\,=\,\frac{dX_1}{X_1} \, \frac{dX_2}{X_2} \cdots \frac{dX_d}{X_d}\,=\,\frac{df_1}{f_1} \, \frac{df_2}{f_2} \cdots \frac{df_d}{f_d}.
\eeq
Expressing the on-shell form in terms of edge weights requires using the GL($1$) redundancies to identify $d$ independent variables. This task is bypassed when using face variables. In the next section, we will develop the generalization of face variables for non-planar graphs.

When the dimension of the graph coincides with the dimension of $Gr_{k,n}$, i.e., $d=k(n-k)$, the on-shell form is said to be top-dimensional and \eqref{eq:edges} becomes equivalent to \eqref{eq:planarG} after including the $\delta$-functions. 
If the dimension of the graph is larger than the dimension of $Gr_{k,n}$, the graph may be reduced, as discussed at length in \sref{sec:reducibility}, into a graph of dimension $d \leq k (n-k)$. If the dimension of the graph is smaller than the dimension of $Gr_{k,n}$, \eqref{eq:edges} arises as certain \textit{residue} of \eqref{eq:planarG}; the residue is taken around the vanishing of those minors which disappear once those graphical degrees of freedom have been turned off.

\bigskip

\subsection{Equivalence Moves and Reductions}

On-shell diagrams form {\it equivalence classes} and can be connected by {\it reductions}. Equivalent on-shell diagrams are related by a sequence of the following {\it equivalence moves}:\footnote{Here we adopt a conservative position and extend the definition of equivalence based on moves from planar graphs to completely general ones.}

\bigskip

\noindent {\bf Merger.} Connected internal nodes of the same color can be merged. A multi-leg black (white) vertex means that all $\tl$'s ($\lambda$'s) connected to it are proportional. Alternatively, whenever two internal nodes of the same color are connected by an edge, we can introduce a 2-valent node of the opposite color between them. Any on-shell diagram can be made bipartite by using these operations. Throughout the rest of the article, we will thus focus almost exclusively on bipartite graphs.\footnote{For this reason, we will use the terms on-shell diagram, diagram, bipartite graph and graph interchangeably.}
Mergers can be used in both directions, to either increase or decrease the valency of nodes. 

\begin{figure}[h]
\begin{center}
\includegraphics[width=9cm]{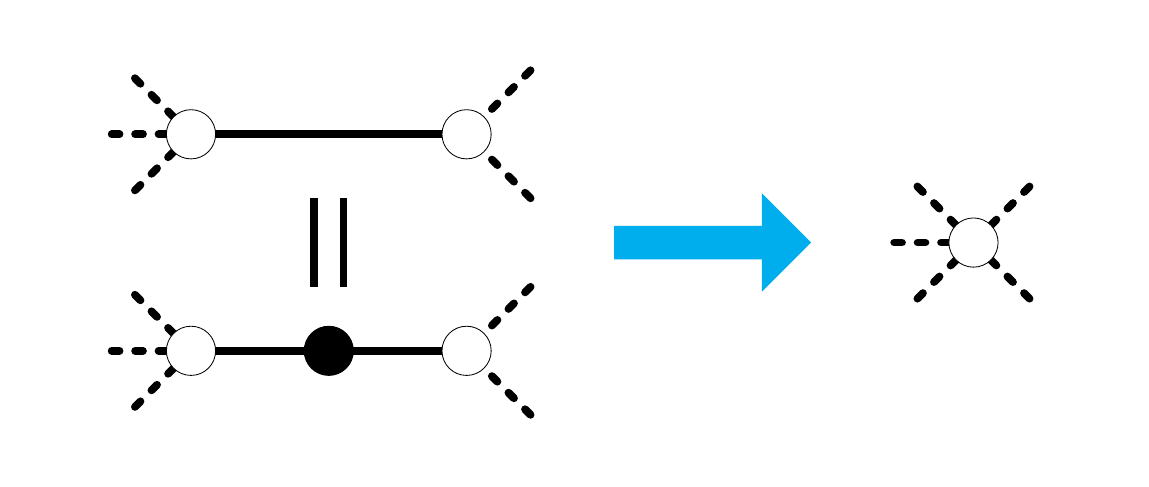}
\vspace{-.5cm}\caption{In a merger move, two connected internal nodes of the same color are condensed. When two internal nodes of the same color are connected by an edge, we can also introduce a 2-valent node of the opposite color between them.}
\label{merger}
\end{center}
\end{figure}

\bigskip

\begin{figure}[h]
\begin{center}
\includegraphics[width=10cm]{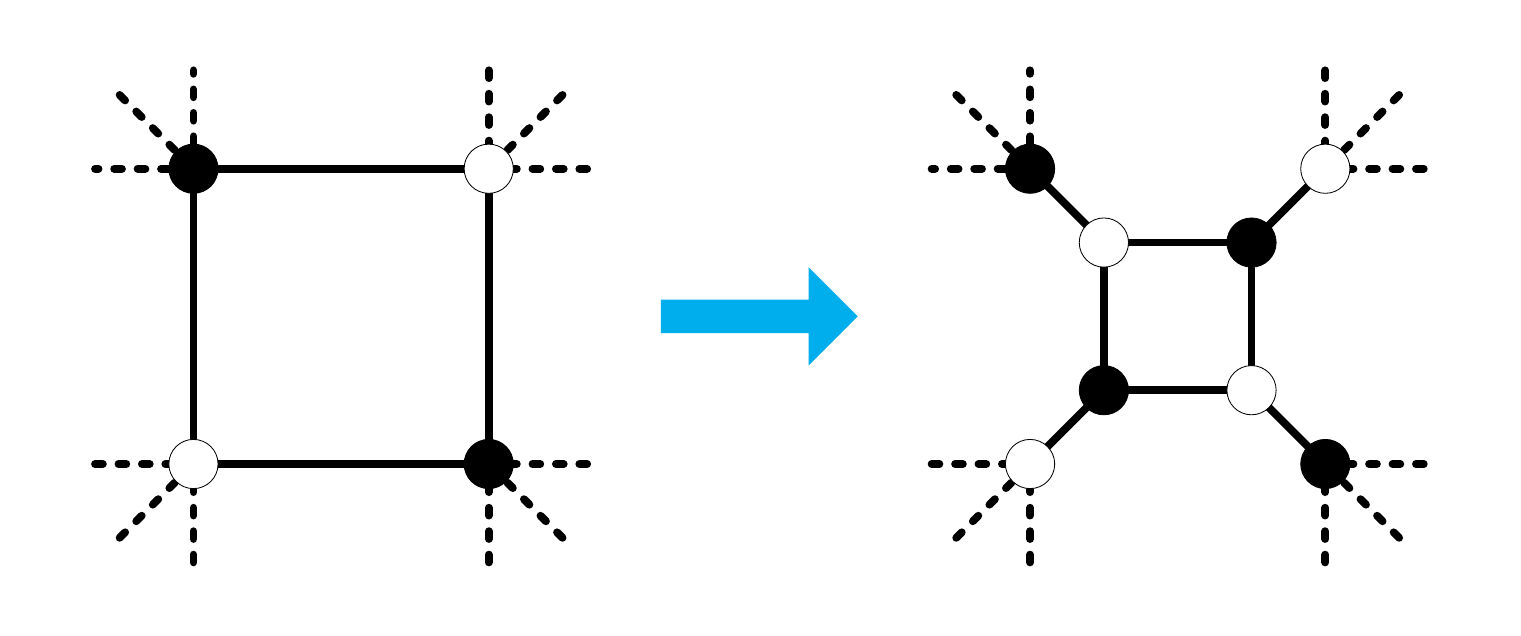}
\vspace{-.5cm}\caption{Square move.}
\label{square_move}
\end{center}
\end{figure}

\noindent {\bf Square Move.} On-shell diagrams are also equivalent under the move shown in \fref{square_move}. We will assume that the square undergoing the move can in fact be any closed loop involving four edges in the graph.

\bigskip

In addition to the equivalence moves discussed above, there is an interesting operation that reduces the number of faces in the graph.

\bigskip

\noindent {\bf Bubble Reduction.} A two-sided face is replaced by a single edge, reducing the number of faces in the graph by one.
\begin{figure}[h]
\begin{center}
\includegraphics[width=10cm]{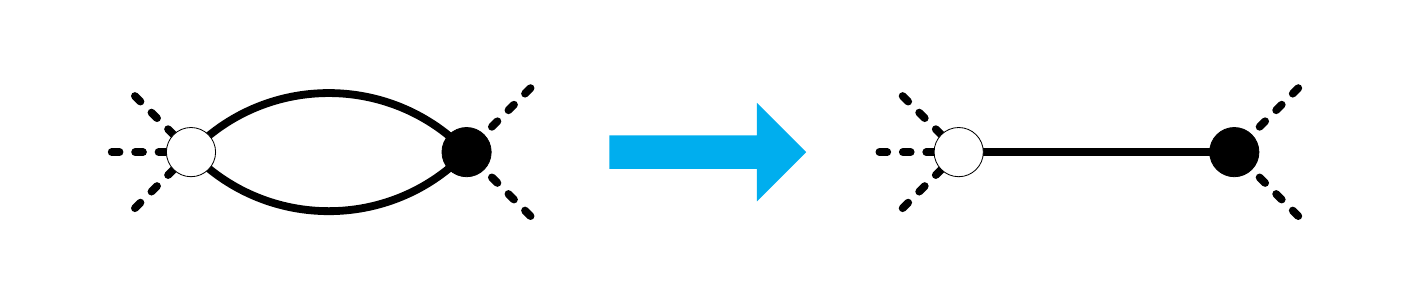}
\vspace{-.5cm}\caption{Bubble reduction.}
\label{bubble_reduction}
\end{center}
\end{figure}

\noindent Bubble reduction reduces the number of degrees of freedom in the diagram by one while preserving the region of the Grassmannian parametrized by it.

\bigskip

More generally, reductions can alternatively be achieved by removing edges. The determination of equivalences and reductions becomes more involved when considering non-planar graphs. For example, a novel feature of non-planar diagrams is that some reductions cannot be achieved by bubble reductions. We will revisit these questions in \sref{section_polytopes} and \sref{sec:reducibility}. 

\bigskip

\subsection{Bipartite Graph Technology and the Boundary Measurement}

Let us now discuss a few additional concepts that are extremely useful in the analysis of bipartite graphs, both planar and non-planar. 

A {\it perfect matching} $p$ is a subset of the edges in the graph, such that every internal node is the endpoint of exactly one edge in $p$ and external nodes belong to one or no edge in $p$. Given a bipartite graph, there is a powerful procedure for obtaining its perfect matchings based on generalized Kasteleyn matrices, certain adjacency matrices of the graph \cite{Franco:2012mm}. 

It is possible to assign orientations to edges in order to produce a {\it perfect orientation}. A perfect orientation is such that each white vertex has a single incoming arrow and each black vertex has a single outgoing arrow. Perfect orientations are in one-to-one correspondence with perfect matchings: the single edge with a special orientation at each internal node is precisely the corresponding edge contained in the perfect matching \cite{2006math09764P,Franco:2012mm}. 

\begin{figure}[h]
\begin{center}
\includegraphics[width=13.5cm]{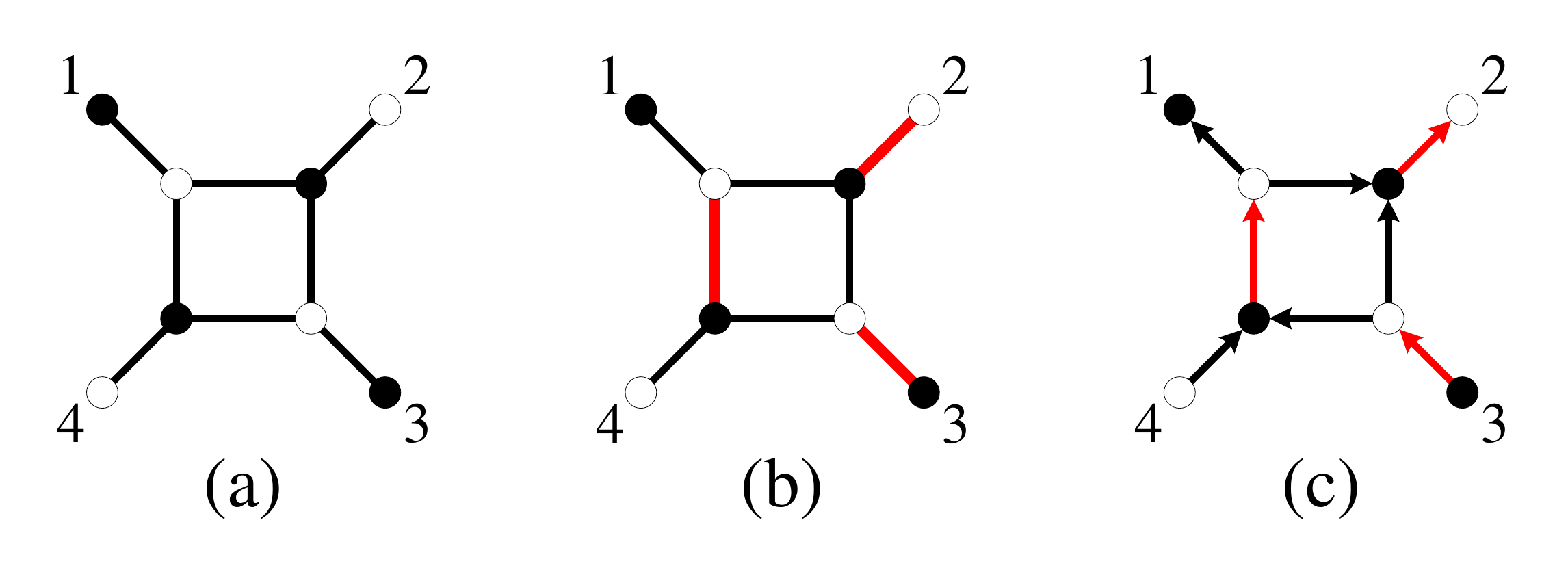}
\vspace{-.4cm}\caption{(a) The graph for $\A^{\text{MHV}}_4$, (b) a choice of a possible perfect matching is shown in red and (c) the perfect orientation associated to it. Here 3 and 4 are the sources while 1 and 2 are the sinks.}
\label{pm_and_perfect_orientation}
\end{center}
\end{figure}

Given a perfect orientation, external nodes are divided into sources and sinks, as shown in the example in \fref{pm_and_perfect_orientation}. We will now explain how bipartite graphs parametrize $Gr_{k,n}$. In this map, $k$ is the number of sources and $n$ is the total number of external nodes in any perfect orientation. This provides us with an alternative way for deriving \eref{k_trivalent} for general graphs. 

We now have all the necessary ingredients for constructing the {\it boundary measurement}, which maps edge weights on the on-shell diagram to a $k\times n$ matrix $C$ in $Gr_{k,n}$ \cite{2006math09764P}. More rigorously, the boundary measurement is constructed in terms of {\it oriented edge weights}; a thorough discussion of this issue appears in \cite{Franco:2013nwa}. The entries of the matrix $C$ are given by
\begin{equation} \label{eq:genericentries}
C_{ij}(X)=\sum_{\Gamma \in \{i \rightsquigarrow j\}}(-1)^{s_\Gamma}\prod_{e\,\in\, \Gamma}X_e\ ,
\end{equation}
where $i$ runs over the sources, $j$ runs over all external nodes and $\Gamma$ is an oriented path from $i$ to $j$. For two sources $i_1$ and $i_2$, this definition results in $C_{i_1 i_2}=\delta_{i_1 i_2}$. Here $X_e$ indicates edge weights oriented along the perfect orientation. In what follows, we will adopt the convention in which oriented edge weights go from white to black nodes. As a result, some edge weights will appear in the numerator or denominator of the previous expressions depending on whether their orientation coincides or opposes that of the corresponding path, respectively. Finally, $(-1)^{s_\Gamma}$ is a crucial sign depending on the details of each path. We postpone its discussion to \sref{sec:general-boundary-measurement}, where we will introduce the boundary measurement for arbitrary graphs, generalizing all cases considered so far in the literature.

\begin{figure}[h]
\begin{center}
\includegraphics[scale=.8]{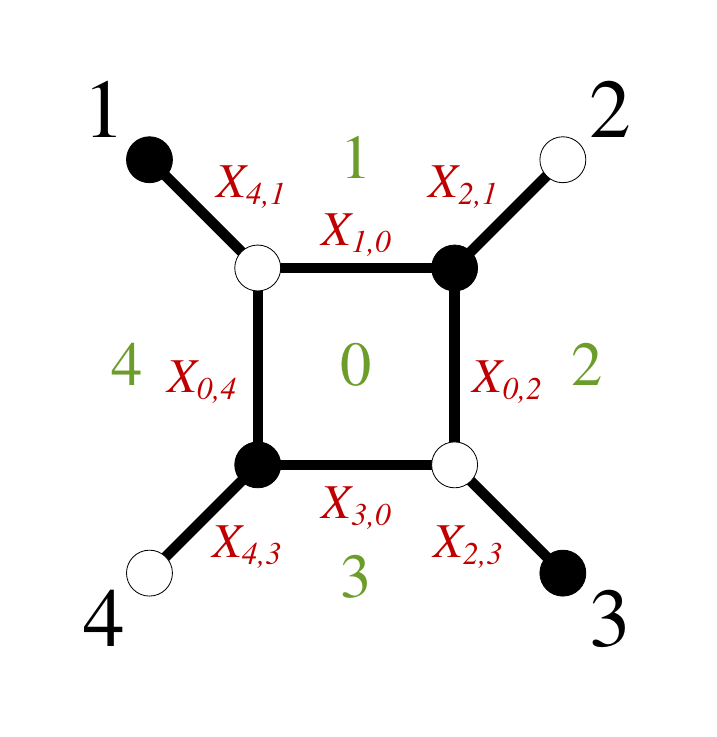}
\vspace{-.3cm}
\caption{On-shell diagram for the tree-level four-point MHV amplitude $\A^{\text{MHV}}_4$. The number of degrees of freedom is $d=4$. Faces are labeled in green, external nodes in black and edges in red.}
\label{G24_edges_and_faces}
\end{center}
\end{figure}

In order to illustrate these ideas, let us consider the simple example shown in \fref{G24_edges_and_faces}. In terms of edge and face variables, the boundary measurement for this graph becomes:
{\small
\begin{align}
\begin{split}
\hspace{-0.5cm}C(X)=\left(\begin{array}{c|cccc} & \ 1 & 2 & 3 & 4 \\ \hline
\Gape[15pt][10pt]
3 \ \ & \dfrac{X_{3,0} X_{4,1}}{X_{2,3} X_{0,4}} & \dfrac{X_{0,2} }{X_{2,3} X_{2,1}} + \dfrac{X_{3,0}X_{1,0}}{X_{2,3} X_{0,4} X_{2,1}} & \ 1 & \ 0 \\ [10pt]
4 \ \ & \ \dfrac{X_{4,3} X_{4,1} }{X_{0,4}} \ & \ \dfrac{X_{4,3} X_{1,0} }{ X_{0,4}  X_{2,1} }  \ & \ 0 & \ 1 \end{array}\right) 
\Rightarrow\, & C(f)=\left(\begin{array}{c|cccc} & \ 1 & 2 & 3 & 4 \\ \hline
3 \ \ & f_0 f_1 f_2  & f_2+f_0 f_2 & \ 1 & \ 0 \\  [.2cm]
4 \ \ & \ f_0 f_1 f_2 f_3 \ & \ f_0 f_2 f_3 \ & \ 0 & \ 1 \end{array}\right)
\end{split}
\end{align}}

As explained above, using the GL($1$) gauge symmetries associated to the the internal nodes, the edge variables in the previous expression can be expressed in terms of $d=4$ independent ones.

\bigskip

\section{Non-Planar On-Shell Diagrams and Non-Adjacent BCFW Shifts} \label{section:non-adjacentBCFW}

Before embarking into a fully general investigation of non-planar on-shell diagrams in the coming sections, we would like to collect a few thoughts about a concrete scenario in which non-planar on-shell diagrams appear and are important: the computation of tree-level amplitudes in $\mathcal{N}=4$ SYM via non-adjacent BCFW shifts \cite{Britto:2005fq}.

\begin{figure}[h]
\begin{center}
\includegraphics[width=14cm]{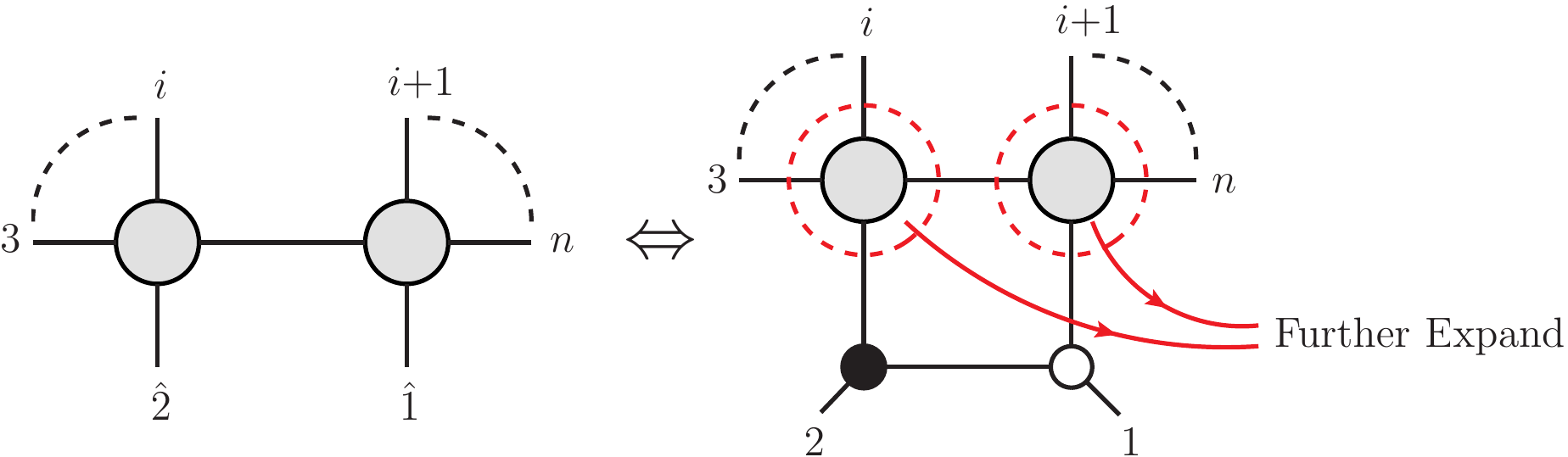}
\vspace{0 cm}\caption{A one-to-one correspondence between a BCFW diagram with an adjacent shift and a two-mass-hard box. The tree-level amplitudes in the two massive corners can be further expanded into two-mass-hard boxes until reaching an on-shell diagram representation of the BCFW diagram.}
\label{fig:adjacentBCFW}
\end{center}
\end{figure}

It is a well known fact that there is a one-to-one correspondence between the quadruple cut of a two-mass-hard box and a BCFW diagram with adjacent shifts \cite{Roiban:2004ix}, as shown in \fref{fig:adjacentBCFW}. In fact, this is how the BCFW recursion relations for tree-level amplitudes were originally derived in \cite{Britto:2004ap}. As emphasized in the figure, one can further recursively express the tree-level amplitudes entering the two massive corners of the box in terms of two-mass-hard boxes, obtaining a representation of the BCFW diagram with adjacent BCFW shifts in terms of on-shell diagrams.

\begin{figure}[h]
\begin{center}
\includegraphics[width=13cm]{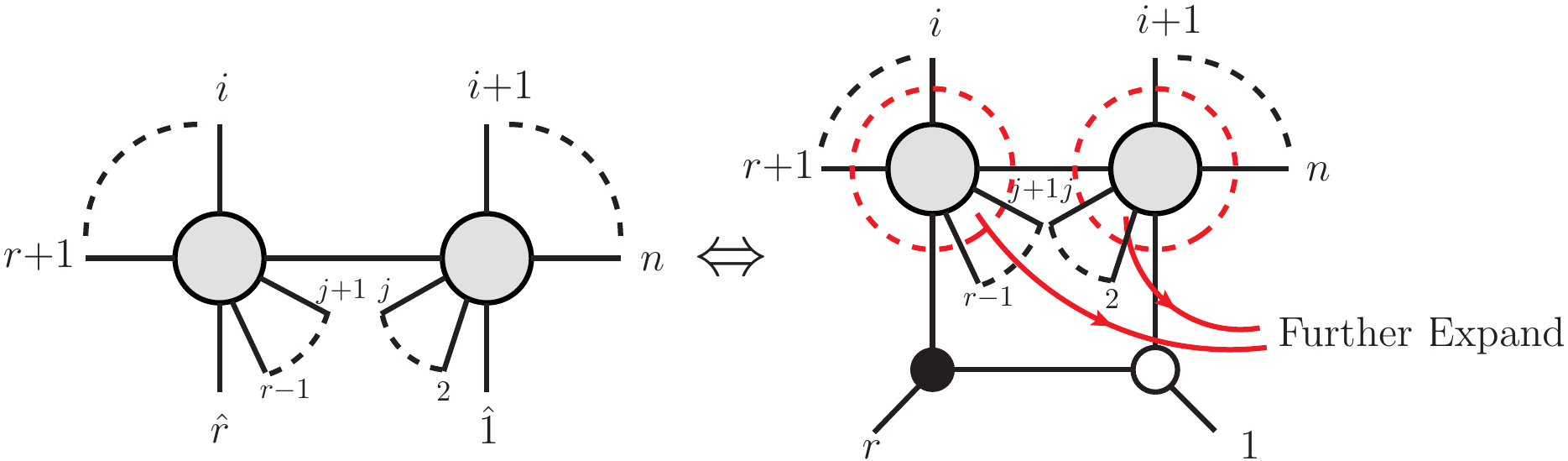}
\vspace{0 cm}\caption{A one-to-one correspondence between a BCFW diagram with non-adjacent shifts and a non-planar two-mass-hard box. The tree-level amplitudes at two massive corners can be further expanded into either non-planar or planar two-mass-hard boxes until reaching an on-shell diagram representation of the BCFW diagram.}
\label{fig:nonadjacentBCFW}
\end{center}
\end{figure}

Since tree-level amplitudes can also be expressed in terms of BCFW diagrams with non-adjacent shifts, it is natural to wonder whether there is a corresponding on-shell diagram representation. Indeed, such a representation exists and the resulting objects are precisely non-planar on-shell diagrams. Similarly to what happens for BCFW diagrams with adjacent shifts, there is a one-to-one correspondence between a BCFW diagram with non-adjacent shifts and a non-planar two-mass-hard box, as shown in \fref{fig:nonadjacentBCFW}. Once again, the tree-level amplitudes in the two massive corners can be further expanded into two-mass-hard boxes, either planar or non-planar. Doing this recursively, we can express any BCFW diagram with non-adjacent shifts in terms of non-planar on-shell diagrams.

\begin{figure}[h]
\begin{center}
\includegraphics[width=14cm]{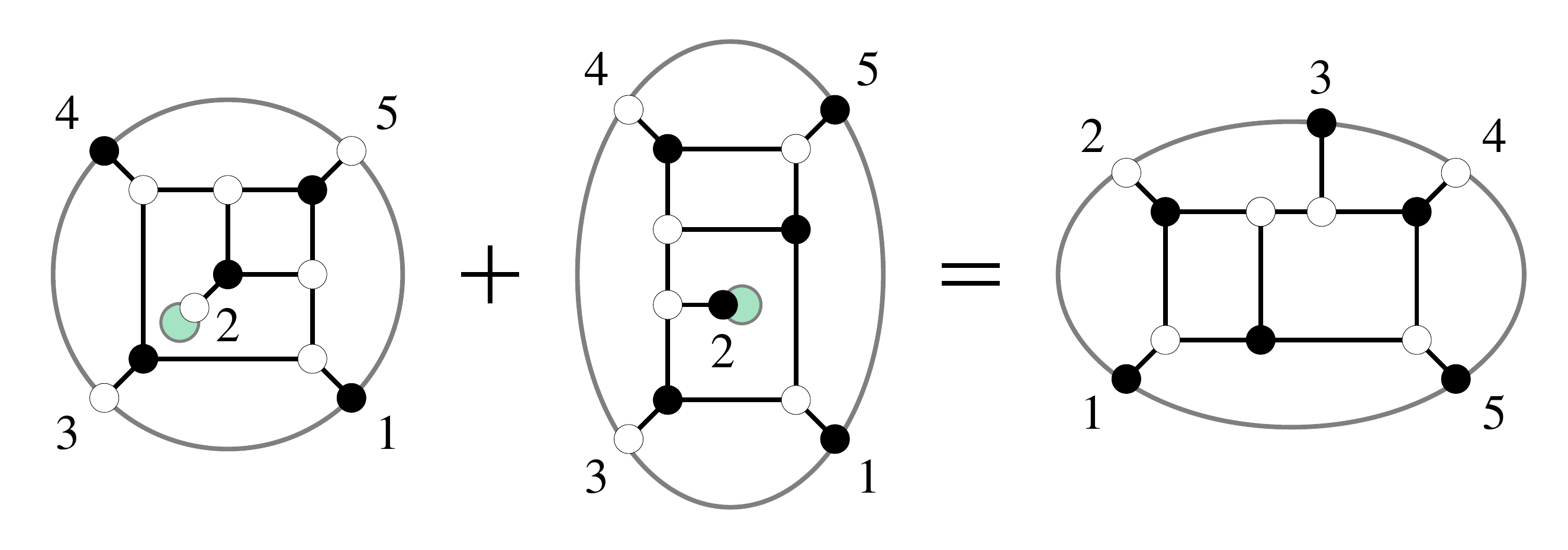}
\vspace{0 cm}\caption{Tree-level five-point MHV amplitude in terms of non-planar on-shell diagrams (left) and a planar on-shell diagram (right).}
\label{fig:fivept}
\end{center}
\end{figure}

It is possible to represent a given amplitude in terms of different on-shell diagrams obtained via different BCFW shifts. This procedure thus generates interesting identities between on-shell diagrams. We present an example of such an identity in \fref{fig:fivept}, where we provide two alternative expressions for the tree-level five-point MHV amplitude $\mathcal{A}^{\rm MHV}_5$. One of the expressions involves two non-planar diagrams and the other one involves a single planar diagram. Furthermore, it is known that there are additional relations between BCFW diagrams with non-adjacent shifts due to the so-called bonus relations \cite{ArkaniHamed:2008gz,Spradlin:2008bu,Feng:2010my}; it would be interesting to explore their application to non-planar on-shell diagrams. Finally, it would be interesting to investigate how general the construction of non-planar on-shell diagrams in terms of non-adjacent BCFW shifts can be.

\bigskip

\section{Generalized Face Variables}

\label{sec:generalized-faces}

In the coming three sections, we will develop new tools for systematically studying non-planar on-shell diagrams. Although many of these ideas have already appeared in the literature in various forms \cite{Franco:2012wv,Franco:2013nwa}, their presentation as a comprehensive set of tools for dealing with non-planar on-shell diagrams is new. 

In this section we begin by introducing canonical variables capturing the degrees of freedom of arbitrary graphs. These variables have the nice property of being invariant under the GL($1$) gauge symmetries associated to all internal nodes, hence being a generalization of the face variables for planar graphs.

\bigskip

\subsection{Embedding Into a Riemann Surface} 

\label{section_embedding}

A useful auxiliary step for identifying generalized face variables is embedding the on-shell diagram into a bordered Riemann surface. While only the connectivity of an on-shell diagram matters, we would like to emphasize that considering such an embedding is very convenient. Given a graph, the choice of embedding is not unique. However we will later see that, as expected, physical results are independent of it.

It is interesting to notice that a choice of embedding is already implicit in the usual discussion of planar diagrams. Indeed, face variables are not an intrinsic property of planar graphs, but arise when imagining them to be embedded on a disk. Similarly, the discussion of zig-zag paths, which are tightly related to the concept of permutations, also depends on assuming planar graphs are embedded on a disk. In fact, as we will see in explicit examples, other embeddings are possible, they lead to different variables, but the final answers remain the same.

In the coming sections, we will present several explicit examples of graph embeddings and their applications.

\bigskip

\subsection{Canonical Variables for Non-Planar Diagrams: Generalized Faces} 

\label{section_generalized_faces}

Generalizing the result for planar graphs, the boundary measurement for generic on-shell diagrams can be constructed in terms of oriented paths in an underlying perfect orientation. Physical answers are independent of the particular choice of perfect orientation. It is convenient to describe such paths in terms of a basis, and this can be done by constructing the {\it generalized face variables} introduced in this section.  Here we will briefly review the ideas introduced in \cite{Franco:2012wv}. The first step, as discussed in \sref{section_embedding}, is to embed the graph into a bordered Riemann surface. Once this is done, we can associate to the the diagram $F$ faces, $B$ boundaries and a genus $g$. These ingredients are sufficient to construct the basis, as follows:

\begin{itemize}

\item {\bf Faces:} A variable $f_i$, $i=1,\ldots, F$, is introduced for every path going clockwise around a face, either internal or external. Face variables satisfy
\beq
\prod_{i=1}^F f_i= 1. \nonumber
\eeq
Hence, one of the face variables can always be expressed in terms of the others. For graphs with boundaries, which are the relevant ones for scattering amplitudes, a useful convention is to discard one of the external faces.

\item {\bf Cuts between boundaries:} For $B > 1$, it is necessary to introduce $B-1$ paths, which we call $b_a$, $a=1,\ldots, B-1$, stretching between different boundary components. The particular choice of these $B-1$ paths, i.e.\ how we chose the pairs of boundaries to be connected by them, is unimportant. We will often refer to them as cuts.\footnote{These cuts have nothing to do with the familiar notion of cutting propagators. We hope the reader is not confused by our choice of terminology.}

\item {\bf Fundamental cycles:} For genus $g$ we need to consider $\alpha_m$ and $\beta_m$ pairs of variables, $m=1,\ldots g$, associated to the fundamental cycles in the underlying Riemann surface.

\end{itemize}
The paths $b_a$, $\alpha_\mu$ and $\beta_\mu$ are expressed as products of oriented edge weights in the same way as for $f_i$.\footnote{It is important to note that the definition of these variables, which correspond to oriented paths, {\it  does not} require an underlying perfect orientation. In fact, the orientation of edges in these paths typically does not agree with the one in any perfect orientation.} Furthermore, they are not unique and can be deformed.

These precisely contain all of the degrees of freedom $d$ of a general on-shell diagram, which is simply determined by 
\beq
d\,=\,F+B+2g-2 \, .
\label{d_general}
\eeq
There is a simple way of understanding the origin of this expression. Notice that for an on-shell diagram with $E$ edges and $N$ internal nodes, we have $d=E-N$. Now, let us consider an embedding of the diagram with Euler characteristic $\chi$ and such that the diagram gives rise to $F$ faces. Since $\chi=F-E+N$, we obtain the compact expression
\beq
d=F-\chi \, ,
\eeq
which agrees with \eref{d_general}.

\bigskip

\subsubsection{The $\boldsymbol{d}$log Form} 

An important feature of on-shell diagrams is the $d\log$ form of the on-shell form, which arises automatically when using generalized face variables, without the need for solving for the GL($1$) redundancies associated to internal nodes when using edge variables.\footnote{The expression of the on-shell form in terms of edge variables \eref{eq:edges} remains valid for non-planar diagrams.} For planar diagrams, it is simply given by
\beq
\Omega = \prod_{i=1}^{F-1} \dfrac{d f_i}{f_i} .
\label{form_face_variables}
\eeq
For arbitrary diagrams, this expression beautifully generalizes to 

\beq
\Omega = \prod_{i=1}^{F-1} \dfrac{d f_i}{f_i} \ \prod_{a=1}^{B-1} \dfrac{d b_a}{b_a} \ \prod_{m=1}^{g} \dfrac{d\alpha_m}{\alpha_m} \ \dfrac{d\beta_m}{\beta_m}
\label{general_integrand_face_variables}
\eeq
when using generalized faces variables. The general form in \eref{general_integrand_face_variables} is an embedding-independent statement, since ultimately it is only the connectivity of the graph which is of importance. 

Appendix \ref{section_simple_example} illustrates embedding independence in a very simple example: a box diagram embedded on a disk and on an annulus. By flipping an external leg, we lose the internal face but give rise to an additional boundary, which in turn produces a new cut. The independent set of generalized face variables would then go from $\{ f_1, f_2, f_3, f_4 \}$ to  $\{ f_1, f_2, f_3, b_1 \}$. The on-shell form, in both sets of variables, becomes

\begin{equation}
\dfrac{d f_1}{f_1}\dfrac{d f_2}{f_2}\dfrac{d f_3}{f_3}\dfrac{d f_4}{f_4} = \dfrac{d f_1}{f_1}\dfrac{d f_2}{f_2}\dfrac{d f_3}{f_3}\dfrac{d b_1}{b_1} \; .
\end{equation}

If instead of using generalized face variables we are interested in expressing the on-shell form in terms of minors of $C$, which is only possible for reduced graphs, it takes the generic form
\begin{equation}
\label{eq:F}
\Omega\,=\, \frac{d^{k\times n} C}{\text{Vol(GL}(k))}\,\frac{1}{(1 \cdots k)(2 \cdots k+1) \cdots (n \cdots k-1)} \times \mathcal{F} ,
\end{equation}
where the non-trivial factor $\mathcal{F}$ accounts for the non-planarity of the on-shell diagram. Explicit examples with non-trivial $\mathcal{F}$ factors will be presented in \sref{sec:computeform}.
\bigskip

\subsection{A Genus-One, $\boldsymbol{B=2}$ Example} \label{genus_1_example_generalized_faces}

\begin{figure}[h]
\begin{center}
\includegraphics[scale=0.65]{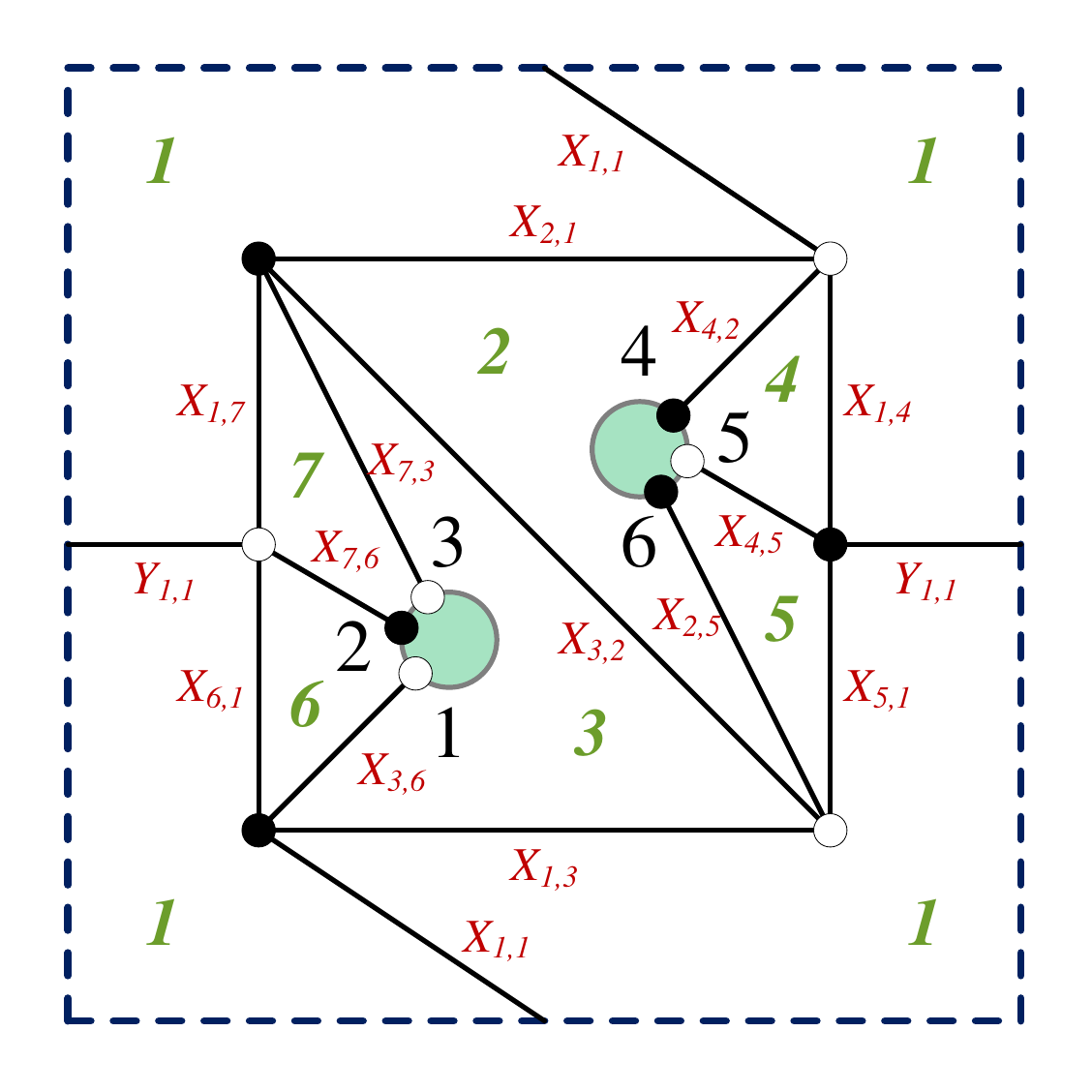}
\vspace{0cm}\caption{A reduced on-shell diagram embedded into a torus with two boundaries. This graph cannot be embedded on any surface with $g=0$. Faces are labeled in green, external nodes in black and edges in red.}
\label{fig:nonplanarizabletorus1}
\end{center}
\end{figure}

In order to understand how generalized face variables work, it is enlightening to study an explicit example. Let us consider the on-shell diagram embedded on a torus with two boundaries shown in \fref{fig:nonplanarizabletorus1}. This diagram does not admit any $g=0$ embedding. Moreover, it is \textit{reduced}, as can be verified using the tools we will present in \sref{sec:reducibility}.

\begin{figure}[h]
\begin{center}
\includegraphics[width=15cm]{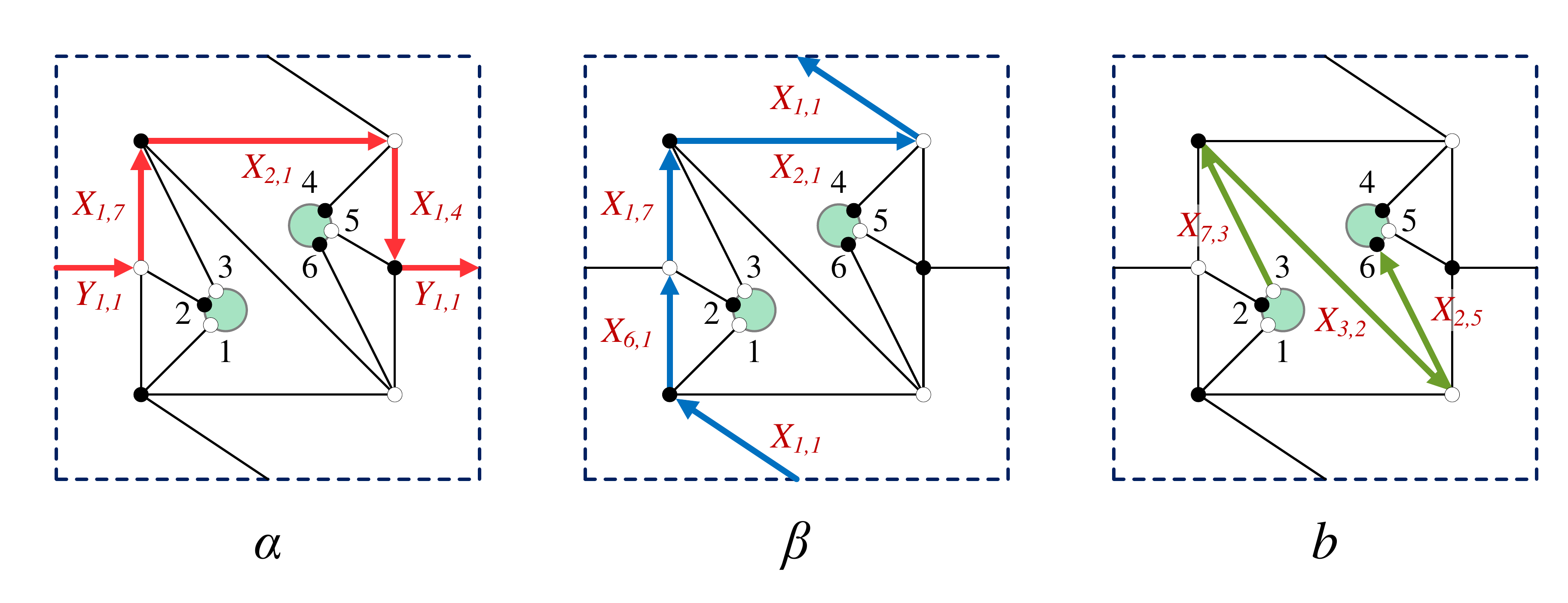}
\vspace{-0.3cm}\caption{Possible choices of the $\alpha$, $\beta$ and $b$ variables.}
\label{alphabetab1}
\end{center}
\end{figure}

This diagram is particularly interesting, since it exhibits the two new types of variables we introduced: cuts and fundamental cycles. Since the diagram is embedded into a torus, there is a pair of variables $\alpha$ and $\beta$ corresponding to its fundamental directions. In addition, there is a cut $b$ connecting the two boundaries. \fref{alphabetab1} shows {\it a possible} set of these variables. As we mentioned earlier, the choice of these paths is not unique. In terms of edges, they are given by
 
 \beq
 \alpha =  \frac{X_{1,7} X_{1,4}}{ Y_{1,1}X_{2,1}} \ \ \ \ \ \ \ \ \beta =  \frac{X_{1,1} X_{1,7} }{ X_{6,1} X_{2,1}} \ \ \ \ \ \ \ \ b =  \frac{X_{7,3} X_{2,5}}{ X_{3,2}} 
 \label{alpha_beta_b_genus_1}
 \eeq

 In addition, the ordinary faces are
 \beq
 \begin{array}{cclccclcccl}
 f_1 & = & \dfrac{X_{2,1} X_{5,1} X_{6,1}}{X_{1,3} X_{1,4} X_{1,7}} & \quad \quad \quad \quad &  f_2 & = & \dfrac{X_{3,2}X_{4,2}}{X_{2,5}X_{2,1}}  & \quad \quad \quad \quad &  f_3 & = & \dfrac{X_{7,3}X_{1,3}}{X_{3,2}X_{3,6}} \\ [.5cm]
f_4 & = & \dfrac{X_{1,4}}{X_{4,2}X_{4,5}}  & & f_5 & = & \dfrac{X_{4,5}X_{2,5}}{X_{5,1}} & &  f_6 & = & \dfrac{X_{3,6}}{X_{6,1}X_{7,6}}  \\ [.5cm]
& & & & f_7 & = & \dfrac{X_{1,7}}{X_{7,6}X_{7,3}} & & & &  
 \end{array}
 \label{faces_genus_1}
 \eeq
 The faces satisfy $\prod_{i=1}^7 f_i=1$ so, without loss of generality, we can discard $f_7$. Interestingly, this example also serves to illustrate some non-trivial feature. Face $f_1$ overlaps with itself over two edges, $X_{1,1}$ and $Y_{1,1}$. This implies that when we circle $f_1$ completely in the clockwise orientation, we transverse each of these edges twice, each time in opposite directions. As a result, the contributions of both edges to $f_1$ cancel out.
 
It is possible to gauge fix the GL(1) redundancies of the 6 internal nodes by setting to 1 one edge for each of them. One consistent way of picking these edges corresponds to setting\footnote{For planar diagrams, this way of fixing the gauge fits nicely into the construction of the diagrams in terms of BCFW bridges \cite{ArkaniHamed:2012nw}. It is interesting to mention that other natural ways of gauge fixing exist. For example, it is possible to treat all edges symmetrically by demanding that the product of edges at every internal node is equal to 1. }
\beq
X_{7,6}=X_{3,6}=X_{4,5}=X_{4,2}=X_{1,3}=X_{1,7}=1.
\label{gauge_fixing_genus_1}
\eeq
The remaining edges are
\beq
X_{1,1}, \, X_{1,4}, \, X_{2,1}, \, X_{2,5}, \, X_{3,2}, \, X_{5,1}, \, X_{6,1}, \, X_{7,3}, \, Y_{1,1} .
\label{free_edges_genus1}
\eeq
We thus conclude that this on-shell diagram has $d=9$ degrees of freedom. Following \sref{sec:generalized-faces}, this counting of course agrees with the one based on generalized face variables; we have: 7 faces (6 of which are independent), an $\alpha$ and a $\beta$ cycle from being on a torus and $B-1=1$ cut. 

After this gauge fixing, the independent generalized face variables become
\beq
 \begin{array}{cclccclcccl}
 f_1 & = & \dfrac{X_{2,1} X_{5,1} X_{6,1}}{X_{1,4}} & \quad \quad \quad \quad &  f_2 & = & \dfrac{X_{3,2}}{X_{2,5}X_{2,1}}  & \quad \quad \quad \quad &  f_3 & = & \dfrac{X_{7,3}}{X_{3,2}} \\ [.5cm]
f_4 & = & X_{1,4}  & & f_5 & = & \dfrac{X_{2,5}}{X_{5,1}} & &  f_6 & = & \dfrac{1}{X_{6,1}}
\\ [.5cm]
\alpha & =  & \dfrac{X_{1,4}}{Y_{1,1}X_{2,1}} & & \beta & = & \dfrac{X_{1,1}}{X_{6,1} X_{2,1}} & & b & = & \dfrac{X_{7,3} X_{2,5}}{X_{3,2}} 
 \end{array}
 \label{generalized_faces_gauge_fixed_genus1}
 \eeq
 If desired, this map can be inverted, obtaining
 \beq
 \begin{array}{cclccclcccl}
X_{1,1}&=&\dfrac{\beta f_1 f_3 f_4 f_5}{b} & \quad \quad \quad \quad & X_{1,4}&=&f_4& \quad \quad \quad \quad & X_{2,1}&=&\dfrac{f_1 f_3 f_4 f_5 f_6}{b} \\ [.25cm]
X_{2,5}&=&\dfrac{b}{f_3} & & X_{3,2}&=&f_1 f_2 f_4 f_5 f_6  & & X_{5,1}&=&\dfrac{b}{f_3 f_5} \\ [.25cm]
X_{6,1}&=&\dfrac{1}{f_6} & & X_{7,3}&=&f_1 f_2 f_3 f_4 f_5 f_6 & & Y_{1,1}&=&\dfrac{b}{\alpha f_1 f_3 f_5 f_6}
 \end{array}
 \eeq
 
Let us now translate the boundary measurement from the edge variables in \eref{free_edges_genus1} to generalized face variables. It becomes
\beq
\begin{array}{ccl}
\Omega &= &\dfrac{d X_{1,1}}{X_{1,1}} \dfrac{d X_{1,4}}{X_{1,4}} \dfrac{d X_{2,1}}{X_{2,1}} \dfrac{d X_{2,5}}{X_{2,5}} \dfrac{d X_{3,2}}{X_{3,2}} \dfrac{d X_{5,1}}{X_{5,1}} \dfrac{d X_{6,1}}{X_{6,1}} \dfrac{d X_{7,3}}{X_{7,3}} \dfrac{d Y_{1,1}}{Y_{1,1}} \\ [.5cm]
 &= &\dfrac{f_1^2 f_2 f_4^4 f_5}{\alpha^2 f_3} \; \dfrac{\alpha}{b \beta f_1^3 f_2^2 f_4^5 f_5^2 f_6} \; d f_1\; d f_2\; d f_3\; d f_4\; d f_5\; d f_6\; d \alpha\; d \beta\; d b  \\ [.5cm]
 &= &\dfrac{d f_1}{f_1} \, \dfrac{d f_2}{f_2} \, \dfrac{d f_3}{f_3} \, \dfrac{d f_4}{f_4} \, \dfrac{d f_5}{f_5} \, \dfrac{d f_6}{f_6} \, \dfrac{d \alpha}{\alpha} \, \dfrac{d \beta}{\beta} \, \dfrac{d b}{b}
\end{array}
\eeq
where, in the middle line, the first factor comes from the Jacobian of the variable transformation and the second factor comes from the product of edge variables. We see that the on-shell form takes the general form in \eref{general_integrand_face_variables}. In other words, generalized variables can be used to directly write the on-shell form in $d\log$ form without having to work through the GL(1) gauge fixing that is necessary for arriving at \eref{free_edges_genus1}.

\begin{figure}[h]
\begin{center}
\includegraphics[width=5cm]{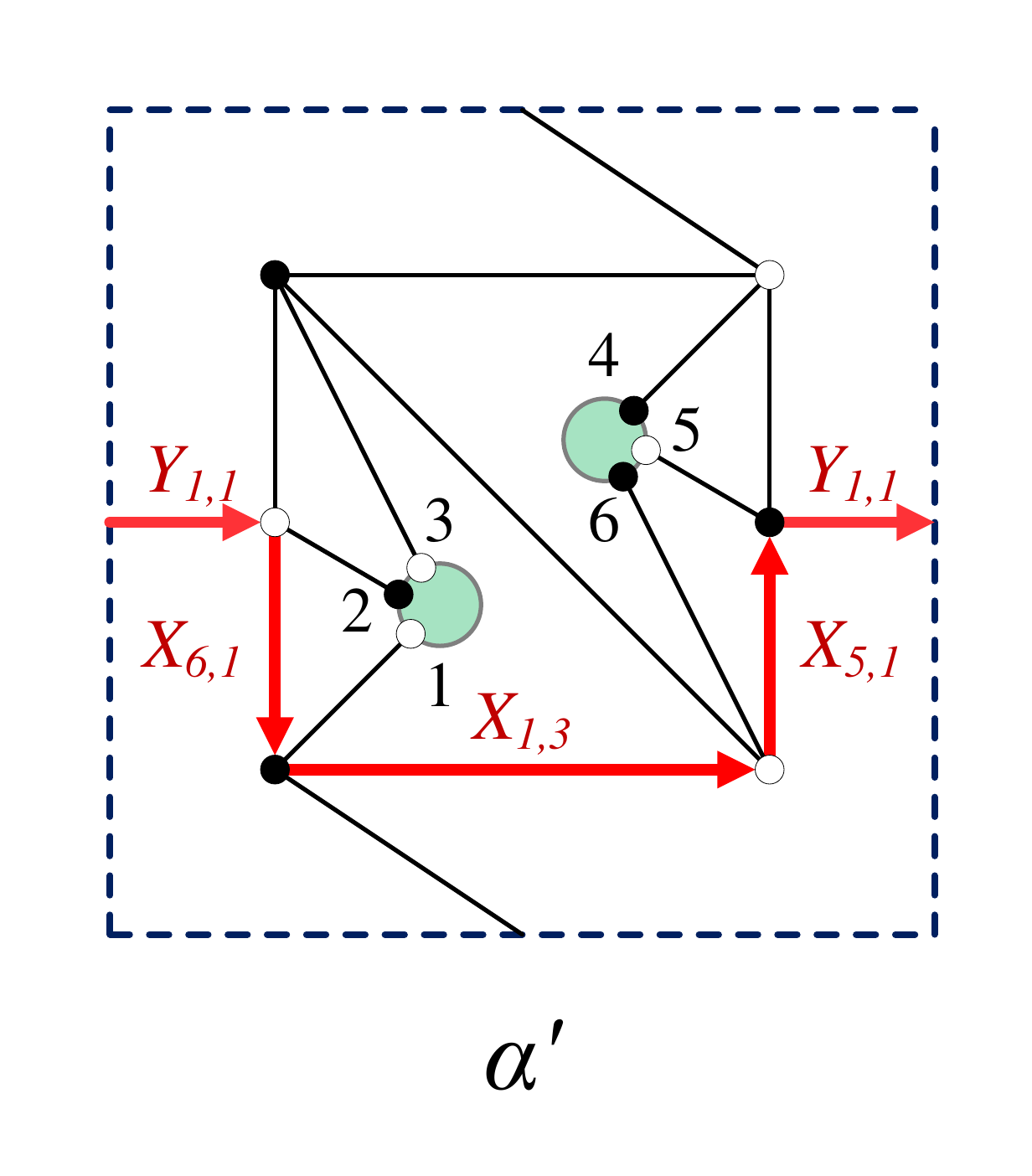}
\vspace{-0.3cm}\caption{An alternative choice for one of the fundamental cycles of the torus. The Jacobian of the change of variables is such that the on-shell form preserves its $d\log$ in terms of generalized face variables.}
\label{alphaprime}
\end{center}
\end{figure}

It is also easy to verify that the $d\log$ form of the on-shell form is independent of the explicit choice of generalized face variables. For example, we could trade $\alpha$ for another path $\alpha'$ also wrapping the torus along the horizontal direction, such as the one shown in \fref{alphaprime}. Once again, the Jacobian of the change of variables is such that the $d\log$ form is preserved.

We will investigate additional aspects of this example in \sref{sref_example_polytopes} and \sref{sec:nonplanarizable}. 

\bigskip

\section{Combinatorial Characterization of Non-Planar Diagrams: Generalized Matching and Matroid Polytopes} \label{sec:matchingmatroid}

\label{section_polytopes}

Finding a combinatorial classification of non-planar on-shell diagrams is a question of crucial importance. In this section we introduce two combinatorial objects, the generalized {\it matching} and {\it matroid polytopes}, which allow us to make substantial progress towards this goal. They are the natural generalizations of the matching and matroid polytopes that appear in the study of planar diagrams \cite{2006math09764P,Postnikovlectures,2007arXiv0706.2501P}. In fact, these objects have been extensively discussed, together with their application to the classification of non-planar diagrams, in \cite{Franco:2012mm,Franco:2012wv,Amariti:2013ija,Franco:2013nwa,Franco:2014nca}. In order to avoid unnecessary repetition, our presentation will be succinct, referring the interested reader to \cite{Franco:2012mm,Franco:2012wv,Franco:2013nwa,Franco:2014nca} for details.

We begin with a constructive definition of the polytopes in the next section and then summarize their more salient features for our purposes.

\bigskip

\subsection{Constructing the Polytopes}

\label{section_constructing_polytopes}

There are multiple ways of constructing the generalized matching and matroid polytopes associated to a given on-shell diagram \cite{Franco:2012wv,Franco:2013nwa}. Here we review two of them. The first one is based on the connection between edges and perfect matchings in the graph. The second method is based on generalized face variables. 

\bigskip

\subsection*{Matching Polytope}

\bigskip

\paragraph{$\bullet$ Method 1.}

As we already mentioned, given an on-shell diagram, all its perfect matchings can be easily found using generalized Kasteleyn matrices. To find the matching polytope, we construct the $(E\times c)$-dimensional  {\it perfect matching matrix} $P$:
\beq
P_{i\mu}=\left\{ \begin{array}{ccccc} 1 & \rm{ if } & X_i  & \in & p_\mu \\
0 & \rm{ if } & X_i  & \notin & p_\mu
\end{array}\right. ,
\label{P_matrix}
\eeq
where $X_i$, $i=1,\ldots, E$, are the edges and $p_\mu$, $\mu=1,\ldots, c$, are the perfect matchings of the diagram. This matrix defines the matching polytope as follows: there is a point for every perfect matching, with a position vector in $\mathbb{Z}^E$ given by the corresponding column vector. Generically, the dimensionality of the matching polytope is lower than $E$. This can be made manifest by e.g.\ row reducing the matrix $P$.\footnote{The sum of all rows in the row-reduction of $P$ is always equal to $(1,\ldots,1)$ \cite{Franco:2013nwa}, so it is possible to discard one of them without losing any information, effectively reducing the dimensionality of the matching polytope by 1. We provide an explicit example of this phenomenon in \sref{sref_example_polytopes}.} Indeed, the dimensionality of the matching polytope is equal to the number of degrees of freedom of the on-shell diagram. This fact becomes more manifest when considering the alternative method for its determination that we present below. 

\bigskip

\paragraph{$\bullet$ Method 2.}

At this point it is useful to introduce the concept of {\it flow}. Given an on-shell diagram and a perfect orientation on it, its flows correspond to all oriented non self-intersecting paths in it. 
Flows can involve more than one disjoint component. Furthermore, these components can connect external nodes or correspond to closed loops. The trivial flow, i.e. the one which does not involve any edge in the graph, is also included. Every flow $\mathfrak{p}_\mu$ is in one-to-one correspondence with a perfect matching $p_\mu$ and is obtained by subtracting from $p_\mu$ the reference perfect matching $p_{\text{ref}}$ that specifies the perfect orientation.

Generalized face variables form a basis in which we can express any oriented path in the graph and, in particular, we can use them to express flows. 
As for perfect matchings, every flow $\mathfrak{p}_\mu$ maps to a point in the matching polytope. Its coordinates are simply given by the vector of powers of the generalized face variables required to specify $\mathfrak{p}_\mu$:
\begin{align} \label{eq:matchingpolytopecoord}
\mathfrak{p}_\mu &= \prod_{i=1}^{F-1} f_i^{x_{i,\mu}} \prod_{j=1}^{B-1} b_j^{y_{j,\mu}} \prod_{m=1}^{g} \alpha_m^{z_{m,\mu}} \beta_m^{w_{m,\mu}}  \nonumber \\
& \longmapsto \quad \begin{array}{c}{\rm \underline{\text{Coordinate in Matching Polytope}}:} \\ (x_{1,\mu},\ldots,x_{F-1,\mu},y_{1,\mu},\ldots,y_{B-1,\mu},z_{1,\mu},\ldots,z_{g,\mu},w_{1,\mu},\ldots,w_{g,\mu}) \end{array}
\end{align}
Since every flow has a unique description in terms of generalized face variables, it becomes clear that every one of them (and hence every perfect matching) gives rise to a unique point in the matching polytope. 

\bigskip

\subsection*{Matroid Polytope}

The matroid polytope is a projection of the matching polytope that only preserves information on how flows connect to external legs of the graph. Below we explain how to attain this when working in terms of edge variables or generalized face variables.

\bigskip

\paragraph{$\bullet$ Method 1.}

Let us consider a diagram with $F_e$ external legs. Calling this number $F_e$ is motivated by the fact that it is equal to the number of external faces. The matrix whose columns encode the position vectors for points in the matroid polytope is simply obtained by starting from the perfect matching matrix $P$ in \eref{P_matrix} and keeping only the rows corresponding to external legs. It turns out that the points generated by this procedure lie on a hyperplane, so one of the rows can be further eliminated, leading to an $F_e-1$ dimensional polytope \cite{Franco:2013nwa}.

\bigskip

\paragraph{$\bullet$ Method 2.}

The projection onto information regarding external legs can similarly be achieved in terms of generalized face variables. To do so, we eliminate all coordinates associated to internal faces, cuts and $\alpha_m$ and $\beta_m$ cycles, preserving only those coming from external faces. Furthermore, since the product of all ordinary faces equals to 1, one of the external faces can be discarded. The projection from flows to the matroid polytope hence takes the form:
\beq\label{eq:matroidpolytopecoord}
\mathfrak{p}_\mu = \prod_{i=1}^{F-1} f_i^{x_{i,\mu}} \prod_{j=1}^{B-1} b_j^{y_{j,\mu}} \prod_{m=1}^{g} \alpha_m^{z_{m,\mu}} \beta_m^{w_{m,\mu}}  \quad \longmapsto \quad \begin{array}{c}{\rm \underline{\text{Coordinate in Matroid Polytope}}:} \\ (x_{1,\mu},\ldots,x_{F_e,\mu}) \end{array} ,
\eeq
where $F_e$ is the total number of external faces and, without loss of generality, we have ordered faces such that the first $F_e-1$ are external.

\bigskip

Typically, different points in the matching polytope are identified when projected down to the matroid polytope. More concretely, perfect matchings that coincide on external legs are identified under this projection. Equivalently, the same happens for flows differing only by internal paths. It is thus clear that points in the matroid polytope can correspond to multiple perfect matchings/flows. In fact, this has an important physical interpretation. As mentioned earlier, points in the matroid polytope are in one-to-one correspondence with \pl coordinates. The \pl coordinates are in turn expressed as linear combinations of flows with coefficients $\pm 1$ through the boundary measurement. The flows associated to the same point in the matroid polytope are precisely all the contributions to the corresponding \pl coordinate \cite{Franco:2013nwa}. The index of a given \pl coordinate, i.e.\ the set of corresponding columns in the matrix $C$, is equal to the common source set of the flows contributing to it. We will expand on these topics in \sref{sec:general-boundary-measurement}, where we introduce a boundary measurement for arbitrary on-shell diagrams.

\bigskip

\subsection{Graph Characterization: Region Matching and Reductions} 

\label{section_equivalence_and_reductions}

There are an infinite number of on-shell diagrams. It is thus desirable to come up with a classification of them, i.e.\ to endow this plethora of diagrams with some structure and order. Specifically, diagrams can be organized into equivalence classes and related by simplifying operations denoted reductions, all of which are defined in terms of the operations discussed in \sref{sec:review-planar}. The spectrum of possibilities becomes far richer when considering non-planar diagrams. Determining whether two diagrams are equivalent by explicitly constructing a sequence of moves connecting them or establishing the reducibility of a diagram can be challenging tasks, even when dealing with relatively small diagrams. It is hence important to develop {\it global} tools for answering such questions directly from the graph.  Methods for achieving this exist for planar diagrams, see \cite{ArkaniHamed:2012nw} and references therein. In this section we will take important steps towards developing a systematic and combinatorial approach, based on the generalized matching and matroid polytopes, to the classification of general on-shell diagrams, including non-planar ones. For this purpose, it is convenient to define:

\smallskip

\begin{itemize}
\item {\bf Region matching:} This term indicates the case in which the regions of the Grassmannian parametrized by different on-shell diagrams coincide. Two necessary conditions in order for two on-shell diagrams to be equivalent are region matching and having the same number of degrees of freedom.

\item {\bf Reduction:} An on-shell diagram $B$ is a reduction of an on-shell diagram $A$, if it is obtained from $A$ by deleting edges and it covers the same region of the Grassmanian as $A$.
\end{itemize}
\smallskip
\noindent Notice that the definition of reduction given above contains, but is more general than, bubble reduction.

A few words of caution are in order when implementing these definitions. For planar diagrams, the region of the Grassmannian covered by the graph is fully determined by specifying the non-vanishing \pl coordinates. This is however more subtle in the non-planar case, since constraints between \pl coordinates beyond \pl relations might exist, as we discuss in detail in \sref{sec:reducibility}.

The idea of reduction leads to the concept of {\it reduced graph}:

\smallskip

\begin{itemize}
\item {\bf Reduced graph:} A graph is reduced if it is impossible to remove edges from it while covering the same region of the Grassmannian.\footnote{Notice that, in particular, this implies that no additional constraints on \pl coordinates can be generated when searching for possible reductions.}
\end{itemize}

\smallskip

The importance of reduced graphs stems from the fact that there are a finite number of them for every scattering process and they contain all information required for addressing certain questions, e.g.\ determining leading singularities.

\bigskip

\subsection{Combinatorial Implementation in Terms of Polytopes} 

The characterization of on-shell diagrams outlined above has a powerful implementation in terms of matching and matroid polytopes. This application for general graphs was introduced in \cite{Franco:2012mm} and further explored in \cite{Franco:2012wv,Franco:2013nwa,Franco:2014nca}.  In this approach, the necessary map between edges and perfect matchings is determined by the matching polytope. Given the correspondence between \pl coordinates and points in the matroid polytope, the previous definitions admit the following combinatorial translations {\it in the absence of additional constraints on \pl coordinates:}\footnote{In \sref{sec:reducibility}, we will discuss in detail how to deal with such constraints. Indeed, this can be done efficiently in terms of matching and matroid polytopes.}
	
\smallskip

\begin{itemize}
\item {\bf Equivalence:} Two on-shell diagrams parameterize the same region of the Grassmannian if they have the same matroid polytope. 
\item {\bf Reduction:} An on-shell diagram $B$ is a reduction of an on-shell diagram $A$ if it is obtained from $A$ by deleting edges and it has the same matroid polytope of $A$.
\end{itemize}

\smallskip

Similarly, 

\smallskip

\begin{itemize}
\item {\bf Reduced graph:} A graph is reduced if it is impossible to remove edges from it while preserving the matroid polytope and not generating additional non-\pl constraints.
\end{itemize}

\smallskip

\noindent This definition can be exploited as a concrete and algorithmic procedure for checking the reducibility of arbitrary on-shell diagrams. We will return to this problem in \sref{section_systematic_reducibility}.

Interestingly, as discussed at length in \cite{Franco:2013nwa} and reviewed in \sref{section_constructing_polytopes}, every point in the matroid polytope has an associated {\it multiplicity} of perfect matchings/flows. Heuristically, reducibility is accompanied by large multiplicities, which reflect a redundancy of oriented paths between external nodes in a perfect orientation contributing to the boundary measurement. A graph is reducible if edges can be deleted without bringing any multiplicity below one, assuming no new constraints are generated in the process. If the removal of an edge causes a point in the matroid polytope to disappear, the corresponding \pl coordinate is set to zero.

Even for planar graphs, matching and matroid polytopes provide the most comprehensive known characterization of on-shell diagrams. For example, unlike the classification of planar graphs based on permutations \cite{ArkaniHamed:2012nw}, this approach does not require the graphs to be reduced.

More generally, matroid and matching polytopes are useful tools for investigating the effect of deleting edges, i.e.\ even in cases in which their removal do not correspond to a reduction. We will consider a detailed example in \sref{sec:reducibility} and refer the reader to \cite{Franco:2012mm,Franco:2012wv,Amariti:2013ija,Franco:2013nwa,Franco:2014nca} for many more.

Finally, let us mention that non-planar diagrams exhibit new features, such as the already mentioned appearance of non-\pl constraints and non-unique reductions \cite{Franco:2012wv,Arkani-Hamed:2014bca}.

\bigskip

\subsection{Examples}

Here we present some explicit examples in order to illustrate the construction of the matroid and matching polytopes and on how to use them for characterizing on-shell diagrams. Since not all readers are familiar with this type of objects, our discussion will be rather meticulous. We refer the reader to \cite{Franco:2012mm,Franco:2012wv,Franco:2013nwa} for several additional examples worked out in exquisite detail. 

\bigskip

\subsubsection{Polytopes for a Genus-One, $\boldsymbol{B=2}$ Diagram}

\label{sref_example_polytopes}

Let us consider again the on-shell diagram in \fref{fig:nonplanarizabletorus1}, which admits an embedding with genus one and two boundaries. This diagram has 34 perfect matchings, which we have determined using the generalized Kasteleyn matrix techniques introduced in \cite{Franco:2012mm}. They can be encoded into the $P$ matrix defined in \eref{P_matrix}, which is given by

{\footnotesize
\be
P=\left(
\begin{array}{c|cccccccccccccccccccccccccccccccccc}
X_{1,1} \ & \ 0 & 0 & 0 & 0 & 0 & 0 & 0 & 0 & 0 & 0 & 0 & 0 & 0 & 1 & 0 & 1 & 0 & 0 & 0 & 0 & 0 & 1 & 0 & 1 & 0 & 0 & 0 & 0 & 0 & 0 & 1 & 1 & 1 & 0 \\
X_{1,3} \ & \ 1 & 1 & 1 & 1 & 1 & 1 & 1 & 0 & 0 & 0 & 0 & 0 & 0 & 0 & 0 & 0 & 0 & 0 & 0 & 0 & 0 & 0 & 0 & 0 & 0 & 0 & 0 & 0 & 0 & 0 & 0 & 0 & 0 & 0 \\
X_{1,4} \ & \ 1 & 1 & 0 & 0 & 0 & 0 & 0 & 1 & 1 & 1 & 1 & 1 & 0 & 0 & 0 & 0 & 0 & 0 & 0 & 0 & 0 & 0 & 0 & 0 & 0 & 0 & 0 & 0 & 0 & 0 & 0 & 0 & 0 & 0 \\
X_{1,7} \ & \ 1 & 0 & 1 & 0 & 0 & 0 & 0 & 1 & 0 & 0 & 0 & 0 & 1 & 1 & 1 & 1 & 0 & 0 & 0 & 0 & 0 & 0 & 0 & 0 & 0 & 0 & 0 & 0 & 0 & 0 & 0 & 0 & 0 & 0 \\
X_{2,1} \ & \ 0 & 0 & 0 & 0 & 0 & 1 & 1 & 0 & 0 & 0 & 0 & 0 & 0 & 0 & 0 & 0 & 0 & 0 & 1 & 1 & 0 & 0 & 1 & 0 & 0 & 0 & 0 & 1 & 0 & 0 & 0 & 0 & 0 & 1 \\
X_{3,2} \ & \ 0 & 0 & 0 & 0 & 0 & 0 & 0 & 0 & 0 & 0 & 1 & 1 & 0 & 0 & 0 & 0 & 0 & 0 & 0 & 0 & 0 & 0 & 0 & 0 & 1 & 0 & 1 & 0 & 1 & 0 & 1 & 0 & 1 & 0 \\
X_{5,1} \ & \ 0 & 0 & 0 & 0 & 0 & 0 & 0 & 0 & 0 & 0 & 0 & 0 & 0 & 0 & 1 & 1 & 0 & 0 & 0 & 0 & 0 & 0 & 0 & 0 & 0 & 1 & 0 & 1 & 0 & 1 & 0 & 1 & 0 & 1 \\
X_{6,1} \ & \ 0 & 0 & 0 & 0 & 0 & 0 & 0 & 0 & 0 & 1 & 0 & 1 & 0 & 0 & 0 & 0 & 0 & 0 & 0 & 0 & 1 & 0 & 1 & 0 & 0 & 0 & 0 & 0 & 1 & 1 & 0 & 0 & 0 & 1 \\
Y_{1,1} \ & \ 0 & 0 & 0 & 0 & 1 & 0 & 1 & 0 & 0 & 0 & 0 & 0 & 0 & 0 & 0 & 0 & 0 & 1 & 0 & 1 & 0 & 0 & 0 & 1 & 0 & 0 & 1 & 0 & 0 & 0 & 0 & 0 & 1 & 0 \\ \hline
X_{3,6} \ & \ 0 & 0 & 0 & 0 & 0 & 0 & 0 & 1 & 1 & 0 & 1 & 0 & 1 & 0 & 1 & 0 & 1 & 1 & 1 & 1 & 0 & 0 & 0 & 0 & 1 & 1 & 1 & 1 & 0 & 0 & 0 & 0 & 0 & 0 \\
X_{7,6} \ & \ 0 & 1 & 0 & 1 & 0 & 1 & 0 & 0 & 1 & 0 & 1 & 0 & 0 & 0 & 0 & 0 & 1 & 0 & 1 & 0 & 0 & 1 & 0 & 0 & 1 & 1 & 0 & 1 & 0 & 0 & 1 & 1 & 0 & 0 \\
X_{7,3} \ & \ 0 & 1 & 0 & 1 & 1 & 0 & 0 & 0 & 1 & 1 & 0 & 0 & 0 & 0 & 0 & 0 & 1 & 1 & 0 & 0 & 1 & 1 & 0 & 1 & 0 & 1 & 0 & 0 & 0 & 1 & 0 & 1 & 0 & 0 \\
X_{4,2} \ & \ 0 & 0 & 1 & 1 & 1 & 0 & 0 & 0 & 0 & 0 & 0 & 0 & 1 & 0 & 1 & 0 & 1 & 1 & 0 & 0 & 1 & 0 & 0 & 0 & 1 & 1 & 1 & 0 & 1 & 1 & 0 & 0 & 0 & 0 \\
X_{4,5} \ & \ 0 & 0 & 1 & 1 & 0 & 1 & 0 & 0 & 0 & 0 & 0 & 0 & 1 & 1 & 0 & 0 & 1 & 0 & 1 & 0 & 1 & 1 & 1 & 0 & 1 & 0 & 0 & 0 & 1 & 0 & 1 & 0 & 0 & 0 \\
X_{2,5} \ & \ 0 & 0 & 0 & 0 & 0 & 0 & 0 & 1 & 1 & 1 & 0 & 0 & 1 & 1 & 0 & 0 & 1 & 1 & 1 & 1 & 1 & 1 & 1 & 1 & 0 & 0 & 0 & 0 & 0 & 0 & 0 & 0 & 0 & 0 
\end{array}
\right)
\label{P_matrix_genus_1}
\eeq
}

\noindent Every column in this matrix corresponds to a perfect matching. We have organized its rows so that the last six of them correspond to the external legs. 

As explained in \sref{section_constructing_polytopes}, the $P$ matrix defines the matching polytope. Concretely, every perfect matching corresponds to a distinct point in the matching polytope, whose position in $\mathbb{Z}^{15}$ is given by the corresponding column. Here 15 is the number of edges in the graph. The matching polytope lives, however, in a lower dimensional subspace of $\mathbb{Z}^{15}$, which can be easily determined by row-reducing $P$. Doing so, we obtain:

\smallskip
{\footnotesize
\beq
P_{\rm red}=\left(
\begin{array}{ccccccccccccccccc}
 \ \ 1 \ \ & \ \ 0 \ \ & \ \ 0 \ \ & -1 & \ \ 0 \ \ & \ \ 0 \ \ & 1 & 0 & -1 & \ \ 0 \ \ & \ \ 0 \ \ & 1 & -1 & 0 & \ \ 0 \ \ & \ \ 1 \ \ & -2 \\
 0 & 1 & 0 & 1 & 0 & 0 & -1 & 0 & 1 & 0 & 0 & -1 & 0 & 0 & 0 & 0 & 1 \\
 0 & 0 & 1 & 1 & 0 & 0 & -1 & 0 & 0 & 0 & 0 & 0 & 1 & 0 & 0 & -1 & 1 \\
 0 & 0 & 0 & 0 & 1 & 0 & 1 & 0 & 0 & 0 & 0 & 0 & 0 & 0 & 0 & 0 & 0 \\
 0 & 0 & 0 & 0 & 0 & 1 & 1 & 0 & 0 & 0 & 0 & 0 & 0 & 0 & 0 & 0 & 0 \\
 0 & 0 & 0 & 0 & 0 & 0 & 0 & 1 & 1 & 0 & 0 & -1 & 1 & 0 & 0 & -1 & 1 \\
 0 & 0 & 0 & 0 & 0 & 0 & 0 & 0 & 0 & 1 & 0 & 1 & 0 & 0 & 0 & 0 & 0 \\
 0 & 0 & 0 & 0 & 0 & 0 & 0 & 0 & 0 & 0 & 1 & 1 & 0 & 0 & 0 & 0 & 0 \\
 0 & 0 & 0 & 0 & 0 & 0 & 0 & 0 & 0 & 0 & 0 & 0 & 0 & 1 & 0 & 1 & 0 \\
\ \ 0 \ \ & \ \ 0 \ \ & \ \ 0 \ \ & \ \ 0 \ \ & \ \ 0 \ \ & \ \ 0 \ \ & \ \ 0 \ \ & \ \ 0 \ \ & \ \ 0 \ \ & \ \ 0 \ \ & \ \ 0 \ \ & \ \ 0 \ \ & \ \ 0 \ \ & \ \ 0 \ \ & \ \ 1 \ \ & \ \ 1 \ \ & 0 
\end{array}
\right. \cdots \nonumber
\eeq}

{\footnotesize
\beq
\cdots \left.
\begin{array}{ccccccccccccccccc}
 -1 & -1 & 0 & -1 & -1 & 0 & 0 & -1 & -1 & 0 & 0 & 0 & 0 & 0 & 0 & 1 & 1 \\
 0 & 0 & -1 & 0 & 1 & -1 & 0 & 0 & 1 & -1 & 0 & -1 & 0 & 0 & 1 & -1 & -1 \\
 0 & 0 & -1 & 1 & 0 & 0 & -1 & 1 & 0 & 0 & -1 & 1 & 0 & 0 & -1 & -1 & -1 \\
 1 & 0 & 1 & 0 & 0 & 0 & 1 & 0 & 0 & 1 & 0 & 0 & 0 & 0 & 0 & 1 & 0 \\
 0 & 1 & 1 & 0 & 0 & 1 & 0 & 0 & 0 & 0 & 1 & 0 & 0 & 0 & 0 & 0 & 1 \\
 1 & 1 & 1 & 0 & 0 & 0 & 0 & 0 & 0 & 0 & 0 & -1 & -1 & -1 & -1 & -1 & -1 \\
 0 & 0 & 0 & 1 & 0 & 1 & 0 & 0 & 0 & 0 & 0 & 1 & 1 & 0 & 0 & 0 & 1 \\
 0 & 0 & 0 & 0 & 0 & 0 & 0 & 1 & 0 & 1 & 0 & 1 & 0 & 1 & 0 & 1 & 0 \\
 0 & 0 & 0 & 0 & 1 & 0 & 1 & 0 & 0 & 0 & 0 & 0 & 0 & 1 & 1 & 1 & 0 \\
 \ \ 0 \ \ & \ \ 0 \ \ & \ \ 0 \ \ & \ \ 0 \ \ & \ \ 0 \ \ & \ \ 0 \ \ & \ \ 0 \ \ & \ \ 0 \ \ & \ \ 1 \ \ & \ \ 0 \ \ & \ \ 1 \ \ & \ \ 0 \ \ & \ \ 1 \ \ & \ \ 0 \ \ & \ \ 1 \ \ & \ \ 0 \ \ & \ \ 1  
\end{array}
\right)
\eeq}

The sum of all rows in the $P_{\rm red}$ matrix is always the vector $(1,\ldots,1)$ \cite{Franco:2013nwa}, so it is possible to drop one of them without losing any information, arriving at a 9-dimensional polytope. This dimensionality nicely agrees with the counting of degrees of freedom presented in \sref{section_constructing_polytopes}.

Let us now construct the matroid polytope. A simple approach consists of starting from the $P$ matrix in \eref{P_matrix_genus_1} and keeping only the rows corresponding to external legs. We thus obtain

{\footnotesize
\beq
G=\left(
\begin{array}{c|cccccccccccccccccccccccccccccccccc}
 X_{3,6} \ & \ 0 & 0 & 0 & 0 & 0 & 0 & 0 & 1 & 1 & 0 & 1 & 0 & 1 & 0 & 1 & 0 & 1 & 1 & 1 & 1 & 0 & 0 & 0 & 0 & 1 & 1 & 1 & 1 & 0 & 0 & 0 & 0 & 0 & 0 \\
 X_{7,6} \ & \ 0 & 1 & 0 & 1 & 0 & 1 & 0 & 0 & 1 & 0 & 1 & 0 & 0 & 0 & 0 & 0 & 1 & 0 & 1 & 0 & 0 & 1 & 0 & 0 & 1 & 1 & 0 & 1 & 0 & 0 & 1 & 1 & 0 & 0 \\
 X_{7,3} \ & \ 0 & 1 & 0 & 1 & 1 & 0 & 0 & 0 & 1 & 1 & 0 & 0 & 0 & 0 & 0 & 0 & 1 & 1 & 0 & 0 & 1 & 1 & 0 & 1 & 0 & 1 & 0 & 0 & 0 & 1 & 0 & 1 & 0 & 0 \\
 X_{4,2} \ & \ 0 & 0 & 1 & 1 & 1 & 0 & 0 & 0 & 0 & 0 & 0 & 0 & 1 & 0 & 1 & 0 & 1 & 1 & 0 & 0 & 1 & 0 & 0 & 0 & 1 & 1 & 1 & 0 & 1 & 1 & 0 & 0 & 0 & 0 \\
 X_{4,5} \ & \ 0 & 0 & 1 & 1 & 0 & 1 & 0 & 0 & 0 & 0 & 0 & 0 & 1 & 1 & 0 & 0 & 1 & 0 & 1 & 0 & 1 & 1 & 1 & 0 & 1 & 0 & 0 & 0 & 1 & 0 & 1 & 0 & 0 & 0 \\
 X_{2,5} \ & \ 0 & 0 & 0 & 0 & 0 & 0 & 0 & 1 & 1 & 1 & 0 & 0 & 1 & 1 & 0 & 0 & 1 & 1 & 1 & 1 & 1 & 1 & 1 & 1 & 0 & 0 & 0 & 0 & 0 & 0 & 0 & 0 & 0 & 0 \\
\end{array}
\right)
\eeq}

\noindent The column vectors in $G$ give the positions of the points in the matroid polytope in $\mathbb{Z}^6$. We notice that several columns are repeated, which means that, as explained in \sref{section_constructing_polytopes}, more than one point in the matching polytope can be projected down to the same point in the matroid polytope. 

An even more compact way of describing the matroid polytope is to construct a new matrix $\tilde{G}$ in which we eliminate the repetition of columns. In this case, we get

{\footnotesize
\beq
\tilde{G}=\left(
\begin{array}{c|cccccccccccccccccccc}
 X_{3,6} \ & \ 0 & 0 & 0 & 0 & 0 & 0 & 1 & 1 & 0 & 1 & 1 & 0 & 1 & 1 & 1 & 1 & 0 & 0 & 1 & 1 \\
 X_{7,6} \ & \ 0 & 1 & 0 & 1 & 0 & 1 & 0 & 1 & 0 & 1 & 0 & 0 & 0 & 1 & 0 & 1 & 0 & 1 & 1 & 1 \\
 X_{7,3} \ & \ 0 & 1 & 0 & 1 & 1 & 0 & 0 & 1 & 1 & 0 & 0 & 0 & 0 & 1 & 1 & 0 & 1 & 1 & 0 & 1 \\
 X_{4,2} \ & \ 0 & 0 & 1 & 1 & 1 & 0 & 0 & 0 & 0 & 0 & 1 & 0 & 1 & 1 & 1 & 0 & 1 & 0 & 1 & 1 \\
 X_{4,5} \ & \ 0 & 0 & 1 & 1 & 0 & 1 & 0 & 0 & 0 & 0 & 1 & 1 & 0 & 1 & 0 & 1 & 1 & 1 & 1 & 0 \\
 X_{2,5} \ & \ 0 & 0 & 0 & 0 & 0 & 0 & 1 & 1 & 1 & 0 & 1 & 1 & 0 & 1 & 1 & 1 & 1 & 1 & 0 & 0 \\ \hline
& \ {\bf 6} & {\bf 2} & {\bf 2} & {\bf 1} & {\bf 2} & {\bf 2} & {\bf 2} & {\bf 1} & {\bf 2} & {\bf 2} & {\bf 1} & {\bf 2} & {\bf 2} & {\bf 1} & {\bf 1} & {\bf 1} & {\bf 1} & {\bf 1} & {\bf 1} & {\bf 1} \\ \hline
\end{array}
\right) \; .
\label{G_tilde_original_genus_1}
\eeq}

\noindent The bold face numbers in the last row indicate the numbers of points in the matching polytope that get projected to every point in the matching polytope. For example, the first column corresponds to 6 perfect matchings: $p_{1}$, $p_{7}$, $p_{12}$, $p_{16}$, $p_{33}$, $p_{34}$. We see that the 34 points of the matching polytope project down to 20 points in the matroid polytope. All the perfect matchings associated to a given point in the matroid polytope represent contributions to the same \pl coordinate. We will see this more explicitly in \sref{sec:nonplanarizable}.

To conclude, let us mention that following our discussion in \sref{section_constructing_polytopes}, it is also possible to construct the matching and matroid polytopes presented above using the generalized face variables we presented in \sref{genus_1_example_generalized_faces}. We leave this as a straightforward exercise for the interested reader.

\bigskip

\subsubsection{Polytopes and Region Matching}

\begin{figure}[h]
\begin{center}
\includegraphics[scale=0.65]{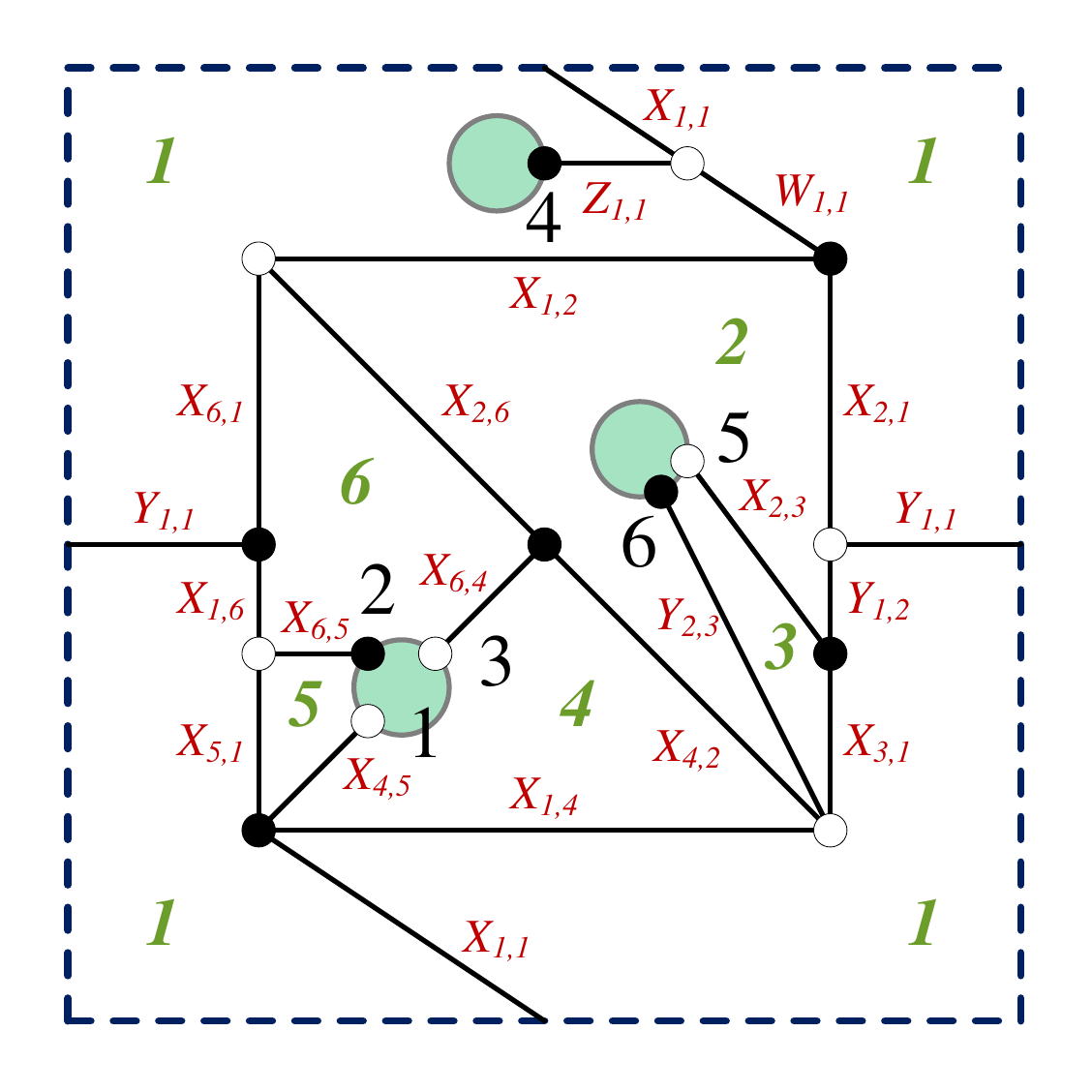}
\vspace{-0.3cm}\caption{A reduced on-shell diagram embedded into a torus with three boundaries. This graph cannot be embedded on any surface with $g=0$. Faces are labeled in green, external nodes in black and edges in red. The labels in this graph are unrelated to those in \fref{fig:nonplanarizabletorus1}.}
\label{torusReduced_SqMove}
\end{center}
\end{figure}

Let us now illustrate how matching and matroid polytopes are used for determining region matching of on-shell diagrams, hence serving as a practical tool for identifying potential equivalences. To do so, let us consider the on-shell diagram shown in \fref{torusReduced_SqMove}, which is shown embedded on a torus with three boundaries.

The matching and matroid polytopes for this diagram can be constructed following the same procedure outlined for the previous section. We will thus be briefer in our discussion and present only the most relevant results.

As before, we begin by determining the perfect matching matrix $P$. This diagram has 19 edges and 42 perfect matchings, which can be determined using generalized Kasteleyn matrices. The matrix $P$ is hence $19\times 42$-dimensional.  Just like for the previous example, the matching polytope is 9-dimensional, i.e.\ this diagram also has $d=9$ degrees of freedom. The degrees of freedom can be alternatively counted using generalized face variables or calculated as 19 edges$-$10 internal nodes$=9$. The matching polytope now has 42 points, one for each perfect matching.

The matroid polytope is constructed by keeping only those rows associated to the external legs. The 42 points in the matching polytope are projected down to 20 points in the matroid polytope, which is encoded in the following matrix

{\footnotesize
\beq
\tilde{G}=\left(
\begin{array}{c|cccccccccccccccccccc}
 X_{4,5} \ & \ 0 & 0 & 0 & 0 & 0 & 0 & 1 & 1 & 0 & 1 & 1 & 0 & 1 & 1 & 1 & 1 & 0 & 0 & 1 & 1 \\
 X_{6,5} \ & \ 0 & 1 & 0 & 1 & 0 & 1 & 0 & 1 & 0 & 1 & 0 & 0 & 0 & 1 & 0 & 1 & 0 & 1 & 1 & 1 \\
 X_{6,4} \ & \ 0 & 1 & 0 & 1 & 1 & 0 & 0 & 1 & 1 & 0 & 0 & 0 & 0 & 1 & 1 & 0 & 1 & 1 & 0 & 1 \\
 Z_{1,1} \ & \ 0 & 0 & 1 & 1 & 1 & 0 & 0 & 0 & 0 & 0 & 1 & 0 & 1 & 1 & 1 & 0 & 1 & 0 & 1 & 1 \\
 X_{2,3} \ & \ 0 & 0 & 1 & 1 & 0 & 1 & 0 & 0 & 0 & 0 & 1 & 1 & 0 & 1 & 0 & 1 & 1 & 1 & 1 & 0 \\
 Y_{2,3} \ & \ 0 & 0 & 0 & 0 & 0 & 0 & 1 & 1 & 1 & 0 & 1 & 1 & 0 & 1 & 1 & 1 & 1 & 1 & 0 & 0 \\ \hline
 & \ {\bf 5} & {\bf 3} & {\bf 3} & {\bf 2} & {\bf 3} & {\bf 3} & {\bf 1} & {\bf 1} & {\bf 2} & {\bf 2} & {\bf 1} & {\bf 2} & {\bf 2} & {\bf 2} & {\bf 1} & {\bf 1} & {\bf 2} & {\bf 2} & {\bf 2} & {\bf 2} \\ \hline
\end{array}
\right),
\label{G_tilde_dual_genus_1}
\eeq}

\noindent where the last row indicates the number of perfect matchings associated to each point of the matroid polytope. Modulo these multiplicities, this matrix is identical to the one in \eref{G_tilde_original_genus_1}!  In hindsight, we organized the external legs in \eref{G_tilde_dual_genus_1} such that they are in the same order as the corresponding ones \eref{G_tilde_original_genus_1}. Failure to do so would have resulted in a permutation of the rows. In that case, comparison between the rows of the two matrices would have determined how to identify the external legs of both diagrams.

In summary, our analysis indicates that the diagrams in \fref{fig:nonplanarizabletorus1} and \fref{torusReduced_SqMove} have the same number of degrees of freedom and the same matroid polytope, i.e. they cover the same region of the Grassmannian. In fact, both diagrams are indeed equivalent; we created this example by starting from the diagram in \fref{fig:nonplanarizabletorus1} and performing a ``square move" on the $\alpha$ loop shown in \fref{alphabetab1}.

Even this simple example illustrates how difficult it can be to find the sequence of moves connecting two equivalent non-planar graphs and the importance of having a {\it global} criterion for characterizing diagrams. This is precisely what matching and matroid polytopes achieve in a systematic way.

\bigskip

\section{Boundary Measurement for Arbitrary On-Shell Diagrams}
\label{sec:general-boundary-measurement}

The purpose of this section is to propose a boundary measurement that is valid for arbitrary on-shell diagrams. The boundary measurement maps edge weights in the diagram to the Grassmannian $Gr_{k,n}$. Its generalization is thus an imperative step for developing the theory of non-planar on-shell diagrams.

Such a map was originally introduced for on-shell diagrams on the disk in \cite{2006math09764P}, extended to the annulus in \cite{2009arXiv0901.0020G} and finally generalized to genus-zero and an arbitrary number of boundaries in \cite{Franco:2013nwa}. Below, we will generalize the boundary measurement to allow for diagrams with arbitrary genus embeddings. Strictly speaking, the boundary measurement is independent of the embedding. However, as in previous sections, considering an explicit embedding will turn out to be a useful tool. More importantly, we can regard on-shell diagrams that do not admit a genus-zero embedding as inherently demanding a higher genus treatment.

\bigskip

\subsection{General Strategy} \label{sec:boundmeasproperties}

As reviewed in \sref{sec:review-planar}, the first step is to pick a perfect orientation of the diagram. Reproducing \eref{eq:genericentries} here for convenience, for $n$ external nodes and $k$ sources in the perfect orientation, the corresponding matrix $C$ in $Gr_{k,n}$ takes the general form

\beq
C_{ij}(X)=\sum_{\Gamma \in \{i \rightsquigarrow j\}}(-1)^{s_\Gamma}\prod_{e\,\in\, \Gamma}X_e ,
\label{bm}
\eeq
where $i$ runs over the sources, $j$ runs over all external nodes and $\Gamma$ is a flow in the perfect orientation going from $i$ to $j$. Moreover, recall that flows are in one-to-one correspondence with perfect matchings. The GL($k$) gauge symmetry of $Gr_{k,n}$ is fixed in this matrix: there is $k\times k$ identity sub-matrix associated to the source nodes. 

For the proposed boundary measurement to blend into the general approach to on-shell diagrams we introduced in earlier sections, it should satisfy two properties. First, planar graphs must parametrize cells in the positive Grassmannian, i.e.\ positive edge weights should give rise to positive \pl coordinates. More generally, we want the boundary measurement to agree with our characterization of on-shell diagrams based on generalized matroid polytopes. In this approach, every point in the matroid polytope corresponds to a \pl coordinate and is associated to a collection of flows (equivalently perfect matchings). For this to happen, we want the \pl coordinates arising from the boundary measurement to be sums of the flows associated to the corresponding point in the matroid polytope. Here and in what follows we use sum of flows as an abbreviation for linear combinations of flows with coefficients $\pm 1$. Notice that while according to \eref{bm} the entries in $C$ are linear combinations of flows, the fact that \pl coordinates, i.e.\ the determinants of its $k\times k$ sub-matrices, are sums of very specific sets of flows is a highly non-trivial property. The latter is the main challenge when generalizing the boundary measurement to arbitrary on-shell diagrams.

\bigskip

\subsection{Signs}

In order to complete the definition of the boundary measurement, it is necessary to provide a prescription for determining the $(-1)^{s_\Gamma}$ sign multiplying every flow in $\eref{bm}$. 

At this point, it is useful to consider an embedding of the graph. As explained in \sref{section_generalized_faces}, for $B$ boundaries we need to consider $B-1$ cuts connecting them. This leads to an ordering of external nodes, determined as follows. Starting from an arbitrary external node, we follow the boundaries and cuts of the graph as done in complex analysis, numbering external nodes as they appear until returning to the original point \cite{Franco:2013nwa}. This generalizes the cyclic ordering of external nodes in diagrams embedded on a disk.

It is convenient to factor the $(-1)^{s_\Gamma}$ signs into two types of contributions, which we explain below.

\bigskip

\subsubsection*{Positivity Signs}

Signs of the first type are common to all flows contributing to a given entry $C_{ij}$. We refer to them as {\it positivity signs} because their effect is to ensure that, for graphs embedded on a disk, positive edge weights result in positive \pl coordinates. They were first introduced by Postnikov for the planar boundary measurement in \cite{2006math09764P}. In fact, these are the only signs present for graphs on disks. To determine them, we need to consider the ordering of external nodes introduced above. All flows in a given entry $C_{ij}$ get an overall positivity sign equal to $(-1)^{s(i,j)}$, where $s(i,j)$ is the number of sources strictly between the external nodes $i$ and $j$, neglecting periodicity.

\bigskip

\subsubsection*{Combinatorial Signs}

A new contribution to the signs needs to be included when considering on-shell diagrams with non-planar embeddings.\footnote{In fact, as we explain below in footnote \ref{footnote_loops_perfect_orientation}, this type of signs also arise for planar diagrams when considering a perfect orientation with closed oriented loops. In general, however, this issue can be avoided in the planar case by choosing a perfect orientation without loops.} The positivity signs typically violate one of the principal requirements of the boundary measurement, i.e.\ the consistency with the determination of region matching for on-shell diagrams based on generalized matroid polytopes expounded in \sref{sec:matchingmatroid}: the cancellations required to ensure that \pl coordinates become sums of flows will generally no longer happen. We refer to the signs that are necessary to correct this problem as {\it combinatorial signs}. In general, combinatorial signs differ among individual terms contributing to a giving entry $C_{ij}$, i.e.\ among individual flows. Combinatorial signs were first introduced in \cite{2009arXiv0901.0020G} for on-shell diagrams on the annulus and extended to general genus-zero embeddings in \cite{Franco:2013nwa}.

In order to determine the combinatorial sign of a flow, we need to turn it into a closed loop as follows. We go from the source to the sink along the flow, and then return to the source following the boundaries and cuts. 
In general, such a loop has self-intersections. The \textit{rotation number} $r$ of the loop is defined as the number of full clockwise revolutions of the loop minus the number of full counter-clockwise revolutions. Equivalently, we can express the rotation number in terms of the parity of the number of self-intersections. The combinatorial sign for the flow is then given by $(-1)^{r+1}$.\footnote{In our previous discussion, we have implicitly assumed that there are no closed oriented loops in the perfect orientation under consideration. In general, it is possible to pick a perfect orientation such that this is the case. Combinatorial signs are controlled by the rotation number and have an additional effect when the perfect orientation contains oriented closed loops. In such cases, there can be an infinite number of contributions to a given entry in the boundary measurement, corresponding to circling around the loop any number of times. Formally summing up the corresponding geometric series gives rise to a non-trivial denominator of the schematic form $1 \pm \mathfrak{p}_{\rm{loop}}$, where $\mathfrak{p}_{\rm{loop}}$ indicates the loop. For planar graphs, where positivity is important, combinatorial signs are such that the denominator picks a plus sign and hence cannot vanish for positive edge weights. More generally, whenever such denominators arise, the fact that \pl coordinates are given by sums of flows is unaltered, after factoring out the denominators. \label{footnote_loops_perfect_orientation}}

\begin{figure}[h]
\begin{center}
\includegraphics[scale=0.4]{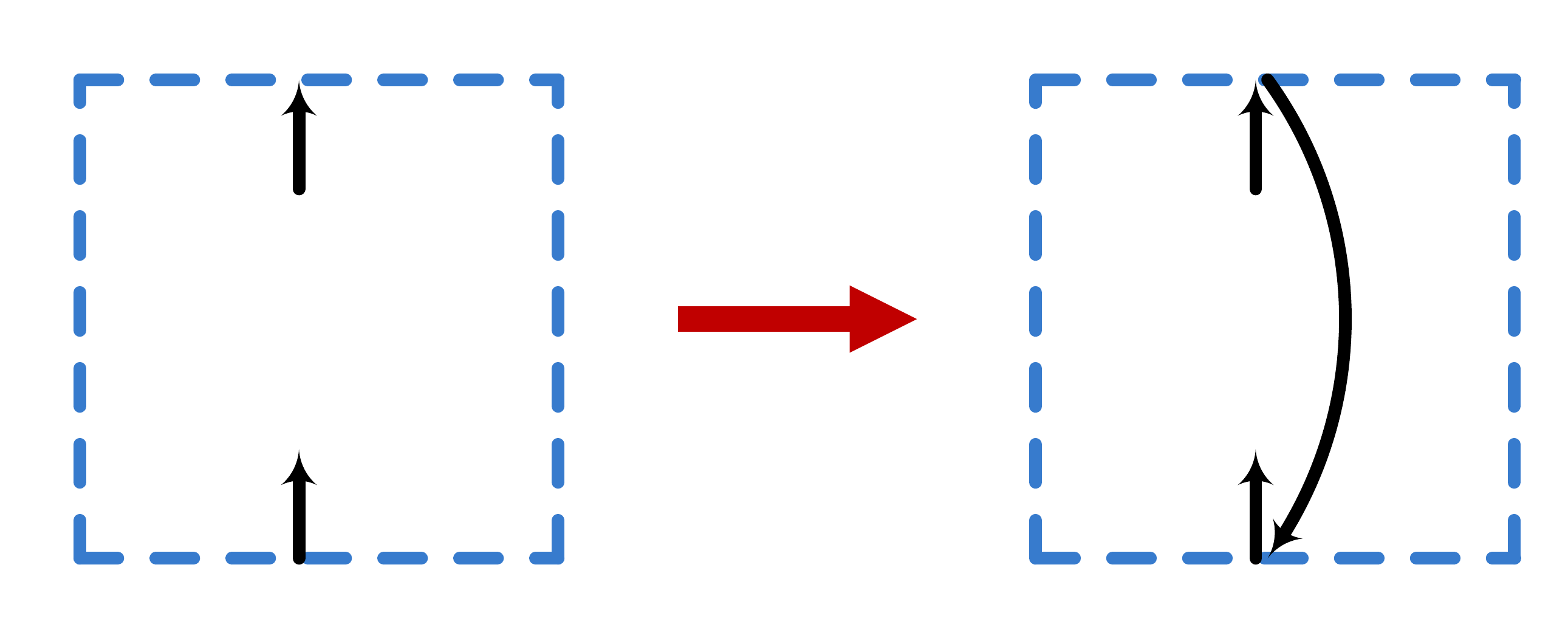}
\caption{A schematic representation of how to close a flow within the unit cell in the case of a torus.}
\label{fig:periodicityrule}
\end{center}
\end{figure}

The discussion above was originally developed for genus-zero embeddings \cite{2006math09764P,2009arXiv0901.0020G,Franco:2013nwa}. In order to extend the boundary measurement to higher genus, we propose an explicit prescription for constructing the loop: it should be closed \textit{within the unit cell}. This is done as follows: every time a flow goes around a non-trivial loop and thus uses the periodicity of the Riemann surface, we connect its exit and entry points of the unit cell. This procedure is illustrated in \fref{fig:periodicityrule} for genus $g=1$. This process creates a closed loop which is entirely contained inside the unit cell, whose rotation number $r$ is used to determine the combinatorial sign $(-1)^{r+1}$ associated to the corresponding flow. 

We emphasize that this prescription is a proposal, and it would be desirable to develop a proof for it and to consider its dependence on things such as the choice of unit cell. In order to arrive to it we have considered several explicit examples, like the ones presented below, and verified it works, as opposed to other ways of determining the rotation numbers. In particular, the parity of the number of self intersections of loops, and hence the combinatorial signs, can change if we do not insist in closing loops within a unit cell, destroying the cancellations which are necessary for \pl coordinates to become sums of all flows associated to a point in the matroid polytope. A detailed example illustrating the dependence on different ways of closing loops, which indeed was used as a guide for constructing our final proposal, is provided in Appendix \ref{sec:motivationexample}. This prescription can have additional interesting consequences such as producing additional signs for flows even in the absence of cuts, as we shall see in \sref{sec:genus2}.

Summarizing our discussion, the signs are factorized into positivity and combinatorial signs as follows
\beq
(-1)^{s_\Gamma}=(-1)^{s(i,j)}(-1)^{r+1} .
\eeq

Our boundary measurement applies to arbitrary genus, reducing to the already known prescription on genus-zero graphs. For illustrative purposes and to provide evidence supporting our proposal, in the coming sections we present $g=1$ and $g=2$ examples.

\bigskip

\subsection{A Genus-One Example} \label{sec:nonplanarizable}

\begin{figure}[h]
\begin{center}
\includegraphics[scale=0.6]{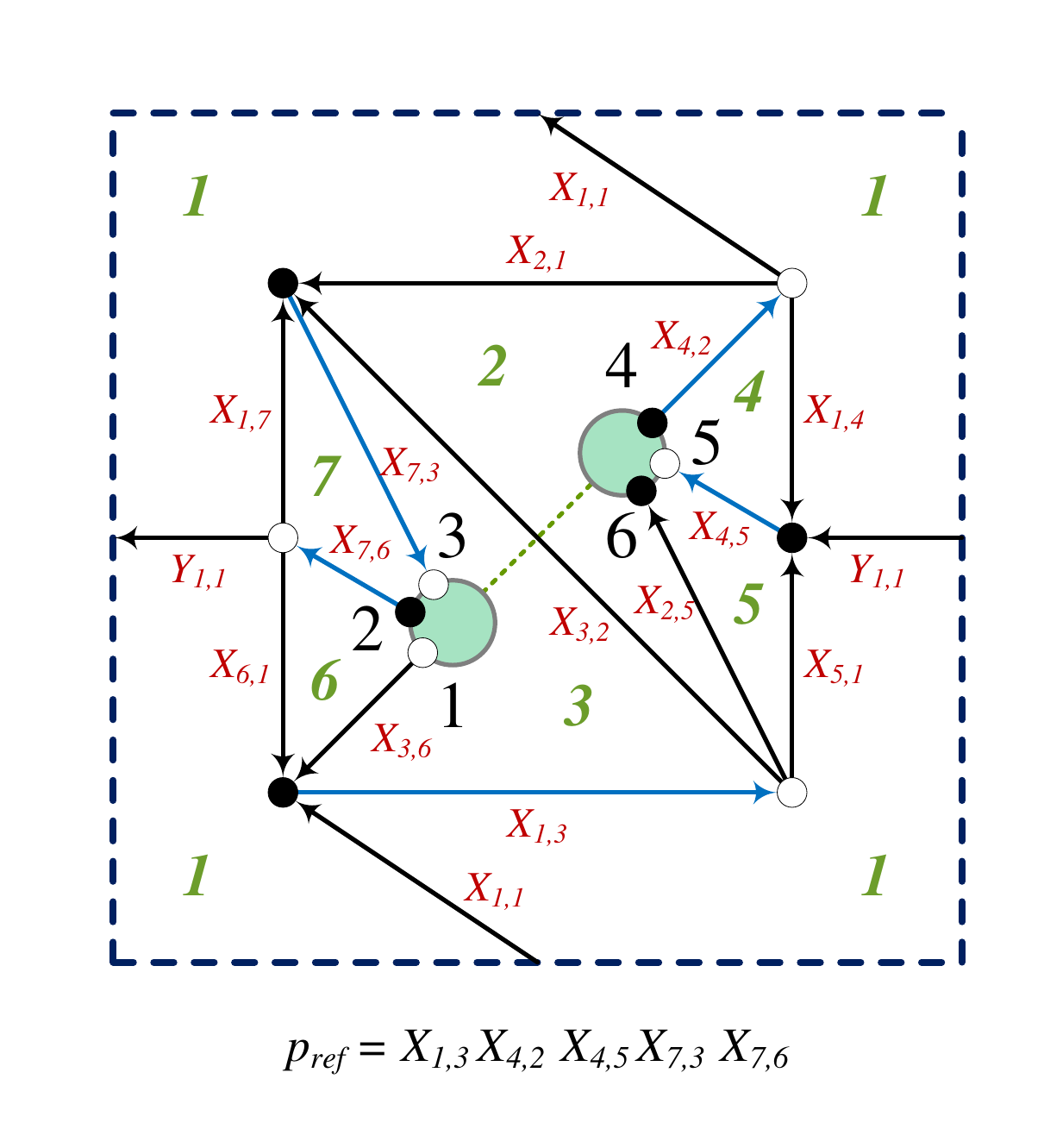}
\vspace{-0.3cm}\caption{A reduced on-shell diagram embedded into a torus with two boundaries. This graph cannot be embedded on any surface with $g=0$. Faces are labeled in green, external nodes in black and edges in red. The dashed line represents the cut.} 
\label{fig:nonplanarizabletorus}
\end{center}
\end{figure}

Let us revisit the on-shell diagram presented in \fref{fig:nonplanarizabletorus1}. As already mentioned, this diagram does not admit a $g=0$ embedding. \fref{fig:nonplanarizabletorus} shows this diagram with the perfect orientation associated to the reference perfect matching $p_4 = X_{1,3} X_{4,2} X_{4,5} X_{7,3} X_{7,6}$.

\noindent This diagram has 34 perfect matchings, which are encoded by the perfect matching matrix given in \eref{P_matrix_genus_1}. The corresponding flows in the perfect orientation under consideration and their source sets are

{\small
\begin{equation}
\begin{array}{cllcll}
 \mathfrak{p}_1=& \frac{X_{1,4} X_{1,7}}{X_{4,2} X_{4,5} X_{7,3} X_{7,6}} & \textcolor{blue}{\{1,3,5\}} \quad \quad \quad \quad & \mathfrak{p}_{18}=& \frac{X_{2,5} X_{3,6} Y_{1,1}}{X_{1,3} X_{4,5} X_{7,6}} & \textcolor{blue}{\{4,5,6\}} \\
 \mathfrak{p}_2=& \frac{X_{1,4}}{X_{4,2} X_{4,5}} & \textcolor{blue}{\{1,2,5\}} & \mathfrak{p}_{19}=& \frac{X_{2,1} X_{2,5} X_{3,6}}{X_{1,3} X_{4,2} X_{7,3}} & \textcolor{blue}{\{2,3,6\}} \\
 \mathfrak{p}_3=& \frac{X_{1,7}}{X_{7,3} X_{7,6}} & \textcolor{blue}{\{1,3,4\}} & \mathfrak{p}_{20}=& \frac{X_{2,1} X_{2,5} X_{3,6} Y_{1,1}}{X_{1,3} X_{4,2} X_{4,5} X_{7,3} X_{7,6}} \quad \quad & \textcolor{blue}{\{3,5,6\}} \\
 \mathfrak{p}_4=& 1 & \textcolor{blue}{\{1,2,4\}} & \mathfrak{p}_{21}=& \frac{X_{2,5} X_{6,1}}{X_{1,3} X_{7,6}} & \textcolor{blue}{\{1,4,6\}} \\
 \mathfrak{p}_5=& \frac{Y_{1,1}}{X_{4,5} X_{7,6}} & \textcolor{blue}{\{1,4,5\}} & \mathfrak{p}_{22}=& \frac{X_{1,1} X_{2,5}}{X_{1,3} X_{4,2}} & \textcolor{blue}{\{1,2,6\}} \\
 \mathfrak{p}_6=& \frac{X_{2,1}}{X_{4,2} X_{7,3}} & \textcolor{blue}{\{1,2,3\}} & \mathfrak{p}_{23}=& \frac{X_{2,1} X_{2,5} X_{6,1}}{X_{1,3} X_{4,2} X_{7,3} X_{7,6}} & \textcolor{blue}{\{1,3,6\}} \\
 \mathfrak{p}_7=& \frac{X_{2,1} Y_{1,1}}{X_{4,2} X_{4,5} X_{7,3} X_{7,6}} & \textcolor{blue}{\{1,3,5\}} & \mathfrak{p}_{24}=& \frac{X_{1,1} X_{2,5} Y_{1,1}}{X_{1,3} X_{4,2} X_{4,5} X_{7,6}} & \textcolor{blue}{\{1,5,6\}} \\
 \mathfrak{p}_8=& \frac{X_{1,4} X_{1,7} X_{2,5} X_{3,6}}{X_{1,3} X_{4,2} X_{4,5} X_{7,3} X_{7,6}} \quad \quad & \textcolor{blue}{\{3,5,6\}} & \mathfrak{p}_{25}=& \frac{X_{3,2} X_{3,6}}{X_{1,3} X_{7,3}} & \textcolor{blue}{\{2,3,4\}} \\
\mathfrak{p}_9=& \frac{X_{1,4} X_{2,5} X_{3,6}}{X_{1,3} X_{4,2} X_{4,5}} & \textcolor{blue}{\{2,5,6\}} & \mathfrak{p}_{26}=& \frac{X_{3,6} X_{5,1}}{X_{1,3} X_{4,5}} & \textcolor{blue}{\{2,4,5\}} \\
 \mathfrak{p}_{10}=& \frac{X_{1,4} X_{2,5} X_{6,1}}{X_{1,3} X_{4,2} X_{4,5} X_{7,6}} & \textcolor{blue}{\{1,5,6\}}  \quad \quad \quad \quad & \mathfrak{p}_{27}=& \frac{X_{3,2} X_{3,6} Y_{1,1}}{X_{1,3} X_{4,5} X_{7,3} X_{7,6}}  \quad \quad & \textcolor{blue}{\{3,4,5\}} \\
 \mathfrak{p}_{11}=& \frac{X_{1,4} X_{3,2} X_{3,6}}{X_{1,3} X_{4,2} X_{4,5} X_{7,3}} & \textcolor{blue}{\{2,3,5\}} & \mathfrak{p}_{28}=& \frac{X_{2,1} X_{3,6} X_{5,1}}{X_{1,3} X_{4,2} X_{4,5} X_{7,3}} & \textcolor{blue}{\{2,3,5\}} \\
 \mathfrak{p}_{12}=& \frac{X_{1,4} X_{3,2} X_{6,1}}{X_{1,3} X_{4,2} X_{4,5} X_{7,3} X_{7,6}} \quad \quad & \textcolor{blue}{\{1,3,5\}} & \mathfrak{p}_{29}=& \frac{X_{3,2} X_{6,1}}{X_{1,3} X_{7,3} X_{7,6}} & \textcolor{blue}{\{1,3,4\}} 
\end{array} \nonumber
\end{equation}
}

{\small
\begin{equation}
\begin{array}{cllcll}
  \mathfrak{p}_{13}=& \frac{X_{1,7} X_{2,5} X_{3,6}}{X_{1,3} X_{7,3} X_{7,6}} & \textcolor{blue}{\{3,4,6\}} & \mathfrak{p}_{30}=& \frac{X_{5,1} X_{6,1}}{X_{1,3} X_{4,5} X_{7,6}} & \textcolor{blue}{\{1,4,5\}} \\
 \mathfrak{p}_{14}=& \frac{X_{1,1} X_{1,7} X_{2,5}}{X_{1,3} X_{4,2} X_{7,3} X_{7,6}} & \textcolor{blue}{\{1,3,6\}} & \mathfrak{p}_{31}=& \frac{X_{1,1} X_{3,2}}{X_{1,3} X_{4,2} X_{7,3}} & \textcolor{blue}{\{1,2,3\}} \\
 \mathfrak{p}_{15}=& \frac{X_{1,7} X_{3,6} X_{5,1}}{X_{1,3} X_{4,5} X_{7,3} X_{7,6}} & \textcolor{blue}{\{3,4,5\}} & \mathfrak{p}_{32}=& \frac{X_{1,1} X_{5,1}}{X_{1,3} X_{4,2} X_{4,5}} & \textcolor{blue}{\{1,2,5\}} \\
 \mathfrak{p}_{16}=& \frac{X_{1,1} X_{1,7} X_{5,1}}{X_{1,3} X_{4,2} X_{4,5} X_{7,3} X_{7,6}} & \textcolor{blue}{\{1,3,5\}} & \mathfrak{p}_{33}=& \frac{X_{1,1} X_{3,2} Y_{1,1}}{X_{1,3} X_{4,2} X_{4,5} X_{7,3} X_{7,6}} \quad \quad & \textcolor{blue}{\{1,3,5\}} \\
 \mathfrak{p}_{17}=& \frac{X_{2,5} X_{3,6}}{X_{1,3}} & \textcolor{blue}{\{2,4,6\}} & \mathfrak{p}_{34}=& \frac{X_{2,1} X_{5,1} X_{6,1}}{X_{1,3} X_{4,2} X_{4,5} X_{7,3} X_{7,6}} & \textcolor{blue}{\{1,3,5\}} 
\end{array} 
\end{equation}
}

\medskip

\noindent It is interesting to remark the beautiful agreement between the multiplicities of flows with the same source sets and the multiplicities of the corresponding perfect matchings for every point in the matroid polytope given in \eref{G_tilde_original_genus_1}. 

\begin{figure}[h]
\begin{center}
\includegraphics[scale=0.3]{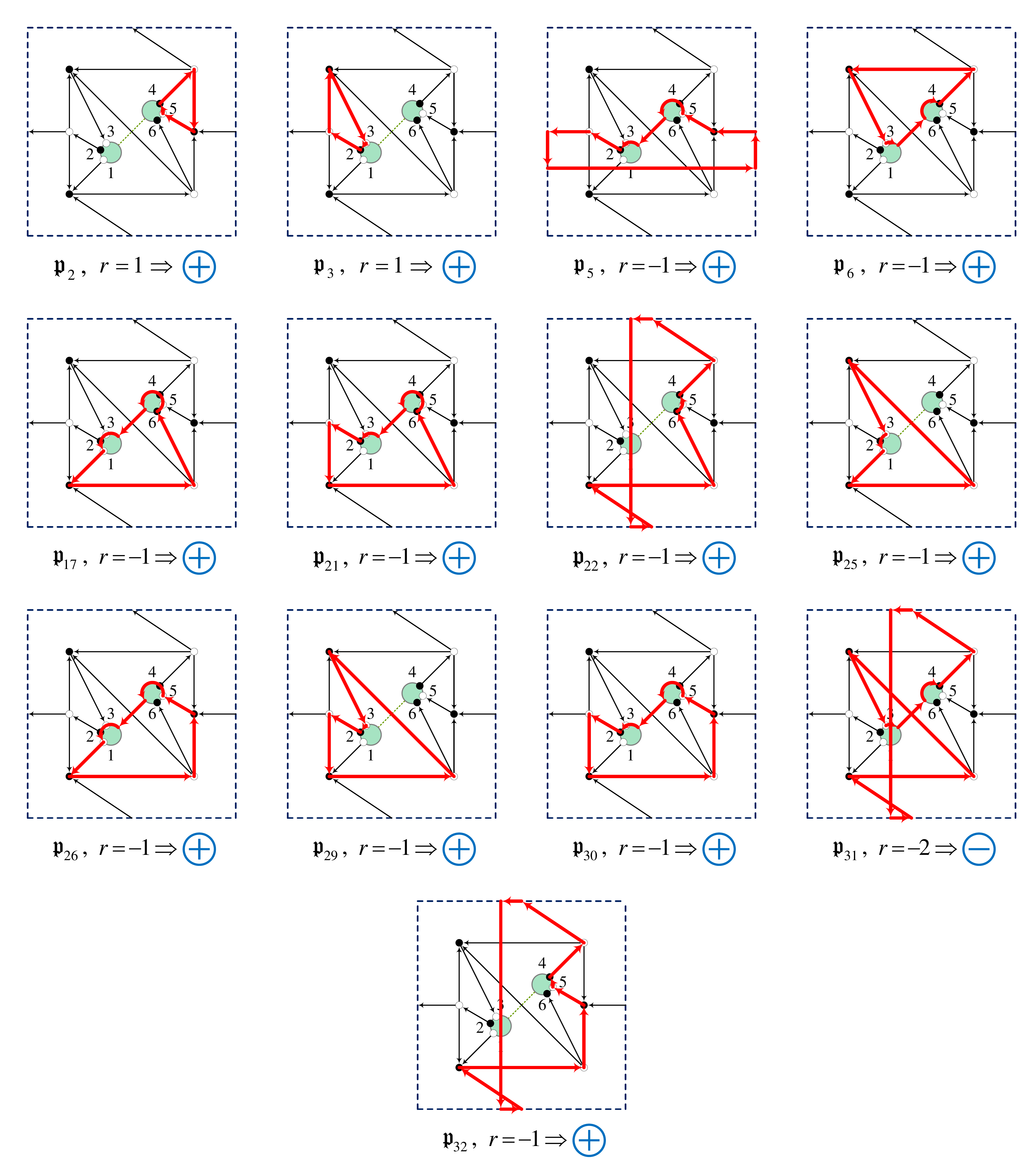}
\vspace{-0.3cm}\caption{Completion of flows into loops inside the unit cell for the example in \fref{fig:nonplanarizabletorus}, their rotation numbers and the resulting signs.}
\label{fig:nonplanarizableflows}
\end{center}
\end{figure}

We are now ready to construct the corresponding element of the Grassmannian. The first step is to introduce the positivity signs $(-1)^{s(i,j)}$. To do so, external nodes must be ordered using cuts to connect different boundaries. According to our definition of generalized face variables, cuts are oriented paths on the graph connecting pairs of boundaries. For the example at hand, an explicit choice of the cut was given in \fref{alphabetab1}. For ordering the external nodes, however, it is sufficient and much more convenient to consider deformations of the cuts that do not necessarily go over the edges of the graph. For this example, such a cut is represented by the dashed line in \fref{fig:nonplanarizabletorus}. In the examples that follow, we will take the same approach when ordering external nodes. Including the positivity signs, we obtain the following preliminary matrix
\begin{equation}
\left(
\begin{array}{c|cccccc}
& \ \ 1 \ \ & \ \ 2 \ \ & 3 & \ \ 4 \ \ & 5 & 6 \\ \hline
1 \ \ & \ \ 1 \ \ & \ \ 0 \ \ & -\mathfrak{p}_{25} & \ \ 0 \ \ & \mathfrak{p}_{26} & \mathfrak{p}_{17} \\
2 \ \ & \ \ 0 \ \ & \ \ 1 \ \ & \mathfrak{p}_3+\mathfrak{p}_{29} & \ \ 0 \ \ & -\mathfrak{p}_5-\mathfrak{p}_{30} & -\mathfrak{p}_{21} \\
4 \ \ & \ \ 0 \ \ & \ \ 0 \ \ & \mathfrak{p}_6+\mathfrak{p}_{31} & \ \ 1 \ \ & \mathfrak{p}_2+\mathfrak{p}_{32} & \mathfrak{p}_{22} \\
\end{array}
\right) .
\label{g1_preliminary_matrix}
\end{equation}
This is not the desired matrix yet, since the combinatorial signs still need to be incorporated. It is straightforward to verify that the minors of this matrix are not linear combinations with coefficients $\pm 1$ of all the flows with the corresponding source sets. The signs in \eref{g1_preliminary_matrix} do not produce the necessary cancellations. \fref{fig:nonplanarizableflows} shows the closed loops associated to each flow and the corresponding combinatorial signs arising from our prescription.

Only the flow $\mathfrak{p}_{31}$ picks up an additional minus sign. After including it, we obtain the Grassmannian matrix

\begin{equation}
C = \left(
\begin{array}{c|ccccccc}
& \ \ 1 \ \ & \ \ 2 \ \ & 3 & \ \ 4 \ \ & 5 & 6 \\ \hline
1 \ \ & \ \ 1 \ \ & \ \ 0 \ \ & -\mathfrak{p}_{25} & \ \ 0 \ \ & \mathfrak{p}_{26} & \mathfrak{p}_{17} \\
2 \ \ & \ \ 0 \ \ & \ \ 1 \ \ & \mathfrak{p}_3+\mathfrak{p}_{29} & \ \ 0 \ \ & -\mathfrak{p}_5-\mathfrak{p}_{30} & -\mathfrak{p}_{21} \\
4 \ \ & \ \ 0 \ \ & \ \ 0 \ \ & \mathfrak{p}_6-\mathfrak{p}_{31} & \ \ 1 \ \ & \mathfrak{p}_2+\mathfrak{p}_{32} & \mathfrak{p}_{22} \\
\end{array}
\right)  .
\end{equation}

This gives rise to the cancellations required to obtain the \pl coordinates in the following form

\bigskip

\begin{equation}
\begin{array}{clcl}
 \Delta _{1,2,3}= & \mathfrak{p}_6-\mathfrak{p}_{31} & \Delta _{2,3,4}= & \mathfrak{p}_{25} \\
 \Delta _{1,2,4}= & 1 & \Delta _{2,3,5}= & \mathfrak{p}_{11}+\mathfrak{p}_{28} \\
 \Delta _{1,2,5}= & \mathfrak{p}_2+\mathfrak{p}_{32} & \Delta _{2,3,6}= & \mathfrak{p}_{19} \\
 \Delta _{1,2,6}= & \mathfrak{p}_{22} & \Delta _{2,4,5}= & \mathfrak{p}_{26} \\
 \Delta _{1,3,4}= & \mathfrak{p}_3+\mathfrak{p}_{29} & \Delta _{2,4,6}= & \mathfrak{p}_{17} \\
 \Delta _{1,3,5}= & \mathfrak{p}_1+\mathfrak{p}_7+\mathfrak{p}_{12}+\mathfrak{p}_{16}-\mathfrak{p}_{33}+\mathfrak{p}_{34} \quad \quad \quad \quad & \Delta _{2,5,6}= & \mathfrak{p}_9 \\
 \Delta _{1,3,6}= & \mathfrak{p}_{14}+\mathfrak{p}_{23} & \Delta _{3,4,5}= & \mathfrak{p}_{15}-\mathfrak{p}_{27} \\
 \Delta _{1,4,5}= & \mathfrak{p}_5+\mathfrak{p}_{30} & \Delta _{3,4,6}= & \mathfrak{p}_{13} \\
 \Delta _{1,4,6}= & \mathfrak{p}_{21} & \Delta _{3,5,6}= & \mathfrak{p}_8+\mathfrak{p}_{20} \\
 \Delta _{1,5,6}= & \mathfrak{p}_{10}-\mathfrak{p}_{24} & \Delta _{4,5,6}= & \mathfrak{p}_{18}
\end{array} 
\label{pl_pm}
\end{equation}

\bigskip

\noindent Our general notation for \pl coordinates will be that $\Delta_{i_1,\ldots,\i_k}$ indicates the minor $(i_1,\ldots,i_k)$ of the boundary measurement $C$. From \eref{pl_pm}, we conclude that in this example every \pl coordinate indeed corresponds to a sum of those flows whose source set is the index of the \pl coordinate, as desired. We note that this example not only is reduced and non-planarizable, but also has multiple boundaries, constituting a rather non-trivial check of our proposal.

\bigskip

\subsubsection*{On-Shell Form in Terms of Generalized Face Variables}

Different variables can be used to describe the flows that contribute to the boundary measurement. In particular, it is instructive to consider how the boundary measurement for this on-shell diagram can be expressed in terms of generalized face variables, which for this example were given in \eref{alpha_beta_b_genus_1} and \eref{faces_genus_1}. Without any gauge fixing, we obtain

{\small
\begin{equation}
C = \left(
\begin{array}{c|cccccc}
& \ \ 1 \ \ & \ \ 2 \ \ & 3 & \ \ 4 \ \ & 5 & 6 \\ \hline
1 \ \ & \ \ 1 \ \ & \ \ 0 \ \ & -\dfrac{1}{f_3} & \ \ 0 \ \ & \dfrac{b}{f_3 f_5} & \dfrac{b}{f_3} \\
2 \ \ & \ \ 0 \ \ & \ \ 1 \ \ & \dfrac{1}{f_1 f_2 f_3 f_4 f_5 f_6}+\dfrac{1}{f_3 f_6} & \ \ 0 \ \ & -\dfrac{b}{f_1 f_3 f_5 f_6 \alpha}-\dfrac{b}{f_3 f_5 f_6} & -\dfrac{b}{f_3 f_6} \\
4 \ \ & \ \ 0 \ \ & \ \ 0 \ \ & \dfrac{1}{f_2 b}-\dfrac{f_1 f_4 f_5 \beta}{b} & \ \ 1 \ \ & f_4+f_1 f_4 \beta & f_1 f_4 f_5 \beta \\
\end{array}
\right) .
\end{equation}}

\noindent It is important to emphasize that going through edge variables is a useful intermediate step but not a necessary one.

\bigskip

\subsection{A Genus-Two Example} \label{sec:genus2}

\begin{figure}[h]
\begin{center}
\includegraphics[scale=0.85]{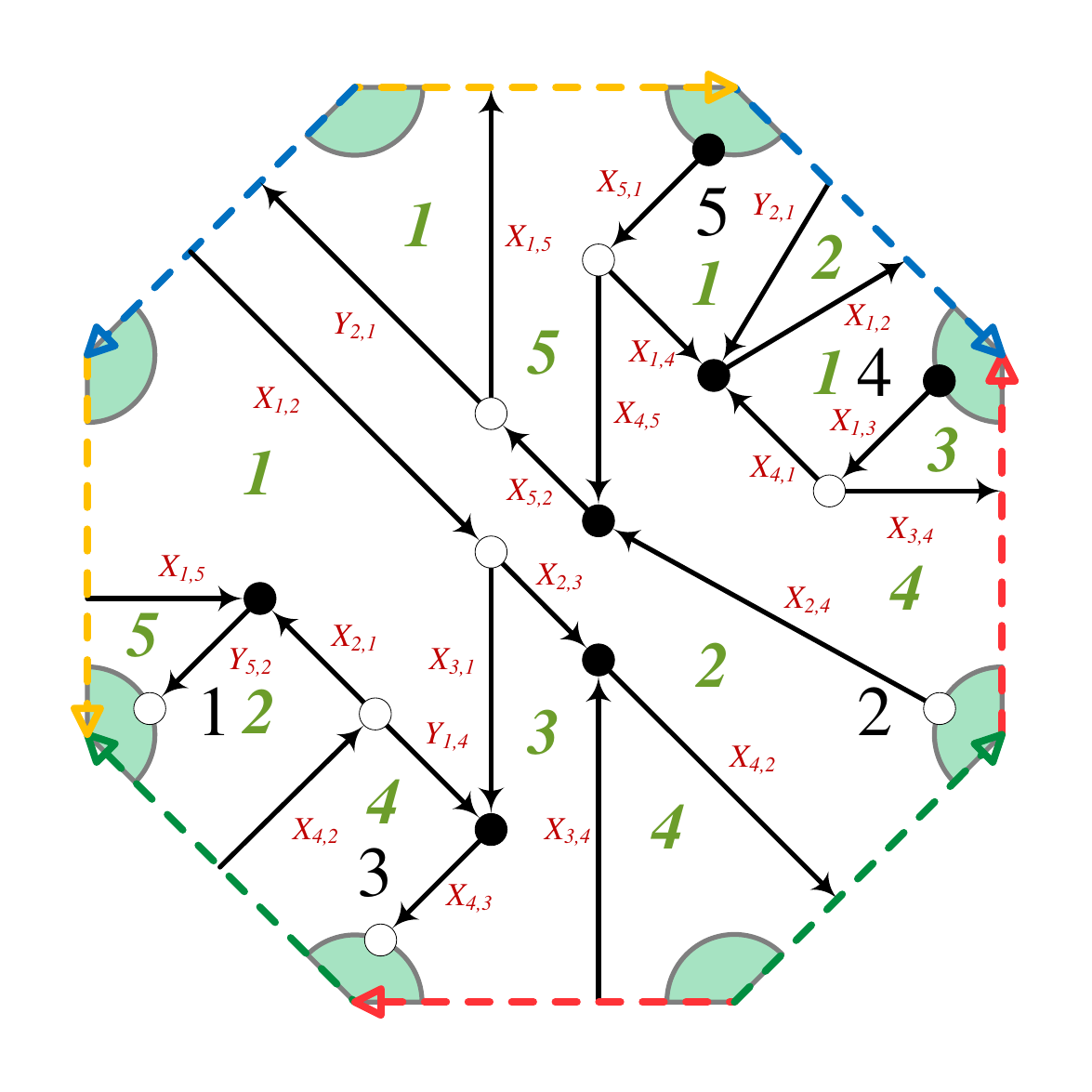}
\vspace{-0.5cm}\caption{An on-shell diagram embedded into a genus-2 surface with a single boundary. The unit cell is an octagon. Dashed arrows of the same color are identified respecting their orientation. Faces are labeled in green, external nodes in black and edges in red.}
\label{fig:genus2graph}
\end{center}
\end{figure}

Let us know apply our boundary measurement prescription to an on-shell diagram embedded into a genus-2 surface with a single boundary.  This example admits an alternative embedding into a genus-0 surface with multiple boundaries, which allows for a non-trivial check of our proposal. Genus-2 surfaces have four fundamental cycles: $\alpha_1$, $\beta_1$, $\alpha_2$, $\beta_2$. The diagram is shown in \fref{fig:genus2graph}, where we present the fundamental cell of the surface and segments on its perimeter are periodically identified according to their color and orientation. We pick a perfect orientation corresponding to the perfect matching $p_{\text{ref}}=X_{1,2} X_{1,3} X_{4,2} X_{4,3} X_{5,1} X_{5,2} Y_{5,2}$.
%

Let us now determine the boundary measurement. To do so, we first list all flows and their source sets.

{\small
\begin{equation}
\begin{array}{cllcll}
 \mathfrak{p}_1 = & \frac{X_{1,5} X_{3,1} X_{4,1} X_{4,5}}{X_{1,2} X_{1,3} X_{4,3} X_{5,1} X_{5,2} Y_{5,2}} & \textcolor{blue}{\{1,2,3\}} \quad \quad \quad \quad & \mathfrak{p}_{15} = & \frac{X_{2,1} X_{3,1} X_{3,4} X_{4,5} Y_{2,1}}{X_{1,2} X_{1,3} X_{4,2} X_{4,3} X_{5,1} X_{5,2} Y_{5,2}} \quad \quad & \textcolor{blue}{\{1,2,3\}} \\
 \mathfrak{p}_2 = & \frac{X_{1,5} X_{2,4} X_{3,1} X_{4,1}}{X_{1,2} X_{1,3} X_{4,3} X_{5,2} Y_{5,2}} & \textcolor{blue}{\{1,3,5\}} & \mathfrak{p}_{16} = & \frac{X_{1,4} X_{2,1} X_{3,1} X_{3,4}}{X_{1,2} X_{1,3} X_{4,2} X_{4,3} X_{5,1} Y_{5,2}} & \textcolor{blue}{\{1,2,3\}} \\
 \mathfrak{p}_3 = & \frac{X_{1,4} X_{1,5} X_{2,4} X_{3,1}}{X_{1,2} X_{4,3} X_{5,1} X_{5,2} Y_{5,2}} & \textcolor{blue}{\{1,3,4\}} & \mathfrak{p}_{17} = & \frac{X_{2,1} X_{2,4} X_{3,1} X_{3,4} Y_{2,1}}{X_{1,2} X_{1,3} X_{4,2} X_{4,3} X_{5,2} Y_{5,2}} & \textcolor{blue}{\{1,3,5\}} \\
 \mathfrak{p}_4 = & \frac{X_{1,5} X_{2,3} X_{4,1} X_{4,5} Y_{1,4}}{X_{1,2} X_{1,3} X_{4,2} X_{4,3} X_{5,1} X_{5,2} Y_{5,2}} \quad  \quad & \textcolor{blue}{\{1,2,3\}} & \mathfrak{p}_{18} = & \frac{X_{2,1} X_{2,3} X_{4,1}}{X_{1,2} X_{1,3} X_{4,2} Y_{5,2}} & \textcolor{blue}{\{1,2,5\}} \\
 \mathfrak{p}_5 = & \frac{X_{1,5} X_{2,3} X_{2,4} X_{4,1} Y_{1,4}}{X_{1,2} X_{1,3} X_{4,2} X_{4,3} X_{5,2} Y_{5,2}} & \textcolor{blue}{\{1,3,5\}} & \mathfrak{p}_{19} = & \frac{X_{2,3} X_{4,1} Y_{1,4}}{X_{1,2} X_{1,3} X_{4,2} X_{4,3}} & \textcolor{blue}{\{2,3,5\}} \\
 \mathfrak{p}_6 = & \frac{X_{1,5} X_{4,5}}{X_{5,1} X_{5,2} Y_{5,2}} & \textcolor{blue}{\{1,2,4\}} & \mathfrak{p}_{20} = & 1 & \textcolor{blue}{\{2,4,5\}} \\
 \mathfrak{p}_7 = & \frac{X_{1,5} X_{2,4}}{X_{5,2} Y_{5,2}} & \textcolor{blue}{\{1,4,5\}} & \mathfrak{p}_{21} = & \frac{X_{2,1} X_{2,3} X_{4,5} Y_{2,1}}{X_{1,2} X_{4,2} X_{5,1} X_{5,2} Y_{5,2}} & \textcolor{blue}{\{1,2,4\}} \\
 \mathfrak{p}_8 = & \frac{X_{1,5} X_{3,4} X_{4,5} Y_{1,4}}{X_{1,3} X_{4,2} X_{4,3} X_{5,1} X_{5,2} Y_{5,2}} & \textcolor{blue}{\{1,2,3\}} & \mathfrak{p}_{22} = & \frac{X_{2,3} X_{4,5} Y_{1,4} Y_{2,1}}{X_{1,2} X_{4,2} X_{4,3} X_{5,1} X_{5,2}} & \textcolor{blue}{\{2,3,4\}} \\
 \mathfrak{p}_9 = & \frac{X_{1,4} X_{1,5} X_{2,3} X_{2,4} Y_{1,4}}{X_{1,2} X_{4,2} X_{4,3} X_{5,1} X_{5,2} Y_{5,2}} & \textcolor{blue}{\{1,3,4\}} & \mathfrak{p}_{23} = & \frac{X_{1,4} X_{2,1} X_{2,3}}{X_{1,2} X_{4,2} X_{5,1} Y_{5,2}} & \textcolor{blue}{\{1,2,4\}} \\
 \mathfrak{p}_{10} = & \frac{X_{1,5} X_{2,4} X_{3,4} Y_{1,4}}{X_{1,3} X_{4,2} X_{4,3} X_{5,2} Y_{5,2}} & \textcolor{blue}{\{1,3,5\}} & \mathfrak{p}_{24} = & \frac{X_{1,4} X_{2,3} Y_{1,4}}{X_{1,2} X_{4,2} X_{4,3} X_{5,1}} & \textcolor{blue}{\{2,3,4\}} \\
 \mathfrak{p}_{11} = & \frac{X_{3,1} X_{4,1}}{X_{1,2} X_{1,3} X_{4,3}} & \textcolor{blue}{\{2,3,5\}} & \mathfrak{p}_{25} = & \frac{X_{2,1} X_{3,4}}{X_{1,3} X_{4,2} Y_{5,2}} & \textcolor{blue}{\{1,2,5\}} \\
 \mathfrak{p}_{12} = & \frac{X_{3,1} X_{4,5} Y_{2,1}}{X_{1,2} X_{4,3} X_{5,1} X_{5,2}} & \textcolor{blue}{\{2,3,4\}} & \mathfrak{p}_{26} = & \frac{X_{2,1} X_{2,3} X_{2,4} Y_{2,1}}{X_{1,2} X_{4,2} X_{5,2} Y_{5,2}} & \textcolor{blue}{\{1,4,5\}} \\
 \mathfrak{p}_{13} = & \frac{X_{1,4} X_{3,1}}{X_{1,2} X_{4,3} X_{5,1}} & \textcolor{blue}{\{2,3,4\}} & \mathfrak{p}_{27} = & \frac{X_{3,4} Y_{1,4}}{X_{1,3} X_{4,2} X_{4,3}} & \textcolor{blue}{\{2,3,5\}} \\
 \mathfrak{p}_{14} = & \frac{X_{2,4} X_{3,1} Y_{2,1}}{X_{1,2} X_{4,3} X_{5,2}} & \textcolor{blue}{\{3,4,5\}} & \mathfrak{p}_{28} = & \frac{X_{2,3} X_{2,4} Y_{1,4} Y_{2,1}}{X_{1,2} X_{4,2} X_{4,3} X_{5,2}} & \textcolor{blue}{\{3,4,5\}} \\
\end{array}
\end{equation}
}

\begin{figure}[h]
\begin{center}
\includegraphics[scale=0.31]{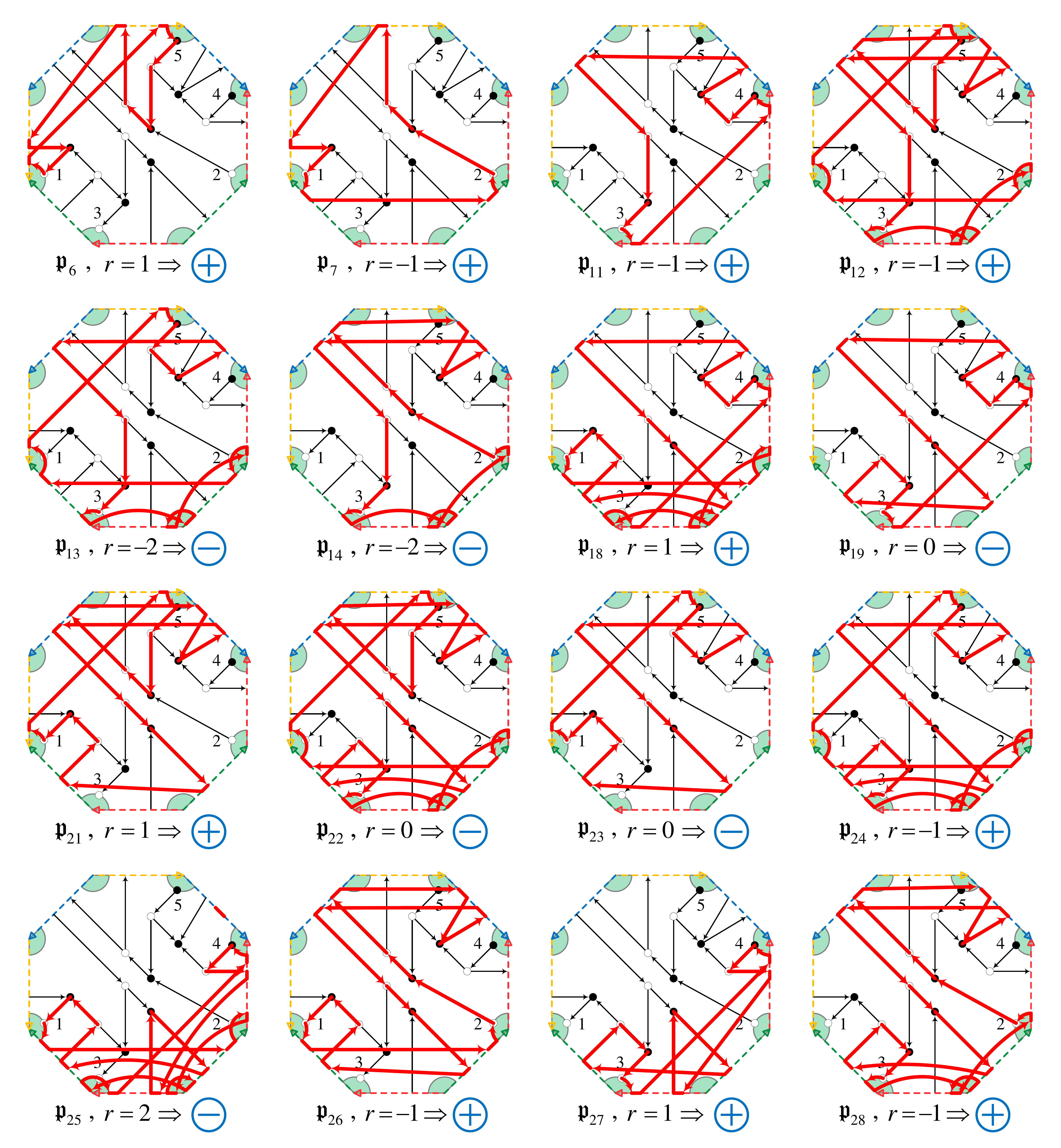}
\vspace{-0.3cm}\caption{Flows contributing to \eref{M_genus_2} completed to loops within the unit cell, the corresponding rotation numbers and the resulting signs.}
\label{fig:genus2flows}
\end{center}
\end{figure}

\noindent Including the positivity signs, we obtain the following matrix
\begin{equation} 
\left(
\begin{array}{c|ccccc}
& 1 & \ \ 2 \ \ & 3 & \ \ 4 \ \ & \ \ 5 \ \ \\ \hline
2 \ \ & \mathfrak{p}_7+\mathfrak{p}_{26} & \ \ 1 \ \ & \mathfrak{p}_{14}+\mathfrak{p}_{28} & \ \ 0 \ \ & \ \ 0 \ \ \\
4 \ \ &  -\mathfrak{p}_{18}-\mathfrak{p}_{25} & \ \ 0 \ \ & \mathfrak{p}_{11}+\mathfrak{p}_{19}+\mathfrak{p}_{27} & \ \ 1 \ \ & \ \ 0 \ \ \\
5 \ \ &  \mathfrak{p}_6+\mathfrak{p}_{21}+\mathfrak{p}_{23} & \ \ 0 \ \ & -\mathfrak{p}_{12}-\mathfrak{p}_{13}-\mathfrak{p}_{22}-\mathfrak{p}_{24} &  \ \ 0 \ \ & \ \ 1 \ \ 
\end{array}
\right) .
\label{M_genus_2}
\end{equation}
As in the previous example, its minors cannot be written as a sum of flows. It is sufficient to determine the combinatorial signs for only those flows appearing in the matrix, which are shown in \fref{fig:genus2flows} along with their respective signs.
%
%
This then yields the Grassmannian matrix
\begin{equation}
 C = 
\left(
\begin{array}{c|ccccc}
& 1 & \ \ 2 \ \ & 3 & \ \ 4 \ \ & \ \ 5 \ \ \\ \hline
2 \ \ & \mathfrak{p}_7+\mathfrak{p}_{26} & \ \ 1 \ \ & -\mathfrak{p}_{14}+\mathfrak{p}_{28} & \ \ 0 \ \ & \ \ 0 \ \ \\
4 \ \ & -\mathfrak{p}_{18}+\mathfrak{p}_{25} & \ \ 0 \ \ & \mathfrak{p}_{11}-\mathfrak{p}_{19}+\mathfrak{p}_{27} & \ \ 1 \ \ & \ \ 0 \ \ \\
5 \ \ & \mathfrak{p}_6+\mathfrak{p}_{21}-\mathfrak{p}_{23} & \ \ 0 \ \ & -\mathfrak{p}_{12}+\mathfrak{p}_{13}+\mathfrak{p}_{22}-\mathfrak{p}_{24} &  \ \ 0 \ \ & \ \ 1 \ \  
\end{array}
\right) . \label{eq:genus2C}
\end{equation}
Interestingly, here we observe a new phenomenon, exclusive of higher genus. For genus-0, in the absence of closed loops in the perfect orientation, all flows whose source and sink lie on the same boundary do not pick up any combinatorial signs. This is because they do not use cuts to be completed into loops, which in this case are the only possible sources of self-intersections. On the contrary, despite the fact that this example has only one boundary, several flows pick up a combinatorial minus sign. This effect is precisely tuned such that the minors of $C$ are subject to important cancellations that result in the simple expressions
\begin{equation}
\begin{array}{clcl}
 \Delta _{1,2,3} = & \mathfrak{p}_1-\mathfrak{p}_4+\mathfrak{p}_8+\mathfrak{p}_{15}-\mathfrak{p}_{16} \quad \quad \quad & \Delta _{1,4,5} = & \mathfrak{p}_7+\mathfrak{p}_{26} \\
 \Delta _{1,2,4} = & \mathfrak{p}_6+\mathfrak{p}_{21}-\mathfrak{p}_{23} & \Delta _{2,3,4} = & \mathfrak{p}_{12}-\mathfrak{p}_{13}-\mathfrak{p}_{22}+\mathfrak{p}_{24} \\
 \Delta _{1,2,5} = & \mathfrak{p}_{18}-\mathfrak{p}_{25} & \Delta _{2,3,5} = & \mathfrak{p}_{11}-\mathfrak{p}_{19}+\mathfrak{p}_{27} \\
 \Delta _{1,3,4} = & \mathfrak{p}_9-\mathfrak{p}_3 & \Delta _{2,4,5} = & 1 \\
 \Delta _{1,3,5} = & \mathfrak{p}_2-\mathfrak{p}_5+\mathfrak{p}_{10}+\mathfrak{p}_{17} & \Delta _{3,4,5} = & \mathfrak{p}_{28}-\mathfrak{p}_{14}
\end{array} \label{eq:genus2pluckers}
\end{equation}

We would like to stress how non-trivial this example is. Not only were we required to introduce signs for paths that start and end on the same boundary, but the signs in \eref{eq:genus2C} seem not to have any particular pattern, yet they magically produce the cancellations required to obtain \eref{eq:genus2pluckers}. Based on the examples presented, it is reasonable to conjecture that we have identified the full set of rules for constructing the boundary measurement for on-shell diagrams embedded on surfaces with arbitrary number of boundaries and genus. It would be interesting to confirm that this is the case and to find a formal derivation of our proposal.

\bigskip

\section{The Non-Planar On-Shell Form} \label{sec:computeform}

We shall now study the differential form associated to each non-planar on-shell diagram. As we have already seen in \sref{sec:generalized-faces} there are multiple ways of expressing it:
\begin{itemize}
\item Using edge variables as in \eref{eq:edges}, which straightforwardly extends to non-planar graphs. This has the advantage of manifestly displaying the $d\log$ form of the on-shell form. A slight disadvantage is that it depends on the choice of GL($1$) gauge at every internal node, which needs to be taken into account to identify $d$ independent edges.
\item Using generalized face variables as in \eref{general_integrand_face_variables}. This approach has the advantage of both displaying the $d\log$ form as well as being independent of the choice of GL($1$)'s. The determination of generalized face variables naturally involves an embedding of the diagram.
\item Using the minors of the Grassmannian, i.e.\ \pl coordinates, such as in \eref{eq:F}. While this representation hides the $d\log$ form and has a GL($k$) redundancy, it has the advantage having a more direct connection to the geometry of $Gr_{k,n}$, naturally expressed in terms of \pl coordinates.
\end{itemize}
In this section we will be primarily concerned with the third point. In particular, the on-shell forms obtained in this section correspond to having non-trivial factors $\mathcal{F}$ in \eref{eq:F}. While the discussion in the previous sections applies to general on-shell diagrams, here we focus on reduced ones. This is physically motivated by being interested in leading singularities, which imply the diagrams are reduced. Formally, it is also required by a dimensionality argument: in order to express the on-shell form in terms of minors, its rank needs to match the number of independent \pl coordinates, implying the diagram must be reduced.

\bigskip

\subsection{From Generalized Face Variables to Minors} \label{sec:integrandfromface}

A possible way of obtaining the on-shell form in term of minors of $C$ is to use generalized face variables and the boundary measurement. More explicitly, starting with the form in \eqref{general_integrand_face_variables}, we can use the boundary measurement introduced in \sref{sec:general-boundary-measurement} to obtain the map between \pl coordinates and generalized face variables. Solving for the generalized face variables will then yield the desired expression:
\begin{equation}
\prod_{i=1}^{F-1}\frac{df_i}{f_i}\prod_{j=1}^{B-1}\frac{db_j}{b_j}\prod_{m=1}^g\frac{d\alpha_m}{\alpha_m}\frac{d\beta_m}{\beta_m}\,=\,|\mathcal{J}|\, d^{\text{dim}} C \prod_{i,j,m}\frac{1}{f_i(\Delta) b_j(\Delta)\alpha_m(\Delta)\beta_m(\Delta)}  \ , 
\end{equation}
where $\Delta$ is the relevant set of \pl coordinates, and $\mathcal{J}$ is the Jacobian for the transformation between entries in the Grassmannian and generalized face variables.\footnote{Of course it is possible to do a similar thing starting from the on-shell form in terms of edge weights and using the boundary measurement to connect it to \pl coordinates. The advantage of using generalized face variables is that they automatically produce the starting point \eqref{general_integrand_face_variables}.} 

\begin{figure}[h]
\begin{center}
\includegraphics[scale=0.6]{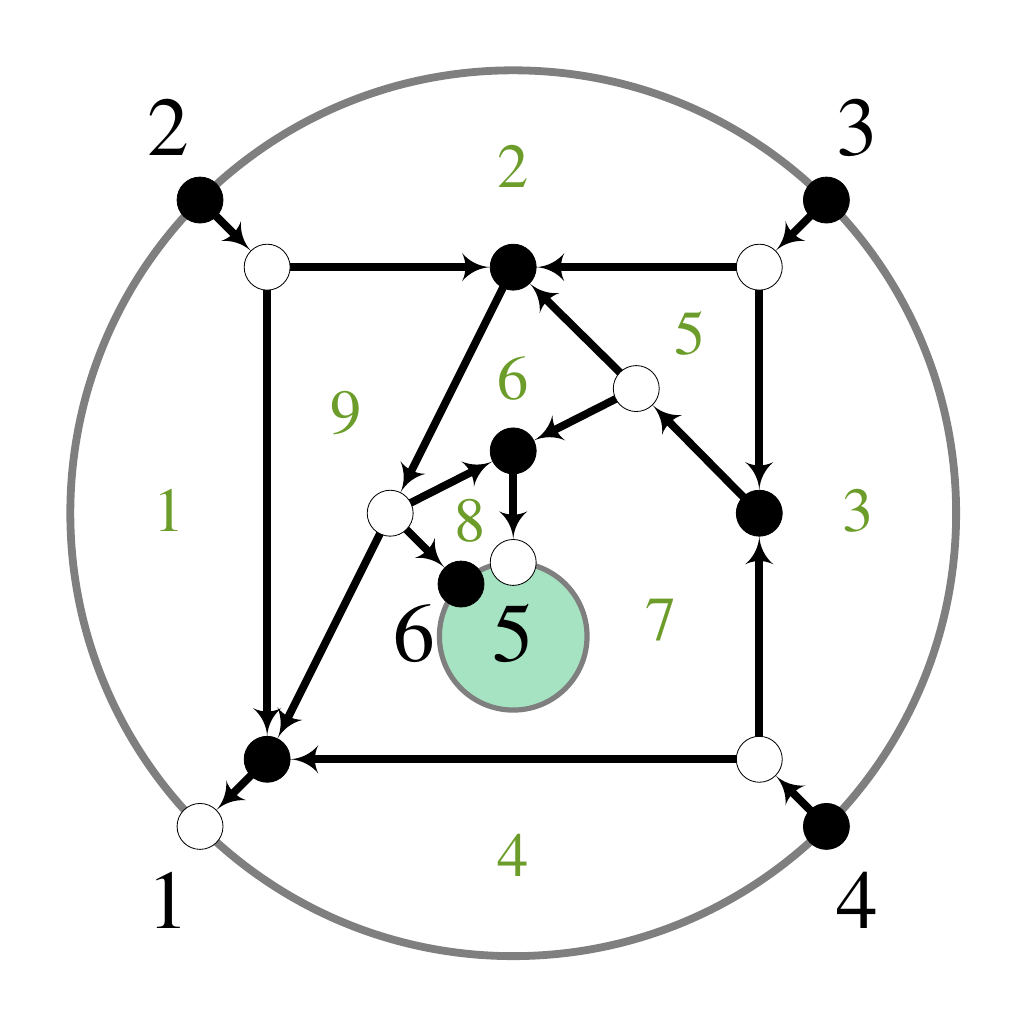}
\caption{A top-dimensional on-shell diagram in $Gr_{3,6}$ embedded on an annulus. The selected perfect orientation has source set $\{2,3,4\}$.}
\label{fig:NMHVex1}
\end{center}
\end{figure}

We shall now illustrate how this works in practice in a top-dimensional example in $Gr_{3,6}$ with two boundaries, shown in \fref{fig:NMHVex1}.

This example has 9 independent generalized face variables: 8 independent $f_i$ variables and one $b_j$. In terms of oriented edge weights, the generalized face variables are given by
\beq
\begin{array}{rlcrlcrlcrlcrl}
f_1=& \dfrac{X_{9,1}}{X_{1,2}X_{1,4}} & \ \ &  f_2= & \dfrac{X_{5,2}X_{1,2}}{X_{2,3}X_{2,9}} & \ \ & f_3= & \dfrac{X_{7,3}X_{2,3}}{X_{3,4}X_{3,5}} & \ \ & f_4= & \dfrac{X_{1,4}X_{3,4}}{X_{4,7}} & \ \ & f_5= & \dfrac{X_{6,5}X_{3,5}}{X_{5,7}X_{5,2}}\\[10pt]
f_6=&\dfrac{X_{7,6}X_{9,6}}{X_{6,8}X_{6,5}} & & f_7= & \dfrac{Y_{8,7}X_{5,7}X_{4,7}X_{8,7}}{X_{7,6}X_{7,3}X_{7,9}} & & f_8= & \dfrac{X_{6,8}}{X_{8,7}Y_{8,7}} & & f_9=& \dfrac{X_{2,9}X_{7,9}}{X_{9,6}X_{9,1}} & & b_1=& \dfrac{X_{1,4}X_{8,7}}{X_{7,9}}
\end{array}
\eeq 
Eliminating $f_4$ using $\prod_{i=1}^9 f_i=1$ we obtain the on-shell form
\begin{equation}
\Omega = \frac{db_1}{b_1} \prod_{i \neq 4}^{9}\frac{df_i}{f_i}  .
\end{equation}

Using the boundary measurement in \sref{sec:general-boundary-measurement}, we obtain the following matrix
{\footnotesize
\beq
C\,=\, \left(
\begin{array}{c|cccccc}
& 1 & \ \ 2 \ \ & \ \ 3 \ \ & \ \ 4 \ \ & 5 & 6 \\ \hline
2 \ \ & f_1 (1 + f_9) &  \ \ 1 \ \ &  \ \ 0 \ \ & \ \ 0 \ \ &  b_1 f_1 f_8 f_9 &  b_1 f_1 f_9\\[10pt]
3 \ \ &  -f_1 f_2 (1 + f_5) f_9 &  
  \ \ 0 \ \ & \ \ 1 \ \ & \ \ 0 \ \ &  -b_1 f_1 f_2 (1 + f_5 + f_5 f_6) f_8 f_9 &  -b_1 f_1 f_2 (1 + 
     f_5) f_9\\[10pt]
4 \ \ &  f_1 f_2 f_3 f_5 (1 + f_6 f_7 f_8) f_9 & \ \ 0 \ \ & \ \ 0 \ \ & \ \ 1 \ \  &  
  b_1 f_1 f_2 f_3 f_5 (1 + f_6) f_8 f_9 &  b_1 f_1 f_2 f_3 f_5 f_9
\end{array} \right) .
\label{eq:CforG36}
\eeq}
The variable transformation from generalized face variables to elements of the above matrix, i.e.\ to $ \prod_{i=1}^9 dc_i \equiv d^9 C$, carries a Jacobian, which can also be expressed in terms of the generalized face variables. 

Using \eref{eq:CforG36} we can express the \pl coordinates in terms of generalized face variables. Solving for the generalized face variables, we obtain the following differential form:
\begin{equation}
\label{eq:example1}
\Omega\,=\, \prod_{i\neq 4}^9 \frac{df_i}{f_i}\frac{db_1}{b_1}\,=\,d^9 C \dfrac{(246)^2}{(234)(345)(456)(612)(124)(146)(236)(256)}\ .
\end{equation}
An important remark is that the resulting expression in terms of minors is independent of the chosen embedding. The simple example in Appendix \ref{section_simple_example} illustrates this point.

\bigskip

\subsection{A Combinatorial Method} 

\label{section_combinatorial_on_shell}

In this section we present an alternative systematic procedure for computing the non-planar on-shell form in terms of \pl coordinates for any MHV degree $k$, which allows us to construct it without the need to compute the boundary measurement. This is a generalization of the method developed in \cite{Arkani-Hamed:2014bca} for general non-planar MHV leading singularities. We will begin by quickly reviewing the procedure in \cite{Arkani-Hamed:2014bca}, and then propose its generalization to any $k$. As a consistency check, all results in this section have also been obtained using the method in \sref{sec:integrandfromface}, providing substantial support for our proposal.

\subsubsection{MHV Leading Singularities}
\label{sec: MHV LS}

\begin{figure}[h]
\begin{center}
\includegraphics[width=6cm]{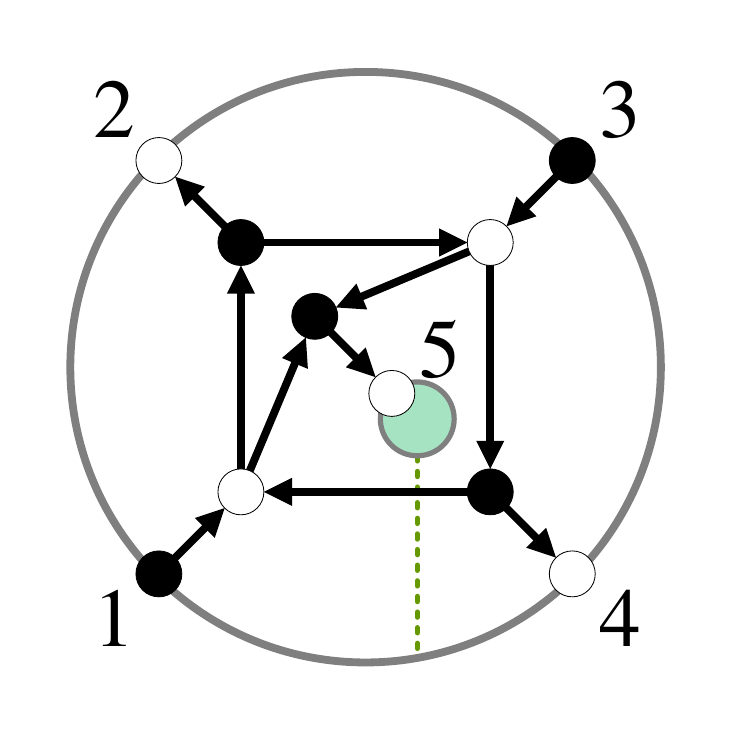}
\caption{A five-point MHV on-shell diagram with two boundaries.}
\label{fig:non-planar-MHV}
\end{center}
\end{figure}

A general method for obtaining non-planar MHV leading singularities was recently introduced in \cite{Arkani-Hamed:2014bca}. We now review this method with a simple example, shown in \fref{fig:non-planar-MHV}.

A general feature of MHV leading singularities is that every internal black vertex can be associated to a set of three external legs. These legs are those that are connected to the black node either directly or through a sequence of edges and internal white nodes. The previous sentence applies to non-necessarily bipartite on-shell diagrams. As explained earlier, every on-shell diagram can be turn into a bipartite one. We will continue focusing on bipartite diagrams, for which it is clear that there can only be at most one internal white node connecting an internal black node to an external leg.\footnote{It is natural to speculate that this basic observation can be turned into a new quantitative characterization of reduced graphs. It seems to suggest that a necessary condition for a bipartite on-shell diagram to be reduced is that all internal black nodes must be at a distance equal or smaller than 2 (as measured in terms of edges following our prescription) from some external node. Not surprisingly, this would tell us that reduced graphs need to be ``small" or ``narrow" in some sense. We leave a more detailed investigation of this thought for future work. \label{footnote_distance_boundary}} The fact that for MHV leading singularities this rule precisely gives rise to three end points for every internal black node is indeed a rather non-trivial graph-combinatorial result.

The procedure for obtaining the differential form is as follows:
\begin{enumerate}
\item For each internal black node, we find the three external legs associated to it. Then we construct a $n_B \times 3$ matrix $T$, where each row contains the labels of the three external nodes associated to each black node. For the example in \fref{fig:non-planar-MHV}, $T$ is given by
\begin{align}
T=\begin{pmatrix}
\ 1 \ & \ 2 \ & \ 3 \ \ \\
\ 1 \ & \ 3 \ & \ 5 \ \ \\
\ 1 \ & \ 3 \ & \ 4 \ \
\end{pmatrix} .
\end{align}
\item Next, we construct an $n_B \times n$ matrix $M$ in the following manner. For each row $\{i,j,k\}$ in $T$ we construct a corresponding row in $M$ by inserting $(i\,j)$ at position $k$, $(j\,k)$ at position $i$, $(k\,i)$ at $j$, and zero for the remaining entries. For our example, we get
\begin{align}
M=\begin{pmatrix}
\ (23) & (31) & (12) & 0 & 0 \\
\ (35) & 0 & (51) &  0 & (13) \\
\ (34) & 0 & (41) & (13) & 0
\end{pmatrix} .
\end{align}
\item We delete two arbitrary columns $a$ and $b$ from the matrix $M$, to obtain the square matrix $\widehat{M}_{a,b}$ of size $n_B \times (n-2) = n_B \times n_B$. We then compute $\det(\widehat{M}_{a,b})/(ab)$. This quantity turns out to be independent of the choice of $a$ and $b$. For the case at hand, we have $\det (\widehat{M}_{a,b}/(ab))=-(13)^2$.
\item Finally, the on-shell form corresponding to a diagram for which
\begin{equation}
\label{eq:Tmatrix}
T=\begin{pmatrix}
i^{(1)}_1 & i^{(1)}_2 & i^{(1)}_{3}\\
i^{(2)}_1 & i^{(2)}_2 & i^{(2)}_{3}\\
\vdots & \vdots & \vdots\\
i^{(n_B)}_1 & i^{(n_B)}_2 &  i^{(n_B)}_{3}
\end{pmatrix}
\end{equation}
is given by
\begin{align}
\label{eq:LS2}
\Omega\,=\,\frac{d^{2\times n}C}{\text{Vol(GL}(2))} \left(\frac{\det(\widehat{M}_{i,j}) }{(i\,j)}\right)^2\frac{1}{\text{PT}^{(1)}\text{PT}^{(2)}\cdots\text{PT}^{(n_B)}} ,
\end{align}
where we denote by $\text{PT}^{(i)}$ the Parke-Taylor-like product corresponding to each row $i$ of $T$; for instance in \eqref{eq:Tmatrix}, $ \text{PT}^{(1)}\, =\, (i^{(1)}_1 i^{(1)}_2) (i^{(1)}_2i^{(1)}_{3})(i^{(1)}_{3} i^{(1)}_{1})$.
For the example in \fref{fig:non-planar-MHV}, the differential form obtained from the above procedure is
\begin{equation}
\Omega\,= \frac{d^{2\times 5}C}{\text{Vol(GL}(2))}\, \frac{ (13)^4 }{ (12)(23)(31)(13)(35)(51) (13)(34)(41) } \ .
\end{equation}
\end{enumerate}

The original rules \cite{Arkani-Hamed:2014bca} are formulated in terms of spinor brackets $\langle i\,j \rangle$; for MHV leading singularities these are equivalent to $(i\,j)$ on the support of the kinematic constraints.  Writing them in terms of minors hints at an appropriate generalization to N$^{k-2}$MHV diagrams, for which the minors are $k \times k$, which we now investigate.

\bigskip

\subsubsection{Generalization to N$^{k-2}$MHV On-Shell Diagrams} \label{sec:rules}

Here we propose a generalization of the procedure shown above to $k>2$. Subsequent sections will illustrate its inner workings with some non-trivial examples. In \sref{sec:proof} we will prove the method for certain subclasses of diagrams. 

MHV leading singularities only require us to take into account on-shell diagrams with trivalent black vertices, but for $k>2$ we will need to consider more general bipartite graphs. The complications arising when $k>2$ are twofold:

\smallskip
\begin{itemize}
\item In order to have $k \times k$ minors we need a $T$ matrix with $k+1$ columns. For $k>2$ it is possible that some internal black nodes do not connect to $k+1$ external legs in the way described for $k=2$.
\item The number of black nodes may exceed $(n-k)$, forcing $\widehat{M}$ to have more rows than columns, thus preventing us from taking its determinant.
\end{itemize}
\smallskip

\begin{figure}[h]
\begin{center}
\includegraphics[scale=0.5]{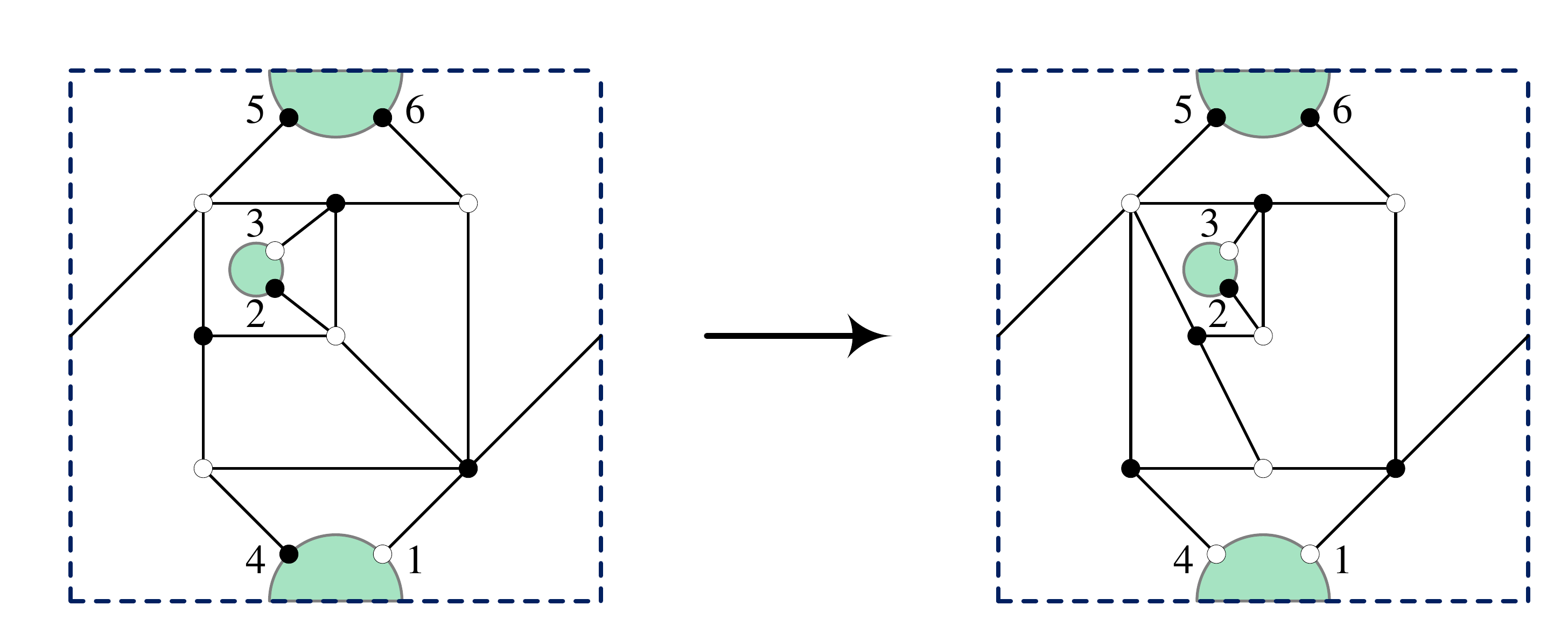}
\caption{On the left, an NMHV diagram where the black node attached to external node 1 has valency $v>k+1$. This is resolved by performing a square move, leading to the diagram on the right, where all nodes have $v \leq k+1$.}
\label{fig:largevalency}
\end{center}
\end{figure}

The first point is related to the valency $v$ of internal black nodes. There are two possible reasons why internal black nodes might fail to connect to $k+1$ external ones. The first one is that the valency of the node is $v>k+1$. Generally, performing a square move changes the valency of nodes in a diagram. In what follows we will assume that it is always possible to perform a series of equivalence moves to turn a diagram into one where every black node has $v\leq k+1$. An example of this procedure is given in \fref{fig:largevalency}.

If, on the other hand, the valency of an internal black node is $v<k+1$, we assign the first entries of the corresponding row in $T$ to the external nodes to which the black node connects to, i.e. $\{i_1,\dots, i_v\}$, 
 and leave the remaining $k+1-v$ entries free, which we denote by $\{*_{v+1}, \dots *_{k+1}\}$. The $(k+1)$-tuple associated to the given black node is then (the order of the labels is irrelevant):

\begin{equation}
\label{eq:*}
\{ i_1, \ldots, i_v, *_{v+1}, \dots *_{k+1}\}\ .
\end{equation}
We then fill these additional entries with external labels, chosen arbitrarily from the set of nodes that do not already appear in the row, i.e.\ $*_j \notin \{i_1, \ldots, i_v\}$. Finally, we need to order the new entries $*_j$ among the $\{i_1, \ldots, i_v\}$, such that $\det(\widehat{M}_{a_1,\dots,a_{k}})/(a_1\cdots a_{k})$ is independent of $\{a_1\cdots a_{k}\}$, up to an overall sign $(-1)^{\sum^k_{j=1} a_j}$. In all cases we have considered, it is always possible to do this, but it would be interesting to understand better how to determine the correct ordering in the $T$ matrix in general.

The second complication listed above, regarding the total number of black nodes, typically arises when the diagram has internal white nodes which are completely surrounded by black nodes. Notice that for bipartite graphs, this is always the case, unless when the internal white nodes are directly connected to some external leg. In the examples we have studied, it appears that $n_B=n-k+\alpha$, where $\alpha$ is the number of such white nodes in the diagram. This issue is resolved by adding an auxiliary external leg to every internal white node contributing to $\alpha$.\footnote{It is interesting to notice that, when thinking in terms of an embedding, this operation can generate new boundary components. In addition, if applied to a reducible graph it can turn it into a reduced one. This is related to our comment in footnote \ref{footnote_distance_boundary}.} Once the form has been obtained, through the generalization of the steps in \sref{sec: MHV LS} which we will outline shortly, we integrate over the extra variables $c_{ij}$, $j=n+1, \ldots, n+\alpha$ around $c_{ij}=0$. We will see this done in detail in several examples.

In summary, the procedure to obtain the differential form for general N$^{k-2}$MHV on-shell diagrams is as follows:
\begin{enumerate}
\item If any internal black node is connected to more than $k+1$ external nodes either directly or through a succession of edges and internal white nodes, perform a series of equivalence moves until all internal black nodes only connect to $k+1$ or fewer external nodes. Also, if $n_B>n-k$, add auxiliary external legs to the internal white nodes which are totally surrounded by internal black nodes, until $n_B=n-k$.
\item Construct the $n_B \times (k+1)$ matrix $T$ where each row corresponds to an internal black node. Every time there is an internal black node that connects to fewer than $k+1$ external nodes, choose the remaining entries freely as described above.\label{item2-rules} 
\item Construct the $n_B \times n$ matrix $M$ in the same way as for the MHV case. For each row $\{i_1,\dots, i_j,\dots,i_{k+1}\}$ in $T$ populate the same row in $M$ as follows. At each position $i_j$, insert the minor $(-1)^{j-1}(i_1 \cdots \hat{i}_{j}\cdots i_{k+1})$ obtained by removing $i_j$ and all other entries are zero.
\label{item3-rules}
\item Remove $k$ columns from $M$, chosen arbitrarily, to form $\widehat{M}_{a_1,\dots,a_{k}}$. Then compute the ratio $(-1)^{\sum\limits_{i=1}^{k} a_i}\det(\widehat{M}_{a_1,\dots,a_{k}})/(a_1\cdots a_{k})$.  We emphasize that this quantity is independent of the choice of $\{a_1,\dots, a_{k}\}$; as will be shown in \sref{sec:proof} different choices of ${a_1,\dots,a_{k}}$ simply correspond to different GL$(k)$ gauge choices.
\label{item4-rules}
\item The on-shell form corresponding to a diagram for which 
\begin{equation}
\label{eq:Tmatrix2}
T=\begin{pmatrix}
i^{(1)}_1 & i^{(1)}_2 & \cdots & i^{(1)}_{k+1}\\
i^{(2)}_1 & i^{(2)}_2 & \cdots & i^{(2)}_{k+1}\\
\vdots & & & \vdots\\
i^{(n_B)}_1 & i^{(n_B)}_2 & \cdots & i^{(n_B)}_{k+1}\\
\end{pmatrix}
\end{equation}
is given by
\begin{align}
\label{eq:LS}
\Omega\,=\,\frac{d^{k\times n}C}{\text{Vol(GL}(k))} \left(\frac{(-1)^{\sum\limits_{i=1}^k a_i}\det(\widehat{M}_{a_1,\dots,a_{k}}) }{(a_1\cdots a_{k})}\right)^k\frac{1}{\text{PT}^{(1)}\text{PT}^{(2)}\cdots\text{PT}^{(n_B)}}\ ,
\end{align}
\noindent where we denote by $\text{PT}^{(i)}$ the Parke-Taylor-like product corresponding to each row $i$ of $T$, for instance in \eref{eq:Tmatrix2}, $ \text{PT}^{(1)}\, =\, (i^{(1)}_1\cdots i^{(1)}_k) (i^{(1)}_2\cdots i^{(1)}_{k+1}) \cdots (i^{(1)}_{k+1}\cdots i^{(1)}_{k-1})$. If there was no need for introducing auxiliary external legs, this is the final answer.
\item In the presence of auxiliary legs, we now need to integrate over the extra variables $C_{ij}$, $j=n+1, \ldots, n+\alpha$ around $C_{ij}=0$. Below we present various examples in which this is done.
\end{enumerate}

An interesting observation is that for every row in $T$ where we have undetermined entries $\{i_1, \ldots ,i_v,*_{v+1}, \ldots ,*_{k+1} \}$, any minor involving the columns $\{i_1, \dots, i_v\}$ vanishes. This will be proven below in \sref{sec:star}. 

\bigskip

\subsection{Examples}

\label{section_method_examples}

We now illustrate the rules introduced in the previous section on various explicit examples.

\bigskip

\subsubsection{NMHV with Low Valency} \label{sec:example1}

\begin{figure}[h]
\begin{center}
\includegraphics[width=5cm]{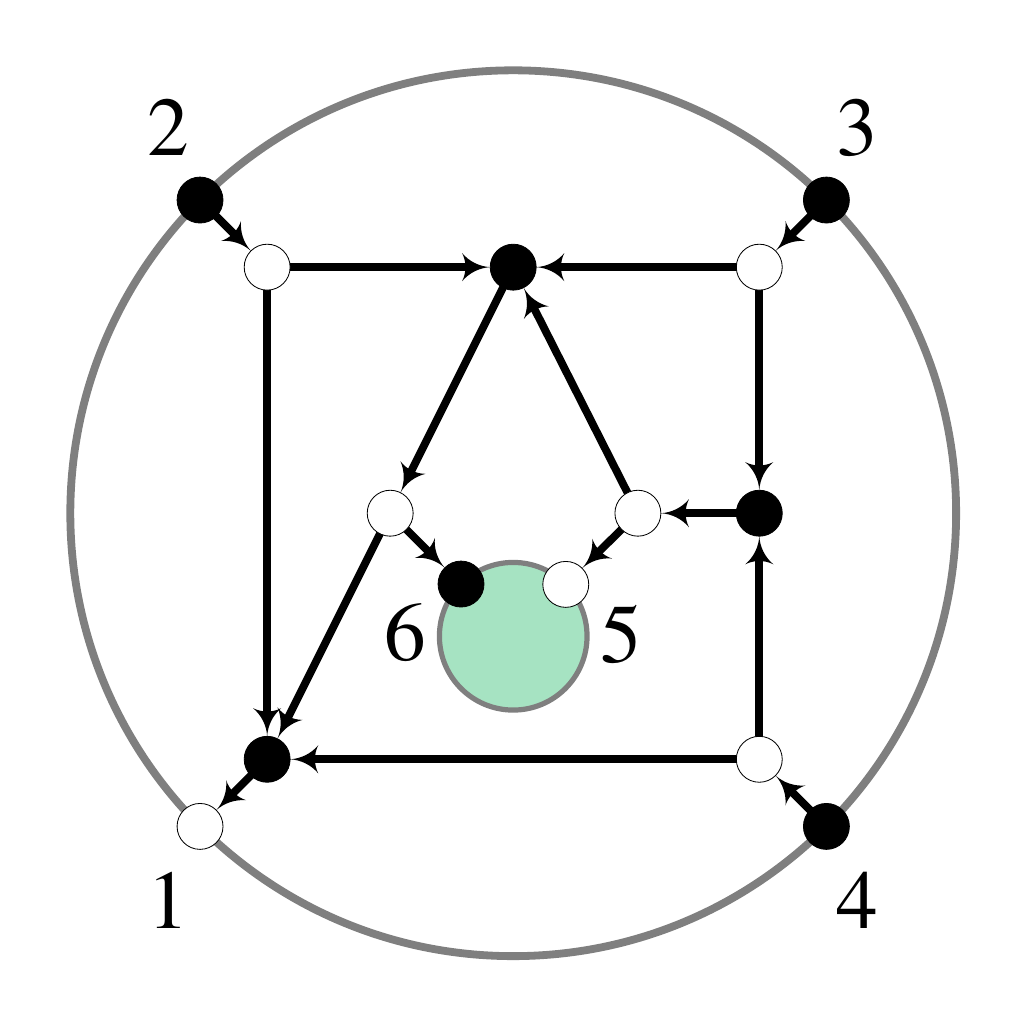}
\vspace{-.3cm}\caption{NMHV leading singularity with $(345)=0$.}
\label{fig:NMHV1}
\end{center}
\end{figure}

Our first example illustrates how to deal with cases when we need to introduce $*$ into the matrix $T$. The diagram is shown in \fref{fig:NMHV1}. We will also show that this diagram is decomposable into a sum of Parke-Taylor factors through the use of Kleiss-Kuijf relations \cite{Kleiss:1988ne}, thus independently confirming the answer.

Since $n_B=n-k$ and all internal black nodes connect to a maximum of $k+1=4$ external nodes, no manipulations of the diagram are required. The $T$ matrix is given by
\begin{align}
\label{eq:T-star}
T\,=\,\begin{pmatrix}
1 & 2 & 6 & 4 \\
2 & 3 & 5 & 6 \\
5 & 3 & 4 & *
\end{pmatrix} ,
\end{align}
where we may choose $*=1,$ $2$ or $6$. The final answer is independent of this choice, and in the following we choose $*=2$. From the bottom row we can also immediately read off that the minor $(345)=0$, as proven in \sref{sec:star}.

We shall now construct the matrix $M$. We have
\begin{align}
T\,=\,\begin{pmatrix}
1 & 2 & 6 & 4 \\
2 & 3 & 5 & 6 \\
5 & 3 & 4 & 2
\end{pmatrix} \; \rightarrow \;
M\,=\,\begin{pmatrix}
(264) & -(164) & 0 & -(126) & 0 & (124) \\
0 & (356) & -(256) & 0 & (236) & -(235) \\
0 & -(534) & -(542) & (532) & (342) & 0
\end{pmatrix} .
\end{align}
Deleting columns $2$, $3$, and $4$ we get
\begin{align}
\widehat{M}_{2,3,4}\,&=\,\begin{pmatrix}
(264) &  0 & (124) \\
0 &  (236) & -(235) \\
0 & (342) & 0
\end{pmatrix} \quad \Rightarrow \quad \frac{\det \widehat{M}_{2,3,4}}{(234)}=-(264)(235) .
\end{align}
Thus, the on-shell form corresponding to the leading singularity in \fref{fig:NMHV1} is given by
\begin{align}
\label{eq:LS1}
\Omega\,=\,\left. \frac{d^{3\times 6}C}{\text{Vol(GL}(3))} \frac{(264)^2(235)}{(126)(641)(412)(356)(562)(623)(342)(425)(345)}\right|_{(345)=0} .
\end{align}

Although we do not have a general proof for the independence of the choice of $*$ and the deleted rows of $M$, this example provides strong evidence to believe this is indeed the case. For the example at hand, we have checked explicitly that this result agrees with the differential form in terms of edge or generalized face variables for any choice of GL($3$) gauge fixing, deleted rows as well as for  $*=1$ or $6$. For this particular example, \eqref{eq:LS1} can be explicitly confirmed to be correct: this leading singularity can be written in terms of planar integrals, with the help of the Kleiss-Kuijf relations \cite{Kleiss:1988ne} on the four-point nodes present in the diagram in \fref{fig:NMHV1}. Explicitly, using \pl relations at the pole $(345)=0$ one may rewrite the ratio in \eqref{eq:LS1} as
\begin{align}
\label{eq:LS1planar}
\begin{split}
 & \left. \frac{(264)^2(235)}{(126)(641)(412)(356)(562)(623)(342)(425)(345)}\right|_{(345)=0} \\
=\, & I(1,6,2,3,5,4) +I(1,6,2,5,3,4) + I(1,2,6,3,5,4)+ I(1,2,6,5,3,4) ,
\end{split}
\end{align}
where $I(i_1, i_2, i_3, i_4, i_5, i_6)$ stands for the planar integrals with ordering indicated by their arguments:
\beq
I(i_1, i_2, i_3, i_4, i_5, i_6) = {1 \over (i_1 i_2 i_3)(i_2 i_3 i_4)(i_3 i_4 i_5)(i_4 i_5 i_6)(i_5 i_6 i_1)(i_6 i_1 i_2)}.
\eeq

For MHV diagrams, \cite{Arkani-Hamed:2014bca} showed that every non-planar leading singularity can be re-expressed as a sum of Parke-Taylor factors with coefficients $+1$. This is not a general feature of N$^{k-2}$MHV leading singularities, as will become clear with the example in \sref{section:complicatedcase}. 
In Appendix \ref{app:highergenus} we present a similar, higher genus, example.

\bigskip

\subsubsection{NMHV with Too Many Black Nodes} \label{sec:example2}

\begin{figure}[h]
\begin{center}
\includegraphics[scale=0.5]{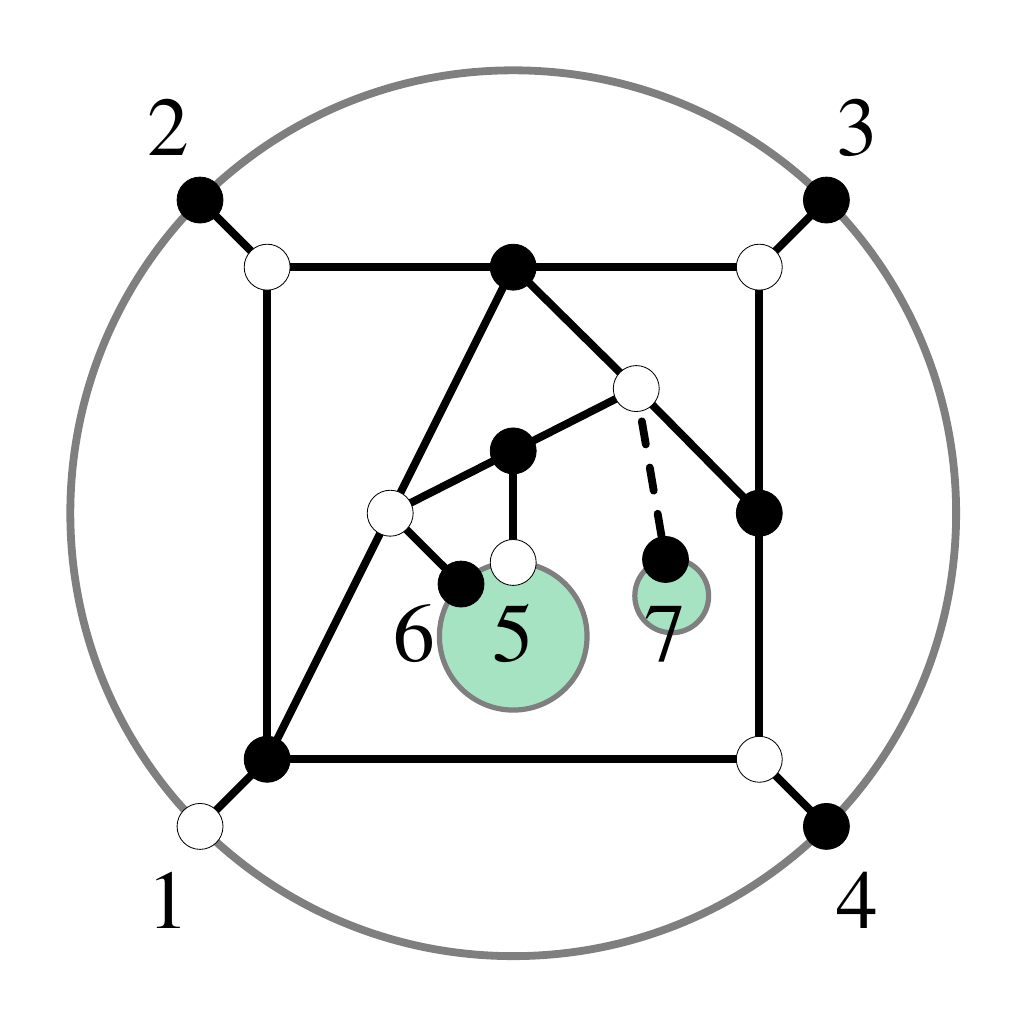}
\caption{NMHV leading singularity with $n_B>n-k$. This requires the introduction of an auxiliary leg, indicated by a dashed line and numbered 7.}
\label{fig:NMHV2}
\end{center}
\end{figure}

Let us now consider diagrams with $n_B > n-k$. An example of this type is provided in \fref{fig:NMHV2}, which is obtained by adding a BCFW bridge to legs 5 and 6 in \fref{fig:NMHV1}. Hence, the two examples must agree on the pole $(345)=0$, which provides us with an additional check of the validity of the procedure in \sref{sec:rules}.

This example has $n_B = n-k+1$. Following \sref{sec:rules}, we must introduce an auxiliary leg as shown in \fref{fig:NMHV2}. This new diagram yields the $T$ matrix
\begin{align}
T\,=\,\begin{pmatrix}
\ 1 \ \ & \ 2 \ \ & \ 6 \ \ & \ 4 \ \ \\
2 & 3 & 7 & 6 \\
7 & 3 & 4 & * \\
5 & 6 & 7 & * \\
\end{pmatrix}\quad\xrightarrow{\text{Choice of }*}\quad T=\begin{pmatrix}
\ 1 \ \ & \ 2 \ \ & \ 6 \ \ & \ 4 \ \ \\
2 & 3 & 7 & 6 \\
7 & 3 & 4 & 2 \\
5 & 6 & 7 & 2 \\
\end{pmatrix}\ .
\end{align}
Notice how from the last two rows of $T$ we learn that $(734)=(567)=0$. 

\noindent This gives the following matrix $M$
\begin{align}
M\,=\,\begin{pmatrix}
\label{eq:M7}
(264) & -(164) & 0 & -(126) & 0 & (124) & 0 \\
0 & (376) & -(276) & 0 & 0 & -(237) & (236) \\
0 & -(734) & -(742) & (732) & 0 & 0 & (342) \\
0 & -(567) & 0 & 0 & (672) & -(572) & (562) 
\end{pmatrix} ,
\end{align}

\noindent which results in the on-shell form
\begin{align}
\Omega\, =\,\frac{d^{3\times 7}C}{\text{Vol(GL}(3))} {(264)^2 \over (126)(641)(412)(623)(234)(256)} \times \left.I\right|_7 ,
\end{align}
where $\left.I\right|_7$ stands for the piece containing the dependence on the auxiliary external node $7$ and must be evaluated at the poles $(347)=(567)=0$. On these poles, it can be recast as
\begin{align}
\left. I\right|_7\,=\, {(256)  \over (456)(347)(567)(725)} .
\end{align}

The final step is to remove the effect of the auxiliary edge. This is done by taking a generic element of the ``extended'' Grassmannian $Gr_{k,n+1}$ and integrating the extra variables $C_{i7}$ around $C_{i7}=0$. To do so, we write a generic $3\times 7$ matrix $C$ and compute the residues of $\left. I\right|_7$ around $C_{i7}=0,\,i=1,2,3$. We obtain
\begin{align}
\Omega\, =\,\frac{d^{3\times 6}C}{\text{Vol(GL}(3))}\dfrac{(246)^2}{(234)(345)(456)(612)(124)(146)(236)(256)} .
\end{align}
As expected, this result agrees with the leading singularity \eqref{eq:LS1} on the support of $(345)=0$.

With the previous two examples, we have illustrated the full set of our tools. As an additional demonstration of the power of this procedure, in Appendix \ref{app:NNMHV} we compute a highly non-trivial N$^2$MHV example.

\bigskip

\subsubsection{NMHV with a New Type of Poles} \label{section:complicatedcase}

\begin{figure}[h]
\begin{center}
\includegraphics[scale=0.75]{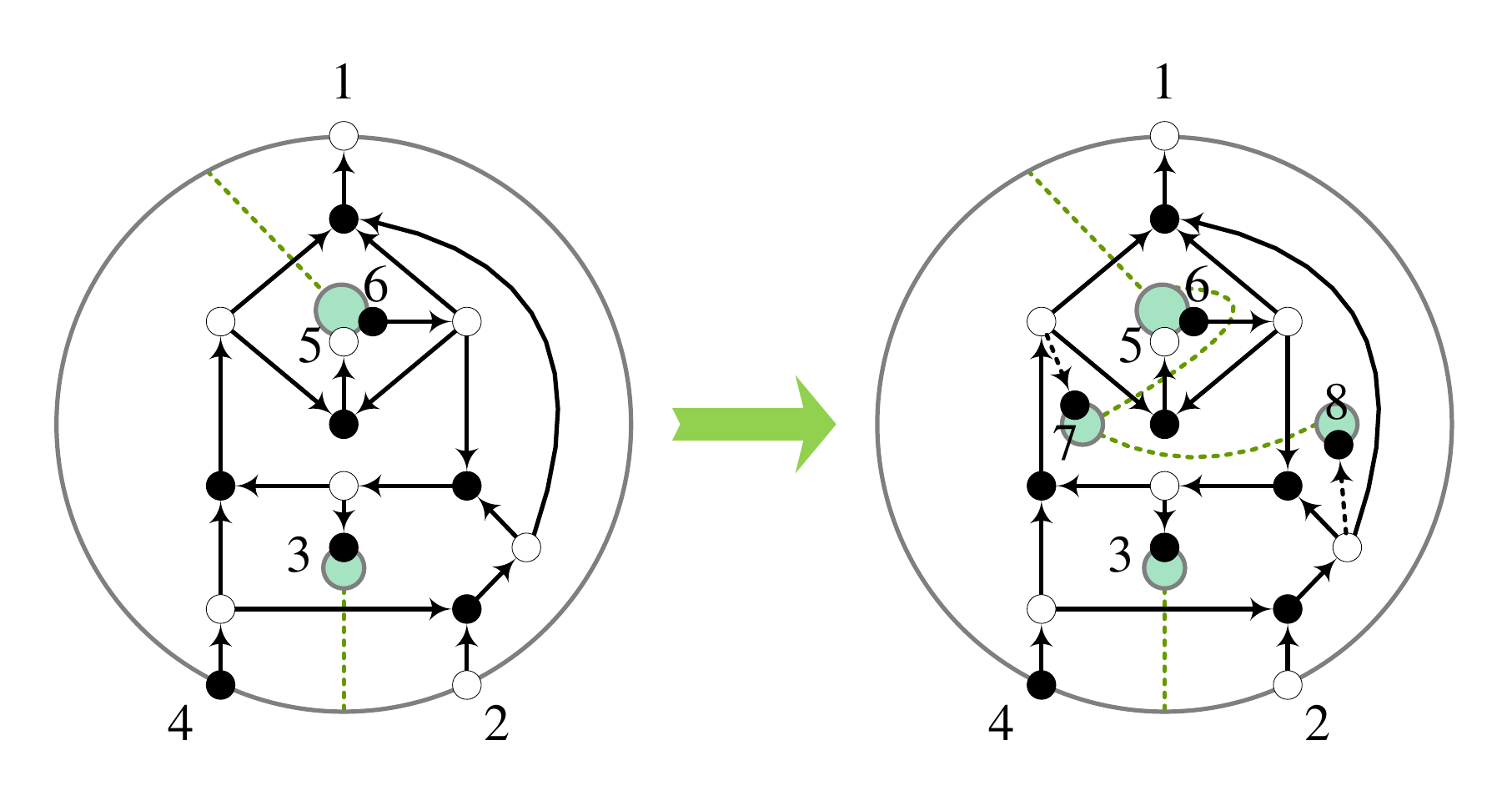}
\caption{Left: an NMHV top-dimensional diagram in $Gr_{3,6}$. Right: this diagram requires the addition of two auxiliary legs, here shown with dashed arrows and terminating on external nodes 7 and 8. This example has a non-standard singularity when $(124)(346)(365)-(456)(234)(136)=0$.} 
\label{fig:ProcedureWeirdPole}
\end{center}
\end{figure}

We shall now apply our tools to computing a top-dimensional example in $Gr_{3,6}$ which exhibits a novel feature: a differential form with a singularity which is not of the form $(ijk)=0$. This fact ultimately prohibits the diagram from being able to be written as a sum of planar terms. The on-shell diagram is shown in \fref{fig:ProcedureWeirdPole}. This example will also be revisited in \sref{sec:reducibility}, where the consequences of such a peculiar differential form will be studied in detail.

The $T$ matrix is
{\small
\begin{align}
T\,=\,\begin{pmatrix}
\ 1 \ \ & \ 8 \ \ & \ 6 \ \ & \ 7 \ \ \\
5 & 6 & 7 & * \\
6 & 8 & 3 & * \\
8 & 2 & 4 & * \\
7 & 3 & 4 & * \\
\end{pmatrix}\quad\xrightarrow{\text{Choice of }*}\quad T=\begin{pmatrix}
\ 1 \ \ & \ 8 \ \ & \ 6 \ \ & \ 7 \ \ \\
5 & 6 & 7 & 2 \\
6 & 8 & 3 & 2 \\
8 & 2 & 4 & 6 \\
7 & 3 & 4 & 2 \\
\end{pmatrix}\ ,
\end{align}
}
from which we can immediately read off that
\begin{align}
(347)=(567)=(368)=(248)=0 . 
\end{align}

\noindent From $T$, we construct the matrix $M$
{\small
\begin{align}
M\,=\,\begin{pmatrix}
\label{eq:M8}
(867) & 0 & 0 & 0 & 0 & (187) & -(186) & -(167) \\
0 & -(567) & 0 & 0 & (672) & -(572) & (562) & 0 \\
0 & -(683) & (682) & 0 & 0 & (832) & 0 & -(632) \\
0 & -(846) & 0 & (826) & 0 & -(824) & 0 & (246) \\
0 & -(734) & -(742) & (732) & 0 & 0 & (342) & 0
\end{pmatrix} .
\end{align}
}

The resulting on-shell form can be simplified on the poles $(567)=(368)=(248)=(347)=0$ to 
\begin{align}
\label{eq:I78}
\Omega\,&=\,\frac{d^{3\times 8}C}{\text{Vol(GL}(3))}  {(346)^2(356) \over (234)(345)(456)(561)(136)(236)} \times \left.I\right|_{7,8}
\end{align}
where $\left.I\right|_{7,8}$ encodes all the dependence on the extra legs $7$ and $8$,
\begin{equation}
\left.I\right|_{7,8}\, =\, {1 \over (781)(567)(368)(248)(347)} .
\end{equation}
As in the previous examples, we now compute the residues of $\left.I\right|_{7,8}$ around $C_{i7}=C_{i8}=0$ for $i=1,2,3$ and obtain
\begin{align}
\label{eq:strangepole}
\left.I\right|_{7,8}\rightarrow {1 \over (124)(346)(365)-(456)(234)(136) } .
\end{align}
Thus we find that the on-shell form of the six-point diagram in \fref{fig:ProcedureWeirdPole} is given by
\begin{align}
\Omega\,=\,\frac{d^{3\times 6}C}{\text{Vol(GL}(3))} {(346)^2(356) \over (234)(345)(456)(561)(136)(236) \left( (124)(346)(365)-(456)(234)(136) \right)} .
\label{Omega_complicated_case}
\end{align}

The appearance of the factor $(124)(346)(365)-(456)(234)(136)$ in the denominator through this process is rather non-trivial and shows that this diagram, unlike the NMHV leading singularity \eqref{eq:LS1}, cannot be written as a linear combination of planar diagrams. This example thus provides concrete evidence for a behavior already announced in \cite{Arkani-Hamed:2014bca}, that already for $k=3$ and $n=6$ not all leading singularities can be expressed as linear combinations of planar ones.

\begin{figure}[h]
\centering
\includegraphics[width=0.8\linewidth]{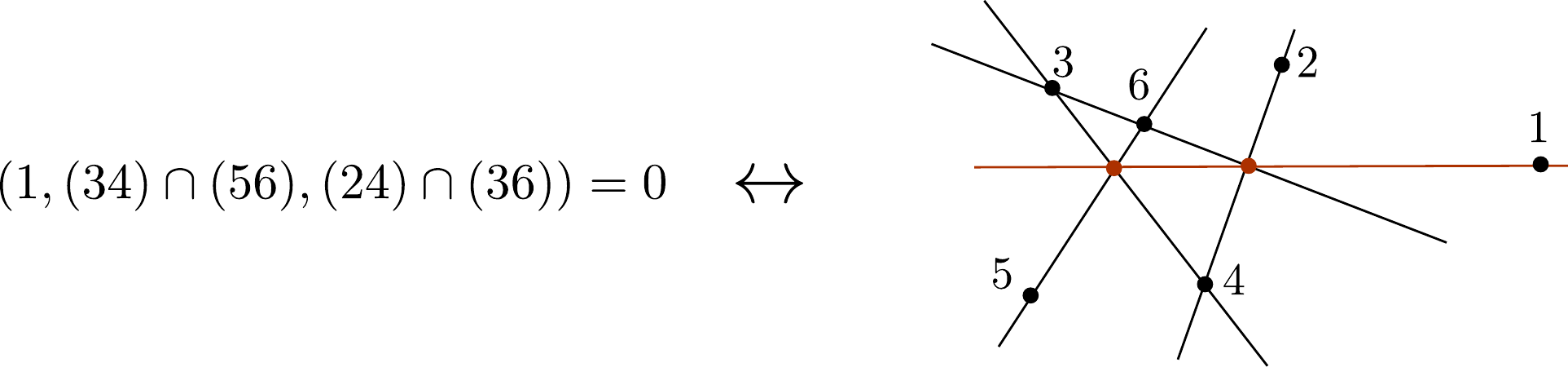}
\caption{\it Configuration of points in $\mathbb{P}^2$ corresponding to the singularity $(124)(346)(365)-(456)(234)(136)=0$.}
\label{fig:strange-pole}
\end{figure}

This singularity can be geometrically seen as follows. Each column $\vec{c}_i$ of $C$ can be thought of as the coordinates of a point in $\mathbb{P}^2$. A usual pole of the form $(ijk)=0$ means that the three points $\vec{c}_i,\,\vec{c}_j$ and $\vec{c}_k$ are on the same line. In contrast with this simple configuration, denoting by $(ij)$ the line defined by points $\vec{c}_i$ and $\vec{c}_j$, the relation between minors at the pole \eqref{eq:strangepole} can be rewritten in a more illuminating way,
\begin{equation}
(124)(346)(365)-(456)(234)(136)\,=\,(1,(34)\cap(56),(24)\cap(36))\ .
\end{equation}
 where $(ij)\cap(kl)$ stands for the point of intersection between the lines $(ij)$ and $(kl)$. The geometrical configuration of points in $\mathbb{P}^2$ is shown in \fref{fig:strange-pole}.

This diagram certainly deserves further study, and we will come back to it in \sref{sec:reducibility-non-planar}. There we will use a matroid polytope perspective to fully understand reducibility in the context of non-planar diagrams. For this diagram we will indeed find an edge which, when removed, does not set any \pl coordinates to zero but instead relates \pl coordinates to each other, i.e.\ it will impose the relation $(124)(346)(365)-(456)(234)(136)=0$. The leading singularity that arises through the removal of this edge is fully computed in Appendix \ref{appendix:ls}.

\bigskip

\subsection{Proof of the combinatorial method}

\label{sec:proof}

In this section we present a proof of the method proposed in \sref{sec:rules} for constructing the on-shell form of N$^{k-2}$MHV in terms of \pl coordinates. We consider the class of on-shell diagrams with $n_B=n-k$ and hence without white nodes surrounded by black nodes. Since the addition of an auxiliary edge on diagrams for which $n_B>n-k$ leads to a graph with $n_B=n-k$ we argue that the proof is valid for these cases as well.

\subsubsection{Top forms with $n_B=n-k$}

Let us consider first on-shell diagrams that are top forms in $Gr_{k,n}$ and $n_B=n-k$. This means that the matrix $T$ has no arbitrary entries $*_i$, so every black node has valency $k+1$ and is associated to a local Grassmannian $Gr_{k,k+1}$. We denote elements of the Grassmannian $Gr_{k,k+1}$ by $\widetilde{C}$ to distinguish them from the elements of the Grassmannian $Gr_{k,n}$ associated with the complete graph. The proof in this section follows the same logic used for MHV leading singularities in \cite{Arkani-Hamed:2014bca}. In the following we will discuss also the case for which $T$ has arbitrary entries.

We start by studying the contribution of each black node to the on-shell form of the full diagram \eqref{eq:F}. For an internal black node associated with the subset $\{i_1,\dots, i_k\} $ of external particles, the corresponding constraint $\delta^{(2)}(\widetilde{C}^{\perp}\cdot\lambda)$ provides a linear relation satisfied by the set $\{\lambda^{i_1},\dots, \lambda^{i_k}\} $ connected to the node. $Gr_{k,k+1}$ has $k$ degrees of freedom, which can be parametrized by the entries of the $1\times (k+1)$ matrix $\widetilde{C}^{\perp}$ modulo $\text{GL}(1)$,
\begin{equation}
\label{eq:c-perp-black}
\widetilde{C}^{\perp}\,=\, \big(\alpha_{i_1} \; \cdots \; \alpha_{i_{k+1}}\big) \ .
\end{equation}
Then, we associate the following form to every internal black node 
\begin{align}
\label{eq:Cperp}
\begin{split}
\{i_1,\dots, i_{k+1}\}\quad \leftrightarrow\;\quad & \frac{1}{\text{Vol(GL}(1))}\prod_{j=1}^{k+1} \frac{d\alpha_{i_j}}{\alpha_{i_j}} \delta^{(2)}\Big(\sum_{j=1}^{k+1} \alpha_{i_j} \lambda^{i_j}\Big) \ .
\end{split}
\end{align}
Recalling that the matrices $\widetilde{C}$ and $\widetilde{C}^{\perp}$ associated to the local $Gr_{k,k+1}$ are complementary matrices, we may equivalently write
\begin{equation}
\label{eq:relation-minors}
\alpha_{i_j}\,=\, (i_j)\Big|_{\widetilde{C}^{\perp}}\, =\, (-1)^{j-1} (i_1\cdots \hat{i}_j\cdots i_{k+1})\Big|_{\widetilde{C}} \ ,
\end{equation}
where $(i_j)\Big|_{\widetilde{C}^{\perp}}$ is a $1\times 1$ minor of $\widetilde{C}^{\perp}$ and $(i_1\cdots \hat{i}_j\cdots i_{k+1})\Big|_{\widetilde{C}}$ is a $k\times k$ minor of $\widetilde{C}$ obtained by deleting the column $i_j$. Using this, \eqref{eq:Cperp} may be recast as
\begin{align}
\label{eq:PTs}
\{i_1,\dots, i_{k+1}\}\; \leftrightarrow\;  \frac{d^{k\times (k+1)}\widetilde{C}}{\text{Vol(GL}(k))} \frac{\delta^{(2)}\Big(\sum_{j=1}^{k+1} (-1)^{j-1}(i_1\cdots \hat{i}_j\cdots i_{k+1}) \lambda_{i_j}\Big)}{(i_1\cdots i_k)(i_2\cdots i_{k+1})\cdots(i_{k+1}\cdots i_{k-1})}\ .
\end{align}
It is clear that the product of $k\times k$ minors in the denominators of the above expression gives rise to the Parke-Taylor-like factors introduced in \eqref{eq:LS}.

The next step is to consider the complete diagram instead of each internal black node separately. We write the matrix $C \in Gr_{k,n}$ as
\begin{equation}
C\,=\,\big(
\vec{c}_1 \; \cdots \; \vec{c}_n \big)\ ,
\end{equation}
where $\vec{c}_i$ are $k$-vectors.
At this point, we recall that the matrix $M$ introduced on item \ref{item3-rules} of \sref{sec:rules} provides a representative of the $(n-k)\times n$ matrix $C^{\perp}$ since
\begin{equation}
\label{eq:Cramer}
\vec{c}_{i_1}(i_2\cdots i_{k+1})- \vec{c}_{i_2}(i_1\cdots i_{k+1})  + \dots + (-1)^{k}\, \vec{c}_{i_{k+1}}\,(i_1\cdots i_{k})\,=\,0\quad \Rightarrow\quad  M\cdot C^{\rm T} \,=\,0\ ,
\end{equation}
where at this point we identified
\begin{equation}
\label{eq:minor-identification}
(i_1\cdots i_k)\Big|_{\widetilde{C}}\,=\, (i_1\cdots i_k)\Big|_{C}\ .
\end{equation}
The next step is to relate $C^{\perp}$ to $M$. In order to do so, we gauge fix the $\text{GL}(k)$ redundancy in $C$ by writing each column as a linear combination of $k$ columns $\{\vec{c}_{a_1},\dots,\vec{c}_{a_k}\}$. This fixes columns $a_1,\dots,a_k$ to the identity matrix. Denoting the matrix gauge fixed in this way by $C_{a_1,\dots,a_k}^{\rm gf}$, the corresponding constraint $\delta^{(2k)}(C\cdot\widetilde{\lambda})$ acquires a Jacobian factor of $\dfrac{1}{(a_1\cdots a_k)^k}$.  This gauge fixing in $C$ induces a gauge fixing in $C^{\perp}$ for which all columns except $a_1,\dots,a_k$ are gauge fixed to the identity matrix, which we denote by $C_{a_1,\dots,a_k}^{\perp \rm gf}$.
Relating $C_{a_1,\dots,a_k}^{\perp \rm gf}$ to $M$ amounts to multiplying $M$ by $\widehat{M}_{a_1,\dots,a_k}^{-1}$, the inverse of $\widehat{M}_{a_1,\dots,a_k}$ defined in item \ref{item4-rules} of \sref{sec:rules}.
Thus, we finally arrive at the result
\begin{align}
\label{eq:jacobianM}
\begin{split}
\frac{\delta^{(2k)}(C\cdot\widetilde{\lambda})\,\delta^{(2(n-k))}(C^{\perp} \cdot\lambda)}{\text{Vol(GL}(k))}\,=\, &\left(\frac{(-1)^{\sum\limits_{i=1}^k a_i}\det(\widehat{M}_{a_1,\dots,a_k})}{(a_1\cdots a_k)}\right)^k  \\
& \times\,\delta^{(2k)}(C_{a_1\cdots a_k}^{\rm gf}\cdot\widetilde{\lambda})\,\delta^{(2(n-k))}(C_{a_1\cdots a_k}^{\perp \rm gf} \cdot\lambda)\ .
\end{split}
\end{align}
Combining \eqref{eq:jacobianM} with the Parke-Taylor denominators of \eqref{eq:PTs} we obtain precisely \eqref{eq:LS}, upon omitting the delta-functions.

\subsubsection{Diagrams with $*$}
\label{sec:star}
We now discuss diagrams for which one or more black nodes have valency $v<k+1$ and thus the matrix $T$ has undetermined entries. This situation corresponds to the case where the diagram is not a top-dimensional form, as will become clear soon.

\begin{figure}[h]
\centering
\includegraphics[width=0.8\linewidth]{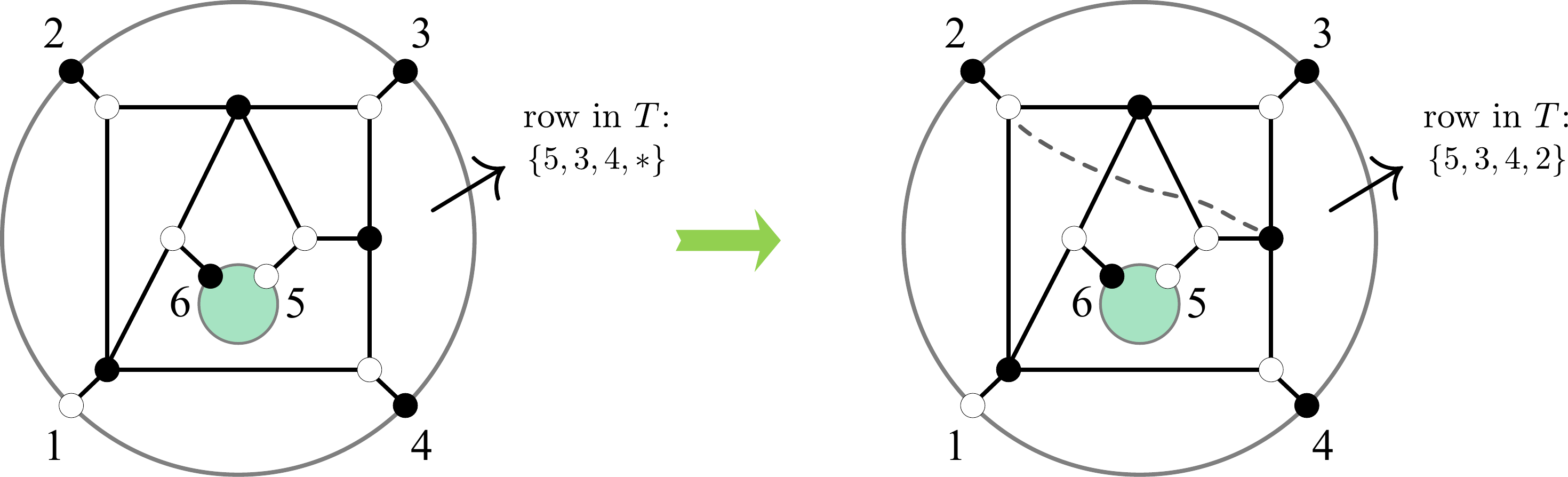}
\caption{\it Addition of an auxiliary edge to a black node with valency $v<k+1$. The grey line fixes the arbitrary entry in the matrix $T$ \eqref{eq:T-star} to be $*=2$. We do not show the new embedding surface since it is not relevant for the computation of the on-shell form and the additional edge is to be deleted in the following step.}
\label{fig:NMHV1extraleg}
\end{figure}

A black node of valency $v$ is associated to the Grassmannian $Gr_{v-1,v}$. Consider for instance a black node for which the corresponding row in $T$ is \begin{equation}
\{i_1,\dots, i_v,*_{v+1},\dots, *_{k+1} \}\ .
\end{equation}
 The first step is to add auxiliary degrees of freedom until the diagram is lifted to a top-cell. This is done by adding extra edges to the black nodes until all of them have valency $k+1$. As a result the analogue of the matrix \eqref{eq:c-perp-black} is
\begin{equation}
\widetilde{C}^{\perp}\,=\, \big( \alpha_{i_1} \; \cdots \; \alpha_{i_v} \; \alpha_{*_{v+1}} \; \cdots \; \alpha_{*_{k+1}} \big)\ .
\end{equation}
The auxiliary edges may connect the black node with any other white node of the graph which is not already connected to it (otherwise the graph would become reducible but not a top-cell). The entries $*_i$ now become labels present in the graph. There are several possible ways to lift the diagram to a top-dimensional cell in $Gr_{k,n}$. Consider for example the diagram from \fref{fig:NMHV1extraleg}, where an auxiliary leg sets the unfixed entry $*=2$.  In this example one could similarly add a leg in a way such that $*=6$ or $*=1$.
The proof now proceeds as if there were no undetermined entries and in the end we remove the auxiliary degrees of freedom by taking residues around $\alpha_{*_v+1},\dots,\alpha_{*_{k+1}}=0$. Notice that this implies that the complementary minors of $C$ vanish, in analogy with \eqref{eq:relation-minors},
\begin{align}
\alpha_{*_i}\,=\, (*_i)\Big|_{\widetilde{C}^{\perp}}\,=\,0\quad\Rightarrow\quad (i_1\cdots \hat{*}_i\cdots *_{k+1})\Big|_{\widetilde{C}}\,=\,0\ .
\end{align}
Taking the residue around all $\alpha_{*_i}=0$ imposes that the columns $\vec{c}_1,\dots,\vec{c}_v$ are linearly dependent vectors after the identification \eqref{eq:minor-identification}.

The independence of the choice of the labels $*_i$ (or in other words how the lift to a top-cell is made) can also be seen in a simple way. Take for instance the example on the left of \fref{fig:NMHV1extraleg} that has a row in $T$ given by $\{5,3,4,*\}$. Since $k=3$, we can choose three linearly independent vectors to form a basis, thus a general redefinition of the column $\vec{c}_*$ of $C$ can be written, for instance, as
\begin{align}
\vec{c}_*\,\rightarrow x\, \vec{c}_* + y\, \vec{c}_3 + z\, \vec{c}_4\ .
\end{align}
Note that we could not choose $\vec{c}_3,\,\vec{c}_4$ and $\vec{c}_5$ to form a basis since $(345)=0$.
The dependence of the general formula \eqref{eq:LS} on $\vec{c}_*$ is through the minors
\begin{align}
(*53)\,\rightarrow\, x\,(*53) + z\,(453)\,,\quad (34*)\,\rightarrow\, x\,(34*)\,,\quad (4*5)\,\rightarrow\, x\,(4*5) + y\,(453)\ .
\end{align}
Since $(345)=0$ every minor involving $\vec{c}_*$ simply gets rescaled. Is it clear that under such a scaling \eqref{eq:LS} transforms as $x^k/x^k$ which guarantees that it is independent of the choice of $\vec{c}_*$.
This completes the proof of the procedure of \sref{section_combinatorial_on_shell} to any on-shell diagram.

\subsubsection{Recursive Proof for Diagrams with Inverse Soft Factors}   \label{sec:ISFproof}

While the previous section presented a proof of the method of \sref{sec:rules} for generic on-shell diagrams, in this section we show a different way to see its validity for a subclass of on-shell diagrams, namely those which can be constructed using inverse soft factors. More precisely, we prove that if the integration measure of a given diagram can be computed by the rules proposed in \sref{sec:rules}, then that of the diagram with an additional leg obtained from the original one via an inverse soft limit can also be calculated using our rules. 

\begin{figure}[h]
\begin{center}
\includegraphics[scale=0.4]{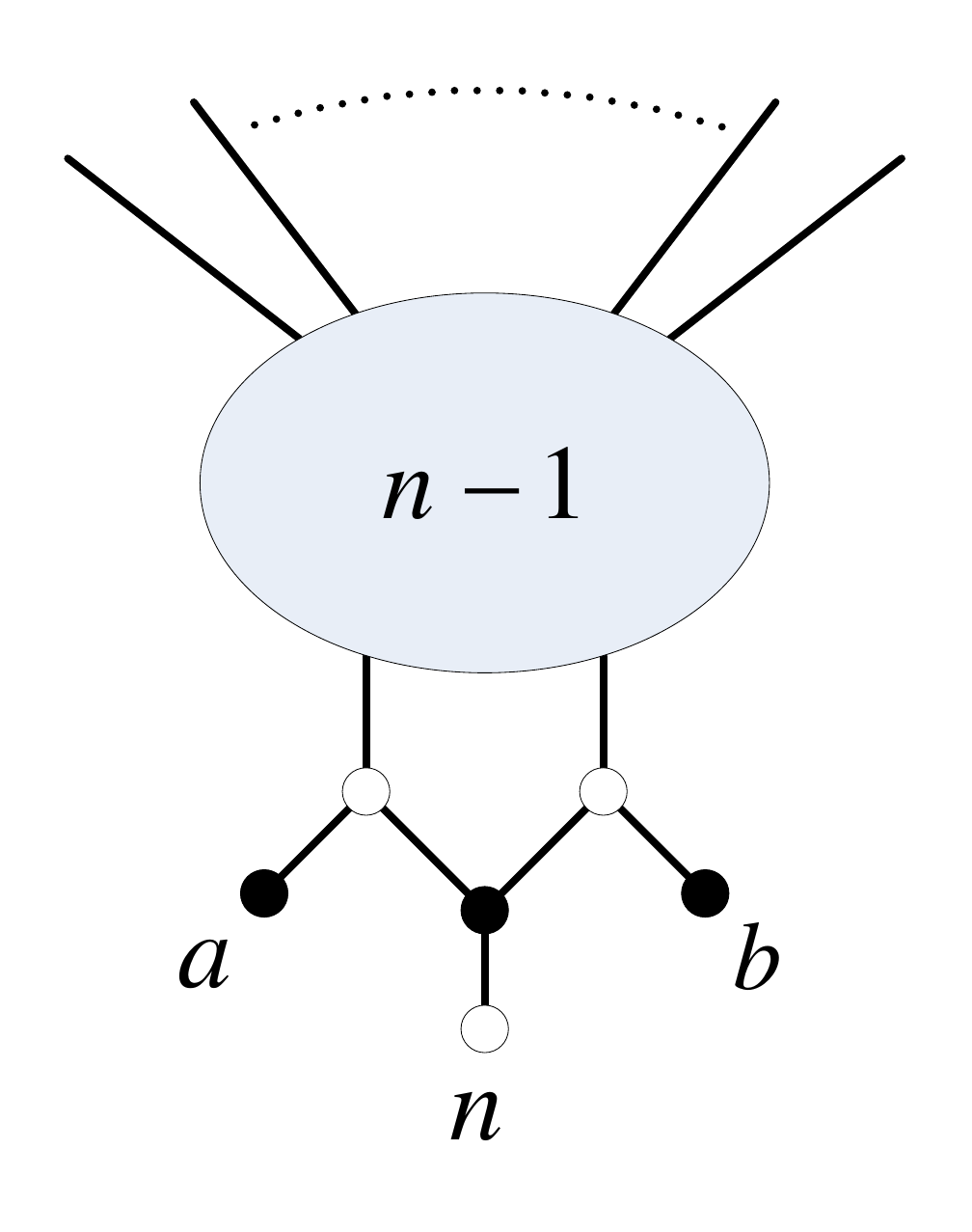}
\vspace{-.3cm}\caption{An on-shell diagram constructed via an inverse soft limit.}
\label{fig:ISL2}
\end{center}
\end{figure}

We shall illustrate the proof with an NMHV diagram constructed via an inverse soft factor, as shown in \fref{fig:ISL2}. 
The internal black vertex from the inverse soft factor generates an additional row in the $T$ matrix with a quadruple $\{b,n,a,*\}$. We shall replace the undetermined entry by a label $i$ and will assume without loss of generality that $a<b<i<n$. The inverse soft factor will modify the $n_B \times (n-1) = (n-1-3) \times (n-1)$ matrix $M^{(n-1)}$ to a matrix $M^{(n)}$ with an additional row and an additional column:

\begin{equation}
\label{eq:MinsideM}
M^{(n)}= \left(
\begin{array}{cccccccc}
  \multicolumn{7}{c|}{} & 0 \\
  \multicolumn{7}{c|}{M^{(n-1)}} & \vdots \\
  \multicolumn{7}{c|}{} & 0 \\
 \hline
 0 \cdots 0 & (ibn) & 0 \cdots 0 & (nai) & 0 \cdots 0 & (bna) & 0 \cdots 0 & (aib)
\end{array}
\right) ,
\end{equation}

Let us now consider the reduced matrices $\widehat{M}_{i_1, i_2, i_3}$, obtained by removing any three columns $\{i_1, i_2, i_3\}$. We shall now consider the case where $n \notin \{i_1, i_2, i_3\}$, the case where $n \in \{i_1, i_2, i_3\}$ will be treated below. Since $\det \left( \widehat{M}^{(n-1)}_{i_1, i_2, i_3} \right) / (i_1 i_2 i_3)$ is independent of $\{i_1, i_2, i_3\}$, we must have
\begin{equation}
\frac{\det \left( \widehat{M}^{(n)}_{i_1, i_2, i_3} \right) }{(i_1 i_2 i_3)} 
= (aib)  \frac{\det \left( \widehat{M}^{(n-1)}_{i_1, i_2, i_3} \right) }{(i_1 i_2 i_3)} ,
\label{eq:relbetweenMhat}
\end{equation}
which is easily verified by selecting $\{i_1, i_2, i_3\}$ as the columns containing $(ibn)$, $(nai)$ and $(bna)$. From this we see that the left-hand side of \eref{eq:relbetweenMhat} must also be independent of $\{i_1, i_2, i_3\}$.

After computing the determinants, we find that the ratio between the $n$-point integration measure and that of the $(n-1)$-point diagram is simply
\begin{equation}
R_{\rm NMHV}^{\text{ISF}}={(aib)^2 \over (bna)(nai)(ibn)}\Big{|}_{(bna)=0} ,
\end{equation}
which is precisely the inverse soft factor for NMHV diagrams, thus proving that the on-shell form we obtain for $n$ external legs is correct. 

We also note that on the pole $(bna)=0$, we may rewrite the ratio as
\begin{equation}
\label{eq:ISF}
R_{\rm NMHV}^{\text{ISF}}={(aib)(ajb) \over (bna)(nai)(jbn)}\Big{|}_{(bna)=0} ,
\end{equation}
and the result is independent of the choice of $i$ and $j$. 

To complete the proof we must consider the case where we remove the columns $\{i_1, i_2, n\}$ in \eref{eq:MinsideM}. Without loss of generality we may choose $i_1=1$ and $i_2=2$, yielding
\begin{eqnarray} \label{eq:removen}
{ {\rm det} \widehat{M}^{(n)}_{1,2,n}   } &=& 
 (-1)^{a+n}  (ibn) \,  {\rm det}  \widehat{M}^{(n-1)}_{1,2,a} +
 (-1)^{b+n}  (nai) \,  {\rm det}  \widehat{M}^{(n-1)}_{1,2,b} \cr
 &+& 
 (-1)^{i+n}  (bna) \,  {\rm det}  \widehat{M}^{(n-1)}_{1,2,i}  \, .
 \end{eqnarray}

Now, if we divide and multiply each $\det  \widehat{M}^{(n-1)}_{1,2,x} $ term by $(12x)$, and using the fact that the ratio of $\det  \widehat{M}^{(n-1)}_{1,2,x} / (12x)$ with $\det  \widehat{M}^{(n-1)}_{1,2,n} / (12n) $ is simply $(-1)^{x-n}$, we obtain
\begin{equation}
\frac{\det \widehat{M}^{(n)}_{1,2,n}  }{(12n)} = \frac{\det \widehat{M}^{(n-1)}_{1,2,n}  }{(12n)} \frac{1}{(12n)} \big[  (12a)(ibn)+
  (12b)(nai) + (12i)(bna) \big] = \frac{\det \widehat{M}^{(n-1)}_{1,2,n}  }{(12n)} (aib)
\end{equation}
where we used the \pl relations to simplify the expression. This is in perfect agreement with the expression \eref{eq:relbetweenMhat}, which completes the proof.

For more general N$^{k-2}$MHV diagrams the proof follows along the same lines, and the procedure yields the corresponding N$^{k-2}$MHV inverse soft factor as expected. As an example, for N$^2$MHV the added leg will generate an additional row in the $T$ matrix with $\{b,n,a,*,*\}$. The free entries can be chosen to be $i$ and $j$. We then need to remove four columns from the $M^{(n)}$ matrix, and regardless of which columns are removed we find
\begin{equation}
{ {\rm det} \widehat{M}^{(n)}_{i_1, i_2, i_3, i_4}  \over (i_1 i_2 i_3 i_4) }
= (aijb)  { {\rm det} \widehat{M}^{(n-1)}_{i_1, i_2, i_3, i_4}   \over (i_1 i_2 i_3 i_4) } \, .
\end{equation}
This yields the ratio between the $n$-point on-shell form and the $(n-1)$-point on-shell form 
\begin{equation}
R_{\rm N^2MHV}^{\text{ISF}}={(aijb)^3 \over (naij)(ijbn)(jbna)(bnai)}\Big{|}_{(*bna)=0}  ,
\end{equation}
which is the correct N$^2$MHV inverse soft factor. Using the fact that $a, b, n$ are on a line, i.e.\ that $(*bna)=0$, the soft factor may be rewritten as
\begin{equation}
R_{\rm N^2MHV}^{\text{ISF}}={(l j b a)(j b a i)( b a i k) \over (naik)(l j b n)(jb na)(bn a i)}\Big{|}_{(*bna)=0} \, ,
\end{equation}
and the result is independent of $i, j, k$ and $l$.

\bigskip

\section{Novel Features of Non-Planar Reductions} 
\label{sec:reducibility}

In \sref{section_equivalence_and_reductions} we discussed reductions and reducibility of on-shell diagrams and introduced a combinatorial framework that can be used to study edge removal. In this section we will work out an example in detail. We will construct its matching and matroid polytopes, establish the precise connection between points in the matroid polytope and \pl coordinates using the boundary measurement and investigate its reducibility. The example has been chosen to illustrate a new phenomenon that can occur when removing an edge from a non-planar diagram: the set of non-vanishing \pl coordinates can remain the same while new non-\pl constraints are generated. This has a direct impact on the issue of reducibility. If a new constraint arises, the new diagram does not cover the same region of the Grassmannian as the original one and hence it is not a reduction.

This story has an interesting counterpart in terms of the on-shell form. The killing of degrees of freedom associated to removing an edge corresponds to taking the residue of the form at the pole where that degree of freedom goes to zero. On-shell forms for planar diagrams have a particularly simple structure; they are just one over a product of \pl coordinates. Every pole of the on-shell form thus corresponds to setting some \pl coordinate to zero. New things can, however, happen for non-planar diagrams: the on-shell form can have poles at which no \pl coordinate vanishes.

Non-\pl constraints should also be taken into account when determining whether two diagrams cover the same region of the Grassmannian. We leave a more detailed investigation of on-shell diagrams with constraints for future work. They certainly arise, as we explicitly show, as limits of more standard diagrams. At present we do not have any argument indicating they are not physical.

\medskip

\subsection{An Example} \label{sec:reducibility-non-planar}

\begin{figure}[h]
\begin{center}
\includegraphics[scale=0.75]{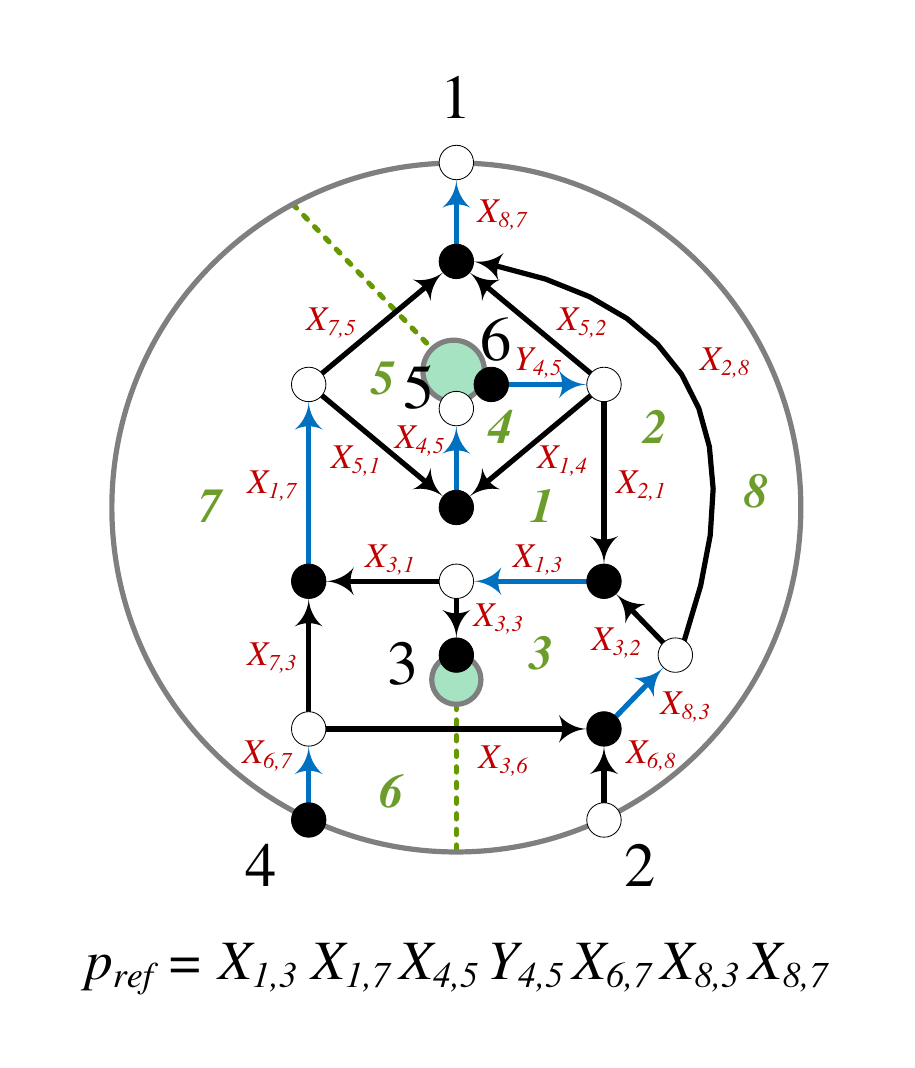}
\vspace{-0.5cm}\caption{An on-shell diagram embedded on a genus-0 surface with three boundaries. Faces are labeled in green, external nodes in black and edges in red.}
\label{fig:hardNonplGuy}
\end{center}
\end{figure}

Let us consider the example on the left of \sref{section:complicatedcase}. For convenience, the on-shell diagram is reproduced in \fref{fig:hardNonplGuy}. The perfect orientation is the one corresponding to the perfect matching $p_{\text{ref}}=X_{1,3} X_{1,7} X_{4,5} X_{6,7} X_{8,3} X_{8,7} Y_{4,5}$. The new possibilities might be anticipated by looking at the denominator of the on-shell form \eref{Omega_complicated_case}, which contains a factor $((124)(346)(365)-(456)(234)(136))$. This means that there is a pole when this factor vanishes, which can be reached without shutting off any \pl coordinate. Furthermore, we expect this can be achieved by deleting edges in the graph. Notice that $((124)(346)(365)-(456)(234)(136))=0$ does not kill any minor but instead imposes a new constraint on them.\footnote{It is interesting to point out that this is very reminiscent of the detailed discussion of boundaries of the amplituhedron presented in \cite{Franco:2014csa}, in which certain boundaries correspond to setting combinations of minors to zero. In that case, too, all boundaries can be mapped to poles of the on-shell form.}  We shall now see how this happens.

The perfect matching matrix $P$ for this graph is
{\footnotesize
\begin{equation} P =
\left(
\begin{array}{c|cccccccccccccccccccccccccccccccccccccccc}
 X_{1,3} & 1 & 1 & 1 & 1 & 1 & 1 & 1 & 1 & 1 & 1 & 1 & 1 & 0 & 0 & 0 & 0 & 0 & 0 & 0 & 0 & 0 & 0 & 0 & 0 & 0 & 0 & 0 & 0 & 0 & 0 & 0 & 0 & 0 & 0 & 0 & 0 & 0 & 0 & 0 & 0 \\
 X_{1,4} & 1 & 1 & 1 & 1 & 0 & 0 & 0 & 0 & 0 & 0 & 0 & 0 & 1 & 1 & 1 & 1 & 1 & 0 & 0 & 0 & 0 & 0 & 0 & 0 & 0 & 0 & 0 & 0 & 0 & 0 & 0 & 0 & 0 & 0 & 0 & 0 & 0 & 0 & 0 & 0 \\
 X_{1,7} & 1 & 1 & 1 & 0 & 1 & 1 & 1 & 1 & 0 & 0 & 0 & 0 & 1 & 1 & 0 & 0 & 0 & 1 & 1 & 1 & 1 & 1 & 1 & 1 & 0 & 0 & 0 & 0 & 0 & 0 & 0 & 0 & 0 & 0 & 0 & 0 & 0 & 0 & 0 & 0 \\
 X_{2,8} & 1 & 1 & 0 & 0 & 1 & 1 & 0 & 0 & 1 & 0 & 0 & 0 & 0 & 0 & 0 & 0 & 0 & 1 & 1 & 0 & 0 & 0 & 0 & 0 & 1 & 1 & 1 & 0 & 0 & 0 & 0 & 0 & 0 & 0 & 0 & 0 & 0 & 0 & 0 & 0 \\
 X_{3,6} & 1 & 0 & 0 & 0 & 1 & 0 & 0 & 0 & 0 & 0 & 0 & 0 & 1 & 0 & 1 & 0 & 0 & 1 & 0 & 1 & 1 & 0 & 0 & 0 & 1 & 0 & 0 & 1 & 1 & 1 & 0 & 0 & 0 & 0 & 0 & 0 & 0 & 0 & 0 & 0 \\
 X_{3,1} & 0 & 0 & 0 & 0 & 0 & 0 & 0 & 0 & 0 & 0 & 0 & 0 & 0 & 0 & 1 & 1 & 0 & 0 & 0 & 0 & 0 & 0 & 0 & 0 & 1 & 1 & 0 & 1 & 1 & 1 & 1 & 1 & 1 & 1 & 1 & 0 & 0 & 0 & 0 & 0 \\
 X_{3,2} & 0 & 0 & 0 & 0 & 0 & 0 & 0 & 0 & 0 & 0 & 0 & 0 & 1 & 1 & 1 & 1 & 1 & 0 & 0 & 1 & 1 & 1 & 1 & 0 & 0 & 0 & 0 & 1 & 1 & 1 & 1 & 1 & 1 & 0 & 0 & 1 & 1 & 1 & 0 & 0 \\
 X_{5,1} & 0 & 0 & 0 & 0 & 0 & 0 & 0 & 0 & 1 & 1 & 1 & 0 & 0 & 0 & 0 & 0 & 0 & 0 & 0 & 0 & 0 & 0 & 0 & 0 & 1 & 1 & 1 & 1 & 1 & 0 & 1 & 1 & 0 & 1 & 0 & 1 & 1 & 0 & 1 & 0 \\
\rowcolor{cyan!90!blue!60} X_{5,2} & 0 & 0 & 0 & 0 & 0 & 0 & 1 & 0 & 0 & 1 & 0 & 0 & 0 & 0 & 0 & 0 & 0 & 0 & 0 & 1 & 0 & 1 & 0 & 0 & 0 & 0 & 0 & 1 & 0 & 0 & 1 & 0 & 0 & 0 & 0 & 1 & 0 & 0 & 0 & 0 \\
 X_{7,5} & 0 & 0 & 0 & 1 & 0 & 0 & 0 & 0 & 0 & 0 & 0 & 1 & 0 & 0 & 1 & 1 & 1 & 0 & 0 & 0 & 0 & 0 & 0 & 0 & 0 & 0 & 0 & 0 & 0 & 1 & 0 & 0 & 1 & 0 & 1 & 0 & 0 & 1 & 0 & 1 \\
 X_{7,3} & 0 & 0 & 0 & 1 & 0 & 0 & 0 & 0 & 1 & 1 & 1 & 1 & 0 & 0 & 0 & 0 & 1 & 0 & 0 & 0 & 0 & 0 & 0 & 0 & 0 & 0 & 1 & 0 & 0 & 0 & 0 & 0 & 0 & 0 & 0 & 1 & 1 & 1 & 1 & 1 \\
 X_{8,3} & 0 & 0 & 1 & 1 & 0 & 0 & 1 & 1 & 0 & 1 & 1 & 1 & 0 & 0 & 0 & 0 & 0 & 0 & 0 & 0 & 0 & 0 & 0 & 1 & 0 & 0 & 0 & 0 & 0 & 0 & 0 & 0 & 0 & 1 & 1 & 0 & 0 & 0 & 1 & 1 \\
 Y_{2,1} & 0 & 0 & 0 & 0 & 0 & 0 & 0 & 0 & 0 & 0 & 0 & 0 & 0 & 0 & 0 & 0 & 0 & 1 & 1 & 0 & 0 & 0 & 0 & 1 & 1 & 1 & 1 & 0 & 0 & 0 & 0 & 0 & 0 & 1 & 1 & 0 & 0 & 0 & 1 & 1 \\
 \hline
 X_{3,3} & 0 & 0 & 0 & 0 & 0 & 0 & 0 & 0 & 0 & 0 & 0 & 0 & 1 & 1 & 0 & 0 & 1 & 1 & 1 & 1 & 1 & 1 & 1 & 1 & 0 & 0 & 1 & 0 & 0 & 0 & 0 & 0 & 0 & 0 & 0 & 1 & 1 & 1 & 1 & 1 \\
 X_{6,7} & 0 & 1 & 1 & 0 & 0 & 1 & 1 & 1 & 0 & 0 & 0 & 0 & 0 & 1 & 0 & 1 & 0 & 0 & 1 & 0 & 0 & 1 & 1 & 1 & 0 & 1 & 0 & 0 & 0 & 0 & 1 & 1 & 1 & 1 & 1 & 0 & 0 & 0 & 0 & 0 \\
 X_{6,8} & 0 & 1 & 0 & 0 & 0 & 1 & 0 & 0 & 1 & 0 & 0 & 0 & 0 & 1 & 0 & 1 & 1 & 0 & 1 & 0 & 0 & 1 & 1 & 0 & 0 & 1 & 1 & 0 & 0 & 0 & 1 & 1 & 1 & 0 & 0 & 1 & 1 & 1 & 0 & 0 \\
 X_{8,7} & 0 & 0 & 1 & 0 & 0 & 0 & 0 & 1 & 0 & 0 & 1 & 0 & 1 & 1 & 0 & 0 & 0 & 0 & 0 & 0 & 1 & 0 & 1 & 1 & 0 & 0 & 0 & 0 & 1 & 0 & 0 & 1 & 0 & 1 & 0 & 0 & 1 & 0 & 1 & 0 \\
 X_{4,5} & 0 & 0 & 0 & 0 & 1 & 1 & 1 & 1 & 0 & 0 & 0 & 1 & 0 & 0 & 0 & 0 & 0 & 1 & 1 & 1 & 1 & 1 & 1 & 1 & 0 & 0 & 0 & 0 & 0 & 1 & 0 & 0 & 1 & 0 & 1 & 0 & 0 & 1 & 0 & 1 \\
 Y_{4,5} & 0 & 0 & 0 & 0 & 1 & 1 & 0 & 1 & 1 & 0 & 1 & 1 & 0 & 0 & 0 & 0 & 0 & 0 & 0 & 0 & 1 & 0 & 1 & 0 & 0 & 0 & 0 & 0 & 1 & 1 & 0 & 1 & 1 & 0 & 0 & 0 & 1 & 1 & 0 & 0 \\
\end{array}
\right) , \label{eq:Pmatrix}
\end{equation}
}

\noindent where we have organized the rows such that the final six correspond to external edges. We have also highlighted the row corresponding to $X_{5,2}$ for future convenience. Recalling the definition of the perfect matching matrix \eref{P_matrix}, every column corresponds to a perfect matching and an entry $P_{i\mu}$ is 1 if the perfect matching $p_\mu$ contains the $i^{th}$ edge, and it is zero otherwise. The columns in $P$ are the coordinate vectors in the matching polytope for the corresponding perfect matchings. Note that every perfect matching maps to a distinct point in the matching polytope. Despite the column vectors in \eref{eq:Pmatrix} are 19-dimensional, it is straightforward to check, e.g.\ by shifting the coordinates such that one of them lies at the origin and then row-reducing $P$, that the matching polytope is a 9-dimensional object. This fact nicely matches the counting in terms of generalized face variables: there are 8 faces $f_i$ (7 of which are independent) and $B-1=2$ variables $b_j$, which totals 9 degrees of freedom. This will become important later.

The matroid polytope is obtained by taking \eref{eq:Pmatrix} and keeping only the 6 coordinates associated to the external edges. Generically, when doing this more than one perfect matching can be projected down to the same point in the matroid polytope. This is the multiplicity we alluded to earlier. The points in the matroid polytope are summarized in \eref{eq:matroidpointsAndMap}, where for every point we list the corresponding perfect matchings and \pl coordinate.

\beq
\begin{array}{c|c|c|c|c|c|c|c|c|c}
\ \ \ \ 0 \ \ \ \ \ & \ \ \ \ 0 \ \ \ \ \ & \ \ \ \ 0 \ \ \ \ \ & \ \ \ \ 0 \ \ \ \ \ & \ \ \ \ 0 \ \ \ \ \ & \ \ \ \ 0 \ \ \ \ \ & \ \ \ \ 0 \ \ \ \ \ & \ \ \ \ 0 \ \ \ \ \ & \ \ \ \ 0 \ \ \ \ \ & \ \ \ \ 1 \ \ \ \ \ \\
0 & 1 & 1 & 0 & 1 & 1 & 1 & 0 & 0 & 0  \\
0 & 1 & 0 & 0 & 1 & 0 & 0 & 1 & 0 & 0  \\
0 & 0 & 1 & 0 & 0 & 0 & 1 & 0 & 1 & 1  \\
0 & 0 & 0 & 1 & 1 & 1 & 1 & 0 & 0 & 0  \\
0 & 0 & 0 & 1 & 1 & 0 & 1 & 1 & 1 & 0  \\ \hline
p_1 , p_4  & p_2 , p_{16} & p_{3} , p_{34} & p_{5} , p_{12} & p_{6} , p_{33} & \textcolor{red}{p_{7}} , p_{35} &  p_{8} &  p_{9} &  p_{11} , p_{29} &  \ p_{13} , p_{39} \\
 \textcolor{red}{p_{10}},p_{15} & p_{26} , \textcolor{red}{p_{31}}  & & p_{30} &  &  &  & &  &  \\
p_{25} , \textcolor{red}{p_{28}} & & & &  & &  & & &  \\ \hline
\Delta_{1,2,5} & \Delta_{1,4,5} & \Delta_{2,4,5} & \Delta_{1,2,6} & \Delta_{1,4,6} & \Delta_{1,2,4} & \Delta_{2,4,6} & \Delta_{1,5,6} & \Delta_{2,5,6} & \Delta_{2,3,5}
\end{array} \nonumber
\eeq

\beq
\begin{array}{c|c|c|c|c|c|c|c|c|c}
\ \ \ \ 1 \ \ \ \ \ & \ \ \ \ 1 \ \ \ \ \ & \ \ \ \ 1 \ \ \ \ \ & \ \ \ \ 1 \ \ \ \ \ & \ \ \ \ 1 \ \ \ \ \ & \ \ \ \ 1 \ \ \ \ \ & \ \ \ \ 1 \ \ \ \ \ & \ \ \ \ 1 \ \ \ \ \ & \ \ \ \ 1 \ \ \ \ \ & \ \ \ \ 1 \ \ \ \ \ \\
1 & 0 & 0 & 1 & 0 & 1 & 1 & 1 & 0 & 0 \\
1 & 1 & 0 & 1 & 0 & 1 & 0 & 1 & 1 & 1 \\
1 & 0 & 0 & 0 & 1 & 1 & 1 & 1 & 1 & 0 \\
0 & 0 & 1 & 1 & 1 & 1 & 1 & 0 & 0 & 1 \\
0 & 0 & 0 & 0 & 1 & 1 & 0 & 1 & 1 & 1 \\ \hline
p_{14} & p_{17} , p_{27} &  p_{18} , \textcolor{red}{p_{20}}  &  p_{19} , \textcolor{red}{p_{22}} & p_{21} & p_{23} & p_{24} & p_{32} & p_{37} & p_{38} \\ 
  & \textcolor{red}{p_{36}} & p_{40} & & & & & & \\ \hline
\Delta_{3,4,5} & \Delta_{1,3,5} & \Delta_{1,2,3} & \Delta_{1,3,4} & \Delta_{2,3,6} & \Delta_{3,4,6} & \Delta_{2,3,4} & \Delta_{4,5,6} & \Delta_{3,5,6} & \Delta_{1,3,6}
\end{array}
\label{eq:matroidpointsAndMap}
\eeq

\bigskip

Using \eref{eq:Pmatrix} it is straightforward to check that there is a single edge, $X_{5,2}$, which can be removed without killing any point in the matroid polytope. Eliminating this edge removes all perfect matchings that contain it, i.e.\ $p_7, p_{10}, p_{20}, p_{22}, p_{28}, p_{31}, p_{36}$, which are shown in red in \eref{eq:matroidpointsAndMap}. Following our previous discussion, none of the \pl coordinates is set to zero. We now investigate what happens to them in more detail, by considering the effect on the boundary measurement.

\bigskip

\paragraph{Boundary Measurement for the Original Diagram.} 

Before removing $X_{5,2}$, the matrix $C$ associated to \fref{fig:hardNonplGuy} is

\beq
C   \equiv  \left(
\begin{array}{c|cccccc}
& \ 1 & \ \ 2 \ \ & 3 & \ \ 4 \ \ & 5 & \ \ 6 \ \ \\ \hline
2 \ \ & \ c_1 & \ \ 1 \ \ & c_2 & \ \ 0 \ \ & c_3 & \ \ 0 \ \ \\
4 \ \ & \ c_4 & \ \ 0 \ \ & c_5 & \ \ 1 \ \ & c_6 & \ \ 0 \ \ \\
6 \ \ & \ c_7 & \ \ 0 \ \ & c_8 & \ \ 0 \ \ & c_9 & \ \ 1 \ \ 
\end{array}
\right) \nonumber
\eeq

\smallskip

{\scriptsize
\beq
 = 
\left( \begin{array}{cccccc}
 \frac{X_{6,8} X_{2,8}}{X_{8,3} X_{8,7}}+\frac{X_{6,8} X_{3,2} X_{3,1} X_{7,5} }{X_{8,3} X_{1,3} X_{1,7} X_{8,7}} & 1 & \frac{X_{6,8} X_{3,2} X_{3,3}}{X_{8,3} X_{1,3}} & 0 & -\frac{X_{6,8} X_{3,2} X_{3,1} X_{5,1}}{X_{8,3} X_{1,3} X_{1,7} X_{4,5}} & 0 \\
 -\frac{X_{3,6} X_{2,8}}{X_{6,7} X_{8,3} X_{8,7}}-\frac{X_{3,6} X_{3,2} X_{3,1} X_{7,5} }{X_{6,7} X_{8,3} X_{1,3} X_{1,7} X_{8,7}}-\frac{X_{7,3} X_{7,5}}{X_{6,7} X_{1,7} X_{8,7}} & 0 & -\frac{X_{3,6} X_{3,2} X_{3,3}}{X_{6,7} X_{8,3} X_{1,3}} & 1 & \frac{X_{7,3} X_{5,1}}{X_{6,7} X_{1,7} X_{4,5}}+\frac{X_{3,6} X_{3,2} X_{3,1} X_{5,1}}{X_{6,7} X_{8,3} X_{1,3} X_{1,7} X_{4,5}} & 0 \\
 \frac{X_{5,2}}{Y_{4,5} X_{8,7}}-\frac{X_{2,1} X_{3,1} X_{7,5}}{Y_{4,5} X_{1,3} X_{1,7} X_{8,7}} & 0 & -\frac{X_{2,1} X_{3,3}}{Y_{4,5} X_{1,3}} & 0 & \frac{X_{1,4}}{Y_{4,5} X_{4,5}}+\frac{X_{2,1} X_{3,1} X_{5,1}}{Y_{4,5} X_{1,3} X_{1,7} X_{4,5}} & 1 \\
\end{array}
\right) 
\eeq
}

All minors of this matrix are generically non-zero:
\smallskip

\begin{equation}
\begin{array}{clcl}
 \Delta_{1,2,3}= & -\mathfrak{p}_{18}-\textcolor{red}{\mathfrak{p}_{20}}-\mathfrak{p}_{40} \quad \quad \quad & \Delta_{2,3,4}= & \mathfrak{p}_{24} \\
 \Delta_{1,2,4}= & \textcolor{red}{\mathfrak{p}_7}-\mathfrak{p}_{35} \quad \quad \quad & \Delta_{2,3,5}= & \mathfrak{p}_{39}-\mathfrak{p}_{13} \\
 \Delta_{1,2,5}= & \mathfrak{p}_1+\mathfrak{p}_4+\textcolor{red}{\mathfrak{p}_{10}}+\mathfrak{p}_{15}+\mathfrak{p}_{25}+\textcolor{red}{\mathfrak{p}_{28}} \quad \quad \quad & \Delta_{2,3,6}= & -\mathfrak{p}_{21} \\
 \Delta_{1,2,6}= & \mathfrak{p}_5+\mathfrak{p}_{12}+\mathfrak{p}_{30} \quad \quad \quad & \Delta_{2,4,5}= & \mathfrak{p}_3+\mathfrak{p}_{34} \\
 \Delta_{1,3,4}= & \mathfrak{p}_{19}+\textcolor{red}{\mathfrak{p}_{22}} \quad \quad \quad & \Delta_{2,4,6}= & 1 \\
 \Delta_{1,3,5}= & \mathfrak{p}_{17}+\mathfrak{p}_{27}+\textcolor{red}{\mathfrak{p}_{36}} \quad \quad \quad & \Delta_{2,5,6}= & \mathfrak{p}_{11}+\mathfrak{p}_{29} \\
 \Delta_{1,3,6}= & \mathfrak{p}_{38} \quad \quad \quad & \Delta_{3,4,5}= & \mathfrak{p}_{14} \\
 \Delta_{1,4,5}= & \mathfrak{p}_2+\mathfrak{p}_{16}+\mathfrak{p}_{26}+\textcolor{red}{\mathfrak{p}_{31}} \quad \quad \quad & \Delta_{3,4,6}= & \mathfrak{p}_{23} \\
 \Delta_{1,4,6}= & \mathfrak{p}_6+\mathfrak{p}_{33} \quad \quad \quad & \Delta_{3,5,6}= & \mathfrak{p}_{37} \\
 \Delta_{1,5,6}= & \mathfrak{p}_9 \quad \quad \quad & \Delta_{4,5,6}= & \mathfrak{p}_{32} \\
\end{array} \label{eq:all20Deltas}
\end{equation}
\smallskip

\noindent Here $\mathfrak{p}_{\mu}$ indicates the flow associated to a perfect matching $p_\mu$. A flow takes the form of a monomial in oriented edge weights. We refer the reader to \cite{Franco:2013nwa} for a thorough discussion of these concepts. The flows for perfect matchings containing $X_{5,2}$ are shown in red. In total we have 9 independent minors, which tells us that $C$ is in the top cell of $Gr_{3,6}$. Thus, we see that if generalized face variables are to parametrize all degrees of freedom of $C$, we cannot lose any $f_i$ or $b_j$, as we already have the minimal number possible to account for a 9-dimensional $C$. Naively, this is in tension with the fact that edge $X_{5,2}$ can be removed without eliminating points in the matroid polytope, i.e.\ without setting \pl coordinates to zero. As we now explain, while this is true, the removal of $X_{5,2}$ does not kill any $\Delta_I$, but it removes a degree of freedom in such a way as to create a new constraint on the $\Delta_I$, independent from the \pl relations. We then conclude, the graph is \textit{not} reducible.

\medskip

\paragraph{Boundary Measurement After Removing $X_{5,2}$.}

Let us understand in detail how the new constraint arises. We will do so from the perspective of the boundary measurement and the matching polytopes. If we remove $X_{5,2}$, i.e.\ set $X_{5,2}=0$, the only entry in $C$ that is affected is $c_7$.\footnote{Let us say a few words on how to eliminate edges that appear in the denominator of entries in the boundary measurement. Once a perfect orientation is chosen, a given oriented edge weight appears either {\it only} in numerators (as it is the case for $X_{52}$ in this example) or denominators. This is determined by whether the perfect orientation coincides or is opposed to the conventional orientation we picked for the edge under consideration. If we want to remove an edge appearing in denominators, all we need to do is to send the corresponding edge weight to infinity. The fact that some edges are removed by sending them to zero while other ones are removed by sending them to infinity is thus a matter of conventions and another reflection of the symmetry of on-shell diagrams under the inversion of the edge weights.} The \pl coordinates now become

\begin{equation}
\begin{array}{clcl}
 \Delta_{1,2,3}= & -\mathfrak{p}_{18}-\mathfrak{p}_{40} \quad \quad \quad & \Delta_{2,3,4}= & \mathfrak{p}_{24} \\
 \Delta_{1,2,4}= & -\mathfrak{p}_{35} \quad \quad \quad & \Delta_{2,3,5}= & \mathfrak{p}_{39}-\mathfrak{p}_{13} \\
 \Delta_{1,2,5}= & \mathfrak{p}_1+\mathfrak{p}_4+\mathfrak{p}_{15}+\mathfrak{p}_{25} \quad \quad \quad & \Delta_{2,3,6}= & -\mathfrak{p}_{21} \\
 \Delta_{1,2,6}= & \mathfrak{p}_5+\mathfrak{p}_{12}+\mathfrak{p}_{30} \quad \quad \quad & \Delta_{2,4,5}= & \mathfrak{p}_3+\mathfrak{p}_{34} \\
 \Delta_{1,3,4}= & \mathfrak{p}_{19} \quad \quad \quad & \Delta_{2,4,6}= & 1 \\
 \Delta_{1,3,5}= & \mathfrak{p}_{17}+\mathfrak{p}_{27} \quad \quad \quad & \Delta_{2,5,6}= & \mathfrak{p}_{11}+\mathfrak{p}_{29} \\
 \Delta_{1,3,6}= & \mathfrak{p}_{38} \quad \quad \quad & \Delta_{3,4,5}= & \mathfrak{p}_{14} \\
 \Delta_{1,4,5}= & \mathfrak{p}_2+\mathfrak{p}_{16}+\mathfrak{p}_{26} \quad \quad \quad & \Delta_{3,4,6}= & \mathfrak{p}_{23} \\
 \Delta_{1,4,6}= & \mathfrak{p}_6+\mathfrak{p}_{33} \quad \quad \quad & \Delta_{3,5,6}= & \mathfrak{p}_{37} \\
 \Delta_{1,5,6}= & \mathfrak{p}_9 \quad \quad \quad & \Delta_{4,5,6}= & \mathfrak{p}_{32} \\
\end{array}
\end{equation}

\bigskip

\noindent These equations can also be directly obtained from \eref{eq:all20Deltas} by removing the red flows. In addition, that same information, up to signs, can be directly obtained from the matroid polytope encoded in \eref{eq:matroidpointsAndMap}.

Here we see the new situation we anticipated from our knowledge of the matroid polytope: no \pl coordinates are shut off despite losing a face variable. 

Let us now consider the generalized face variables. In addition to the ordinary faces, we will use the cuts
\beq
b_1 = \dfrac{X_{1,3} X_{8,3}}{X_{3,3} X_{3,2} X_{6,8}} \quad \quad \quad \quad b_2 = \dfrac{X_{4,5} X_{7,5}}{X_{5,1} X_{8,7}}.
\eeq
At this point, a natural question is whether it is even possible to express all paths in the matrix $C$ using the generalized face variables that remain at our disposal. The answer is yes. We have
\begin{equation}
\begin{array}{clclcl}
c_1= & \dfrac{1}{f_3 f_6 f_7} + \dfrac{1}{f_1 f_3 f_4 f_{2/5} f_6 f_7} \quad \quad & c_4= & -\dfrac{1}{f_7} - \dfrac{1}{f_3 f_7} - \dfrac{1}{f_1 f_3 f_4 f_{2/5} f_7} \quad \quad & c_7= & -b_2 f_1 f_4 \\ [.4cm]
c_2= & \dfrac{1}{b_1} \quad \quad & c_5= & -\dfrac{f_6}{b_1} \quad \quad & c_8= & -\dfrac{b_2 f_1 f_3 f_4 f_6 f_7}{b_1} \\ [.4cm]
c_3= & -\dfrac{1}{b_2 f_3 f_6 f_7} \quad \quad & c_6= & \dfrac{1}{b_2 f_3 f_7}+\frac{1}{b_2 f_7} \quad \quad & c_9= & f_1 f_4+f_4
\end{array}
\end{equation}
We see that only the 8 variables $f_1$, $f_{2/5}$, $f_3$, $f_4$, $f_6$, $f_7$, $b_1$ and $b_2$ are used, where $f_{2/5}\equiv f_2 f_5$ indicates the combination of $f_2$ and $f_5$. It is possible to invert this map without using $c_7$, obtaining 
\begin{equation}
\begin{array}{clclclcl}
f_1= & \dfrac{c_3 c_8}{c_2 c_9 - c_3 c_8} \quad \quad & f_4= & \dfrac{c_2 c_9 - c_3 c_8}{c_2} \quad \quad & f_6= & -\dfrac{c_5}{c_2} \quad \quad & b_1= & \dfrac{1}{c_2} \\ [.4cm]
f_3= & \dfrac{c_2 c_6 - c_3 c_5}{c_3 c_5} \quad \quad & f_{2/5}= & \dfrac{c_1 c_5 - c_2 c_4}{c_8 (c_3 c_4 - c_1 c_6)} \quad \quad & f_7= & \dfrac{c_2}{c_1 c_5 - c_2 c_4} \quad \quad & b_2= & \dfrac{c_1 c_5-c_2 c_4}{c_2 c_6-c_3 c_5} \\
\end{array}
\end{equation}
This implies that $c_7$ can indeed be expressed in terms of the other $c_i$'s as follows
\begin{equation} \label{eq:Cdependence}
c_7 = -b_2 f_1 f_4 = \frac{c_8 c_3 (c_2 c_4 - c_1 c_5)}{c_2 (c_2 c_6 - c_3 c_5)} .
\end{equation}

We have just shown that although it appears that all 9 entries of the matrix $C$ are independent, this is not the case. This condition can be translated into a constraint on the \pl coordinates, by noting that
\begin{equation}
\begin{array}{clclcl}
c_1= & \Delta_{1,4,6} \quad \quad \quad & c_4= & -\Delta_{1,2,6} \quad \quad \quad & c_7= & \Delta_{1,2,4} \quad \quad \\
c_2= & \Delta_{3,4,6} \quad \quad  \quad & c_5= & \Delta_{2,3,6} \quad \quad \quad & c_8= & -\Delta_{2,3,4} \quad \quad \\
c_3= & -\Delta_{4,5,6} \quad \quad \quad & c_6= & \Delta_{2,5,6} \quad \quad \quad & c_9= & \Delta_{2,4,5} \quad \quad \\
\end{array}
\end{equation}
Hence, \eref{eq:Cdependence} becomes 

\beq
\Delta_{1,2,4} = \frac{\Delta_{2,3,4}\Delta_{4,5,6}(-\Delta_{3,4,6}\Delta_{1,2,6}-\Delta_{1,4,6}\Delta_{2,3,6})}{\Delta_{3,4,6}(\Delta_{3,4,6}\Delta_{2,5,6}+\Delta_{4,5,6}\Delta_{2,3,6})} = -\frac{\Delta_{2,3,4}\Delta_{4,5,6}(\Delta_{1,3,6}\Delta_{2,4,6})}{\Delta_{3,4,6}(\Delta_{3,5,6}\Delta_{2,4,6})} \nonumber 
\eeq
\beq
\Leftrightarrow \quad \Delta_{1,2,4}\Delta_{3,4,6}\Delta_{3,5,6} = - \Delta_{2,3,4}\Delta_{4,5,6}\Delta_{1,3,6}  \label{eq:secreterelation1}
\eeq

\noindent where we used two \pl relations to simplify the expression. This constraint is equivalent to the one we expected from the denominator $((124)(346)(365)-(456)(234)(136))$.

We then see a novel and interesting feature appearing in non-planar graphs: by removing an edge we have created a constraint on the \pl coordinates that is independent of the \pl relations. We conclude that the original graph was indeed reduced. Irreducibility can manifest when deleting edges as the vanishing of \pl coordinates (as for planar graphs) or as the emergence of new constraints on them.

This constraint can alternatively be simply determined by using \eref{eq:Pmatrix} and \eref{eq:matroidpointsAndMap}, because it just reflects the linear dependencies of vectors in the matching polytope. From \eref{eq:matroidpointsAndMap} we see that \eref{eq:secreterelation1} is 
\begin{align}
p_{35} \, p_{23} \, p_{37} &= p_{24} \, p_{32} \, p_{38} \nonumber \\  \nonumber \\
\Leftrightarrow \quad
{\scriptsize
\left( \begin{array}{c} 0 \\ 0 \\ 0 \\ 0 \\ 0 \\ 1 \\ 0 \\ 0 \\ 0 \\ 1 \\ 0 \\ 1 \\ 1 \\ 0 \\ 1 \\ 0 \\ 0 \\ 1 \\ 0 \\ \end{array} \right) + 
\left( \begin{array}{c} 0 \\ 0 \\ 1 \\ 0 \\ 0 \\ 0 \\ 1 \\ 0 \\ 0 \\ 0 \\ 0 \\ 0 \\ 0 \\ 1 \\ 1 \\ 1 \\ 1 \\ 1 \\ 1 \\ \end{array} \right) + 
\left( \begin{array}{c} 0 \\ 0 \\ 0 \\ 0 \\ 0 \\ 0 \\ 1 \\ 1 \\ 0 \\ 0 \\ 1 \\ 0 \\ 0 \\ 1 \\ 0 \\ 1 \\ 1 \\ 0 \\ 1 \\ \end{array} \right) }&{\scriptsize =
\left( \begin{array}{c} 0 \\ 0 \\ 1 \\ 0 \\ 0 \\ 0 \\ 0 \\ 0 \\ 0 \\ 0 \\ 0 \\ 1 \\ 1 \\ 1 \\ 1 \\ 0 \\ 1 \\ 1 \\ 0 \\ \end{array} \right) + 
\left( \begin{array}{c} 0 \\ 0 \\ 0 \\ 0 \\ 0 \\ 1 \\ 1 \\ 1 \\ 0 \\ 0 \\ 0 \\ 0 \\ 0 \\ 0 \\ 1 \\ 1 \\ 1 \\ 0 \\ 1 \\ \end{array} \right) + 
\left( \begin{array}{c} 0 \\ 0 \\ 0 \\ 0 \\ 0 \\ 0 \\ 1 \\ 0 \\ 0 \\ 1 \\ 1 \\ 0 \\ 0 \\ 1 \\ 0 \\ 1 \\ 0 \\ 1 \\ 1 \\ \end{array} \right)
}
\label{eq:secretrel1PM}
\end{align}
Now we understand how the new constraint arises. While \eref{eq:secretrel1PM} is always true, we need to set $X_{5,2}=0$ in order to translate it into a constraint on \pl coordinates $\Delta_I$. Phrased differently, before removing $X_{5,2}$, \eref{eq:secreterelation1} would imply that $(p_{35}-p_7) \, p_{23} \, p_{37} = p_{24} \, p_{32} \, p_{38}$, which is not true. Once $X_{5,2}$ has been removed, however, $p_7$ disappears and \eref{eq:secreterelation1} becomes equivalent to the known relation among perfect matchings \eref{eq:secretrel1PM}.

\bigskip

\subsection{A Systematic Approach to Reducibility}

\label{section_systematic_reducibility}

One lesson we should draw from the previous section is that for non-planar graphs the preservation of the matroid polytope under edge removal is a necessary but not sufficient condition for reducibility. It is nonetheless possible to establish a systematic procedure for determining whether a non-planar graph is reducible or not, which goes as follows. Simply remove as many edges as possible while preserving the matroid polytope, and count the degrees of freedom of the generalized face variables $f_i$ and $b_j$ in the resulting graph.\footnote{Generically, multiple combinations of removed edges are possible at this step. In addition, these combinations might involve different numbers of edges.} This number should be compared to the expected number of degrees of freedom based on the surviving points of the matroid polytope, i.e.\ a naive counting of dimensions of $C$ that assumes the absence of constraints other than the \pl relations. Two scenarios may occur:

\smallskip

\begin{itemize}
\item The surviving points of the matroid polytope suggest a dimension that is \textit{equal} to the number of independent generalized face variables. This means that the graph is now maximally reduced, and there are no new constraints on the $\Delta_I$.
\item The surviving points of the matroid polytope suggest a dimension that is \textit{larger} than the number of independent generalized face variables. This means that the collection of removed edges, which did not affect the matroid polytope, have reduced the graph ``more than the maximal amount''. The difference $\delta$ between the naive and actual dimensions gives the number of new constraints on non-vanishing \pl coordinates which have been generated. Whenever $\delta > 0$, it means that too many edges have been removed and the graph was already reduced after deleting a subset of them.
\end{itemize}

\smallskip

For illustration, let us reconsider the graph in \fref{fig:hardNonplGuy}. As we saw, it is possible to remove the edge $X_{5,2}$ while preserving the matroid polytope. The number of points in the matroid polytope after this operation is 20, which for $Gr_{3,6}$ suggests a naive dimension equal to 9 (i.e.\ as many dimensions as the top cell). However, we only have $6+2=8$ independent generalized face variables, so $\delta=9-8=1$. We conclude that the original graph was already reduced and by deleting $X_{5,2}$ we generate a new constraint on \pl coordinates.

These operations are very simple to implement algorithmically on a computer and thus provide a quick check for whether a graph is reduced or not. 

\bigskip

\subsection{Discovering Non-\pl Constraints}

As mentioned above, $\delta>0$ indicates the existence of constraints on the $\Delta_I$ that are independent from the \pl relations. It is natural to want to find these constraints. To this end, we suggest the following strategy:

\begin{itemize}
\item Solve the linear relations among column vectors in $P$ to obtain all constraints on linear combinations of these vectors.
\item Solve the \pl relations.
\item Rewrite the perfect matchings in terms of \pl coordinates, by inverting the map in \eref{eq:all20Deltas}.
\item Plug the expressions of perfect matchings into the constraints obtained from the first point, to obtain the corresponding constraints in terms of \pl coordinates.
\item Insert the solution of the \pl relations into these constraints. The number of new constraints that do not trivialize should be $\delta$.
\end{itemize}

\bigskip

\section{Conclusions}

We have established several concepts and machinery to undertake the study of non-planar on-shell diagrams. Some of our main results are: the introduction of generalized face variables, the construction of generalized matching and matroid polytopes, their application to the questions of region matching and reductions, the proposal of a boundary measurement for general on-shell diagrams, a study of reducibility of non-planar diagrams and a generalization of the prescription of \cite{Arkani-Hamed:2014bca} for obtaining the on-shell form in terms of minors that applies beyond the MHV case.

The natural goal of this general program is to achieve a level of understanding of non-planar diagrams similar to the existing one for planar diagrams. As we have repeatedly witnessed in this paper, the non-planar realm is far richer.

In addition, there are several concrete questions for future investigation, and we now mention a few of them. First, it would be interesting to investigate in further detail the interplay between our combinatorial tools and the classification of diagrams based on equivalence moves. For example, a concrete problem is to classify the on-shell diagrams associated to all permutation inequivalent top-dimensional cells for various $Gr_{k,n}$'s \cite{Bourjaily:2016mnp}. It would be interesting to find an algorithm that starting from a generalized matroid polytope constructs an on-shell diagram, perhaps a reduced representative, associated to it. Similar methods exist for constructing planar on-shell diagrams from permutations \cite{ArkaniHamed:2012nw} and for constructing dimer models (i.e.\ bipartite graphs on a torus without boundaries) from toric diagrams \cite{Hanany:2005ss,Feng:2005gw}. It would be worth studying whether the stratification of non-planar on-shell diagrams hints at some interesting topologies of the associated geometries and, if so, what its physical significance is.\footnote{Here we have in mind the approach to stratification introduced in \cite{Franco:2013nwa}, based on the generalized matching and matroid polytopes.} It is also natural to investigate whether there are non-planar counterparts for some of the objects which followed on-shell diagrams in planar $\mathcal{N}=4$ SYM, such as deformed on-shell diagrams \cite{Ferro:2012xw,Ferro:2013dga,Beisert:2014qba,Kanning:2014maa,Broedel:2014pia}\footnote{Deformed amplitudes have been studied in \cite{Broedel:2014hca,Ferro:2014gca}.} and the amplituhedron \cite{Arkani-Hamed:2013jha,Arkani-Hamed:2013kca}. Another question to explore is whether there is a non-planar generalization of the connection between scattering amplitudes in the 3d ABJM theory \cite{Aharony:2008ug} and the positive orthogonal Grassmannian \cite{Huang:2013owa,Huang:2014xza}.

\bigskip

\bigskip

\section*{Acknowledgements}

We would like to thank N. Arkani-Hamed, J. Bourjaily, A. Brandhuber, S. He, P. Mattioli, J. McGrane, D. Meidinger, B. Spence and G. Travaglini for very useful and enjoyable discussions. D. G. and S. F. would like to thank Walter Burke Institute for Theoretical Physics at Caltech for hospitality during the completion of this work and the participants of the ``Grassmannian Geometry of Scattering Amplitudes" workshop for enjoyable exchanges.
The work of D. G. is supported by the U.K. Science and Technology Facilities Council (STFC). The work of B. P. is supported by the Science and Technology Facilities Council Consolidated Grant ST/L000415/1 ``String theory, gauge theory and duality".

\bigskip

\appendix

\section{Embedding Independence}

\label{section_simple_example}

\begin{figure}[h]
\begin{center}
\includegraphics[width=6cm]{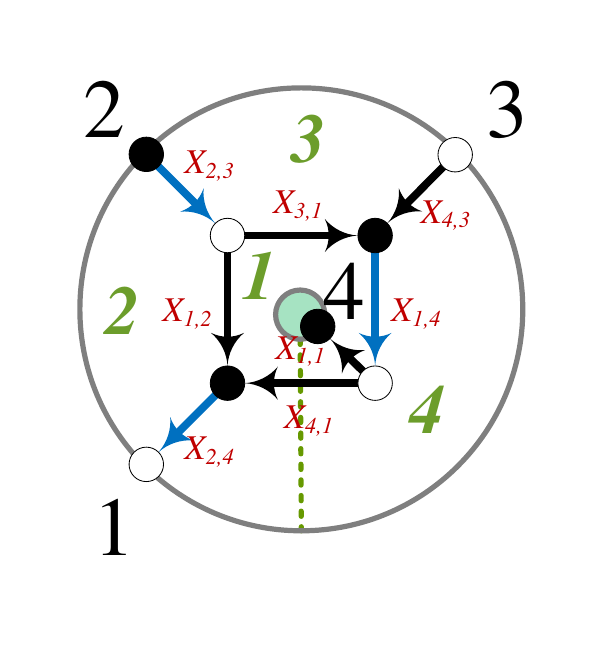}
\vspace{-0.6cm}\caption{An on-shell diagram on an annulus. This particular graph can be planarized by flipping the $X_{1,1}$ edge. Faces are labeled in green, external nodes in black and edges in red.}
\label{fig:sqbcrossed}
\end{center}
\end{figure}

Here we illustrate the independence on the embedding of the on-shell diagram with the simple example shown in \fref{fig:sqbcrossed}. It is clear that the non-planarity of this diagram is fake, since it can be embedded on a disk by flipping $X_{1,1}$. 

%
%

Here we have four face variables, three of which are independent, and one cut. In terms of oriented edge weights, they are given by
\begin{equation}
f_1=\frac{X_{3,1} X_{4,1}}{X_{1,2} X_{1,4}} \; , \quad f_2=\frac{X_{1,2}}{X_{2,3} X_{2,4}} \; , \quad f_3=\frac{X_{2,3} X_{4,3}}{X_{3,1}} \; , \quad b_1=\frac{X_{4,1}}{X_{1,1} X_{2,4}} \; .
\end{equation}

Let us consider the perfect orientation corresponding to the reference perfect matching $p_{\text{ref}} = X_{1,4} X_{2,3} X_{2,4} $, which has source set $\{2,3\}$. Using our prescription for the boundary measurement, we obtain the Grassmannian matrix

{\small
\begin{equation}
C = \left(
\begin{array}{c|cccc}
 & 1 & 2 & 3 & 4\\
 \hline
2 \ \ & \dfrac{X_{1,2}}{X_{2,3} X_{2,4}}+\dfrac{X_{3,1} X_{4,1}}{X_{1,4} X_{2,3} X_{2,4}} & 1 & 0 & \Gape[3pt][0pt]{-\dfrac{X_{1,1} X_{3,1}}{X_{1,4} X_{2,3}}} \\
3 \ \ &  -\dfrac{X_{4,1} X_{4,3}}{X_{1,4} X_{2,4}} & 0 & 1 & \dfrac{X_{1,1} X_{4,3}}{X_{1,4}} \\
\end{array}
\right) = \left(
\begin{array}{c|cccc}
 & 1 & 2 & 3 & 4 \\
 \hline
2 \ \ & f_1 f_2+f_2 & 1 & 0 & \Gape[3pt][0pt]{-\dfrac{f_1 f_2}{b_1}} \\
3 \ \ & -f_1 f_2 f_3 & 0 & 1 & \dfrac{f_1 f_2 f_3}{b_1} \\
\end{array}
\right) .
\end{equation}}

\noindent The on-shell form becomes
\begin{equation}
\Omega\,=\,\frac{df_1}{f_1} \frac{df_2}{f_2} \frac{df_3}{f_3} \frac{db_1}{b_1} .
\end{equation}
In terms of minors, it becomes 
\begin{equation}
\Omega\,=\, {d^{2\times 4} C \over \text{Vol(GL}(2))} \frac{1 }{(12)(23)(34)(41)} ,
\end{equation}
which is simply the form for the planar embedding, i.e.\ the ordinary square box in \fref{G24_edges_and_faces}. This illustrates the independence of the on-shell form on the embedding and shows that the generalized face variables maintain a $d\log$ form regardless of its choice.

\bigskip

\section{Combinatorial Signs: a Torus Example} \label{sec:motivationexample}

\begin{figure}[h]
\begin{center}
\includegraphics[width=7cm]{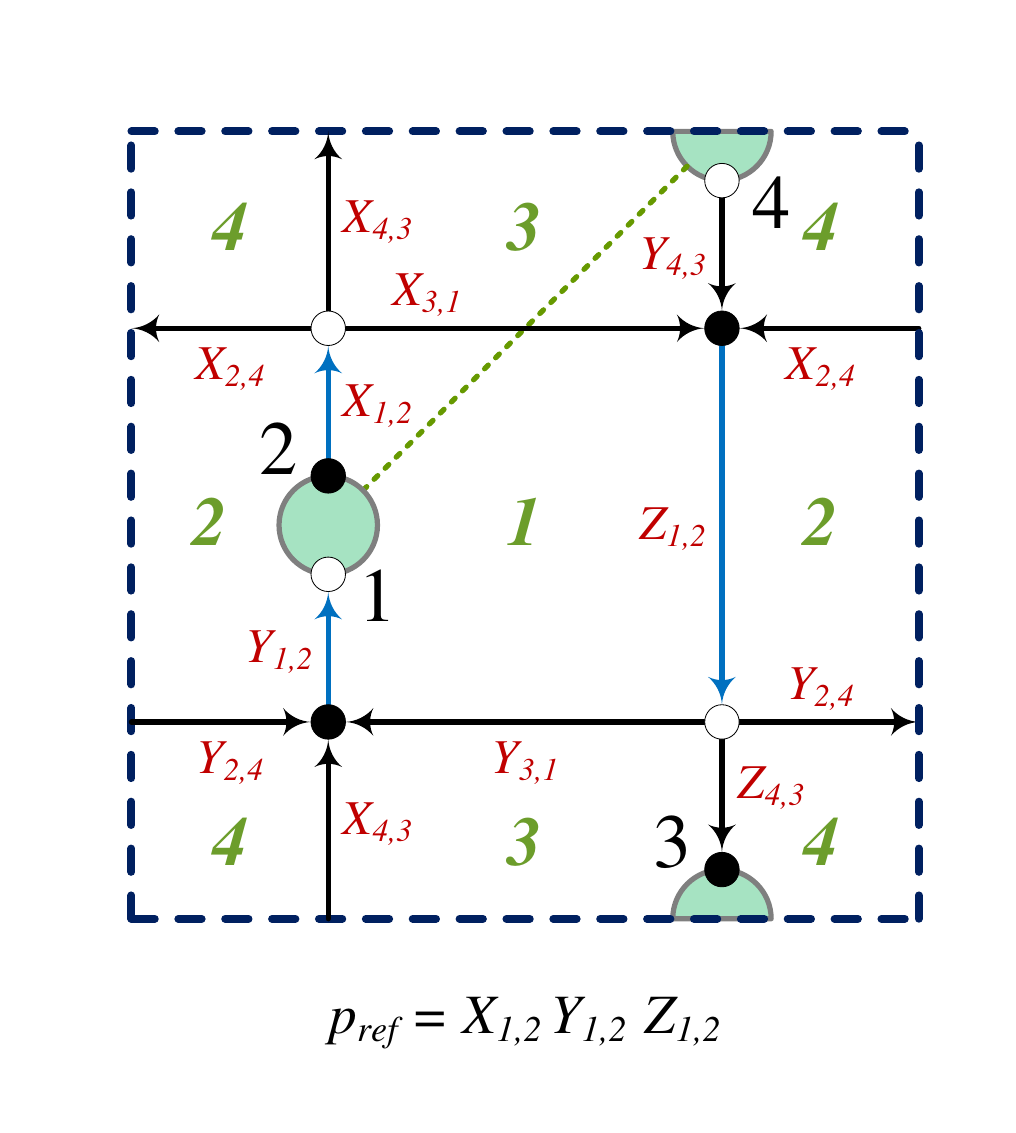}
\vspace{-0.5cm}\caption{An on-shell diagram with two boundaries and four external nodes on the torus. The blue edges are those in the reference perfect matching.}
\label{fig:torus1}
\end{center}
\end{figure}

To illustrate the main issues concerning the combinatorial signs that arise when trying to generalize the boundary measurement to higher genus, let us consider the on-shell diagram on the torus shown in \fref{fig:torus1}. We pick the reference perfect matching $p_{\text{ref}}=X_{1,2}Y_{1,2}Z_{1,2}$, which gives the source set $\{2,4\}$. External nodes have been ordered according to the prescription in \sref{sec:general-boundary-measurement}.

The matrix $C$ takes the schematic form:
\begin{equation}
C = \left(
\begin{array}{c|cccc}
 & 1 & 2 & 3 & 4 \\
 \hline
2 \ \ & \ast & 1 & \ast & 0 \\
4 \ \ & -\ast & 0 & \ast & 1 \\
\end{array}
\right) , \label{eq:schematicC}
\end{equation}
where we have already included the positivity signs $(-1)^{s(i,j)}$. 

\begin{figure}[h]
\begin{center}
\includegraphics[scale=0.5]{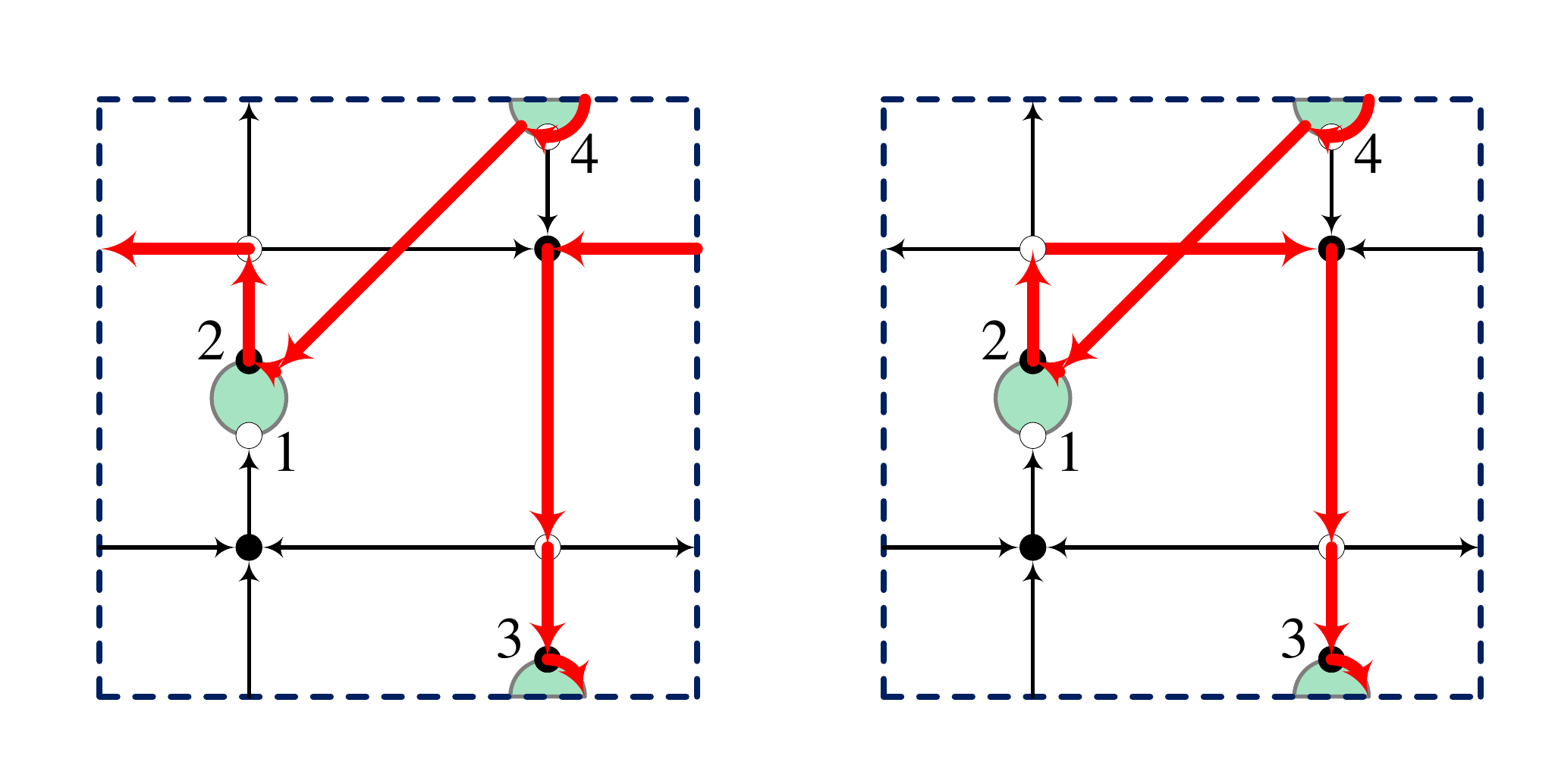}
\vspace{-0cm}\caption{The two flows contributing to $C_{23}$, completed into closed loops using the boundaries and the cut.}
\label{fig:torus1naive}
\end{center}
\end{figure}

As mentioned in \sref{sec:boundmeasproperties}, for non-planar diagrams the individual flows are also subject to combinatorial signs. Let us now investigate what happens if we naively extend the genus-0 prescription to higher genus, i.e.\ if we do not insist in closing flows into loops within the unit cell. Focusing on the example at hand, \fref{fig:torus1naive} shows the two flows that contribute to the entry $C_{23}$, which run from source 2 to sink 3. We then use the boundaries and cuts to form a loop, as for the genus-0 cases. The first flow gives rise to a single loop with no self-intersections, since the flow does not contain any edge that crosses the cut.  The second flow contains a self-intersection since one of its edges crosses the cut. The first flow gets no sign while the second one picks a $(-1)$. 

Similarly, two flows contribute to $C_{41}$. Both of them are analogous to the first flow in \fref{fig:torus1naive}, in that neither of them has self-intersections. We would thus expect no combinatorial signs for them. All other flows connect pairs of nodes on the same boundary, and hence never utilize the cut and do not self-intersect. As a result, we do not give any additional signs to them.

The matrix corresponding to this sign prescription becomes

{\footnotesize
\begin{align}
\hspace{-10pt} K &= \left(
\begin{array}{cccc}
 \dfrac{X_{4,3}}{X_{1,2} Y_{1,2}}+\dfrac{X_{2,4} Y_{2,4}}{X_{1,2} Y_{1,2} Z_{1,2}}+\dfrac{X_{3,1} Y_{2,4}}{X_{1,2} Y_{1,2} Z_{1,2}}+\dfrac{X_{2,4} Y_{3,1}}{X_{1,2} Y_{1,2} Z_{1,2}}+\dfrac{X_{3,1} Y_{3,1}}{X_{1,2} Y_{1,2} Z_{1,2}} & \,1\, & \dfrac{X_{2,4} Z_{4,3}}{X_{1,2} Z_{1,2}}-\dfrac{X_{3,1} Z_{4,3}}{X_{1,2} Z_{1,2}} & 0 \\
 &&&\\
 -\dfrac{Y_{2,4} Y_{4,3}}{Y_{1,2} Z_{1,2}}-\dfrac{Y_{3,1} Y_{4,3}}{Y_{1,2} Z_{1,2}} & \, 0\, & \dfrac{Y_{4,3} Z_{4,3}}{Z_{1,2}} &  1  \\
\end{array}
\right) \nonumber \\ \nonumber \\
&= \left(
\begin{array}{cccc}
 \mathfrak{p}_1+\mathfrak{p}_2+\mathfrak{p}_4+\mathfrak{p}_6+\mathfrak{p}_{12} & 1 & \mathfrak{p}_3-\mathfrak{p}_7 & 0 \\
 -\mathfrak{p}_5-\mathfrak{p}_8 & 0 & \mathfrak{p}_9 & 1 \\
\end{array}
\right) . \label{eq:naiveC}
\end{align}}
We have denoted it $K$ to differentiate it from the true Grassmannian matrix $C$, which we will write shortly. We note that the only signs that have been introduced are the overall $(-1)^{s(i,j)}$ positivity signs for entire entries and the one given to the second flow in \fref{fig:torus1naive}, i.e.\ to $\mathfrak{p}_7=\frac{X_{3,1} Z_{4,3}}{X_{1,2} Z_{1,2}}$. The minors that arise from \eref{eq:naiveC} are:
\begin{equation}
\begin{array}{cl}
 k_{12} = & \dfrac{Y_{2,4} Y_{4,3}}{Y_{1,2} Z_{1,2}}+\dfrac{Y_{3,1} Y_{4,3}}{Y_{1,2} Z_{1,2}} \\
 k_{13} = & \dfrac{X_{4,3} Y_{4,3} Z_{4,3}}{X_{1,2} Y_{1,2} Z_{1,2}}+2 \, \dfrac{X_{2,4} Y_{2,4} Y_{4,3} Z_{4,3}}{X_{1,2} Y_{1,2} Z_{1,2}^2}+2\, \dfrac{X_{2,4} Y_{3,1} Y_{4,3} Z_{4,3}}{X_{1,2} Y_{1,2} Z_{1,2}^2} \\
 k_{14} = & \dfrac{X_{4,3}}{X_{1,2} Y_{1,2}}+\dfrac{X_{2,4} Y_{2,4}}{X_{1,2} Y_{1,2} Z_{1,2}}+\dfrac{X_{3,1} Y_{2,4}}{X_{1,2} Y_{1,2} Z_{1,2}}+\dfrac{X_{2,4} Y_{3,1}}{X_{1,2} Y_{1,2} Z_{1,2}}+\dfrac{X_{3,1} Y_{3,1}}{X_{1,2} Y_{1,2} Z_{1,2}} \\
 k_{23} = & \dfrac{Y_{4,3} Z_{4,3}}{Z_{1,2}} \\
 k_{24} = & 1 \\
 k_{34} = & \dfrac{X_{2,4} Z_{4,3}}{X_{1,2} Z_{1,2}}-\dfrac{X_{3,1} Z_{4,3}}{X_{1,2} Z_{1,2}} \\
\end{array} \label{eq:naivepluckers}
\end{equation}
The only minor that requires delicate cancellations in order to achieve the desired map between \pl coordinates and perfect matchings is $k_{13}$. We see straight away that this choice of signs did not work: we would like to see a single term contributing to $\Delta_{13}$, corresponding to the only perfect matching with source set $\{1,3\}$, i.e.\ $\mathfrak{p}_{11}=\frac{X_{4,3} Y_{4,3} Z_{4,3}}{X_{1,2} Y_{1,2} Z_{1,2}}$. The two surplus terms in \eref{eq:naivepluckers} should have been subject to a cancellation rather than doubling up, and for this reason appear with a coefficient of 2.

\begin{figure}[h]
\begin{center}
\includegraphics[scale=0.55]{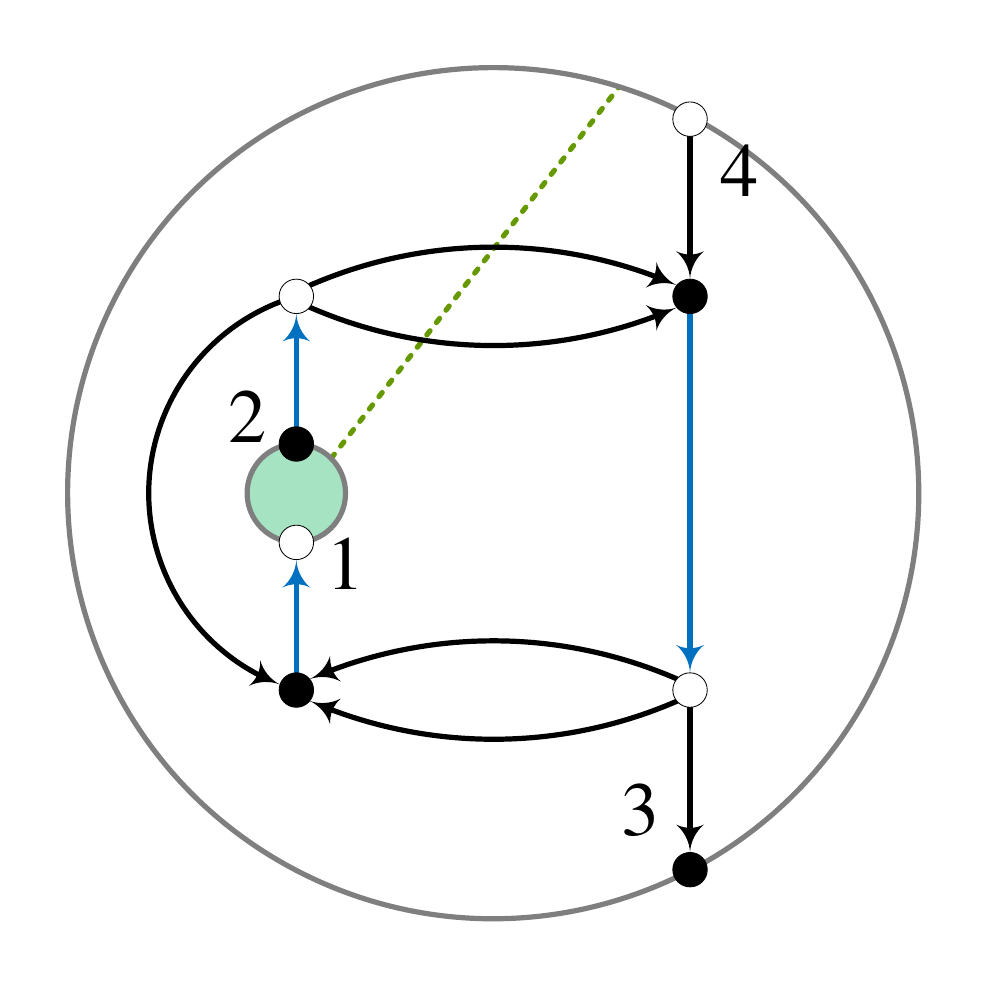}
\vspace{-0cm}\caption{An alternative embedding of the graph in \fref{fig:torus1} into an annulus.}
\label{fig:planarizedtorus}
\end{center}
\end{figure}

The graph under consideration can also be embedded on a genus-zero surface, as shown in \fref{fig:planarizedtorus}. Considering this alternative embedding is useful for identifying the source of the failure. 
%
%

In this case, it becomes clear that both flows in \fref{fig:torus1naive} cross the cut, and hence should receive a minus sign. Indeed, giving $\mathfrak{p}_3 = \frac{X_{2,4} Z_{4,3}}{X_{1,2} Z_{1,2}}$ a minus sign, all minors become well behaved:
\begin{equation}
\begin{array}{clccl}
 \Delta_{12}= & \mathfrak{p}_5+\mathfrak{p}_8 \quad \quad & \ \ \ & \Delta_{23}= & \mathfrak{p}_9 \\
 \Delta_{13}= & \mathfrak{p}_{11} \quad \quad & & \Delta_{24}= & 1 \\
 \Delta_{14}= & \mathfrak{p}_1+\mathfrak{p}_2+\mathfrak{p}_4+\mathfrak{p}_6+\mathfrak{p}_{12} \quad \quad & & \Delta_{34}= & -\mathfrak{p}_3-\mathfrak{p}_7 \\
\end{array} \label{eq:correctedpluckers}
\end{equation}
The lesson is simple: it appears it is possible to avoid self-intersections by looping around the torus. Simply giving a minus sign to all paths that go around the torus does not work either, as is easy to verify for this example --- we precisely want to give a minus sign to those paths that use the periodicity of the torus to avoid self-intersections. Again, we stress that this example should be regarded only as motivation for the prescription in \sref{sec:general-boundary-measurement}, which directs us to close loops within the unit cell. The prescription works more generally, such as in the genus-two example in \sref{sec:genus2} which contains no cuts at all.

\begin{figure}[h]
\begin{center}
\includegraphics[scale=0.36]{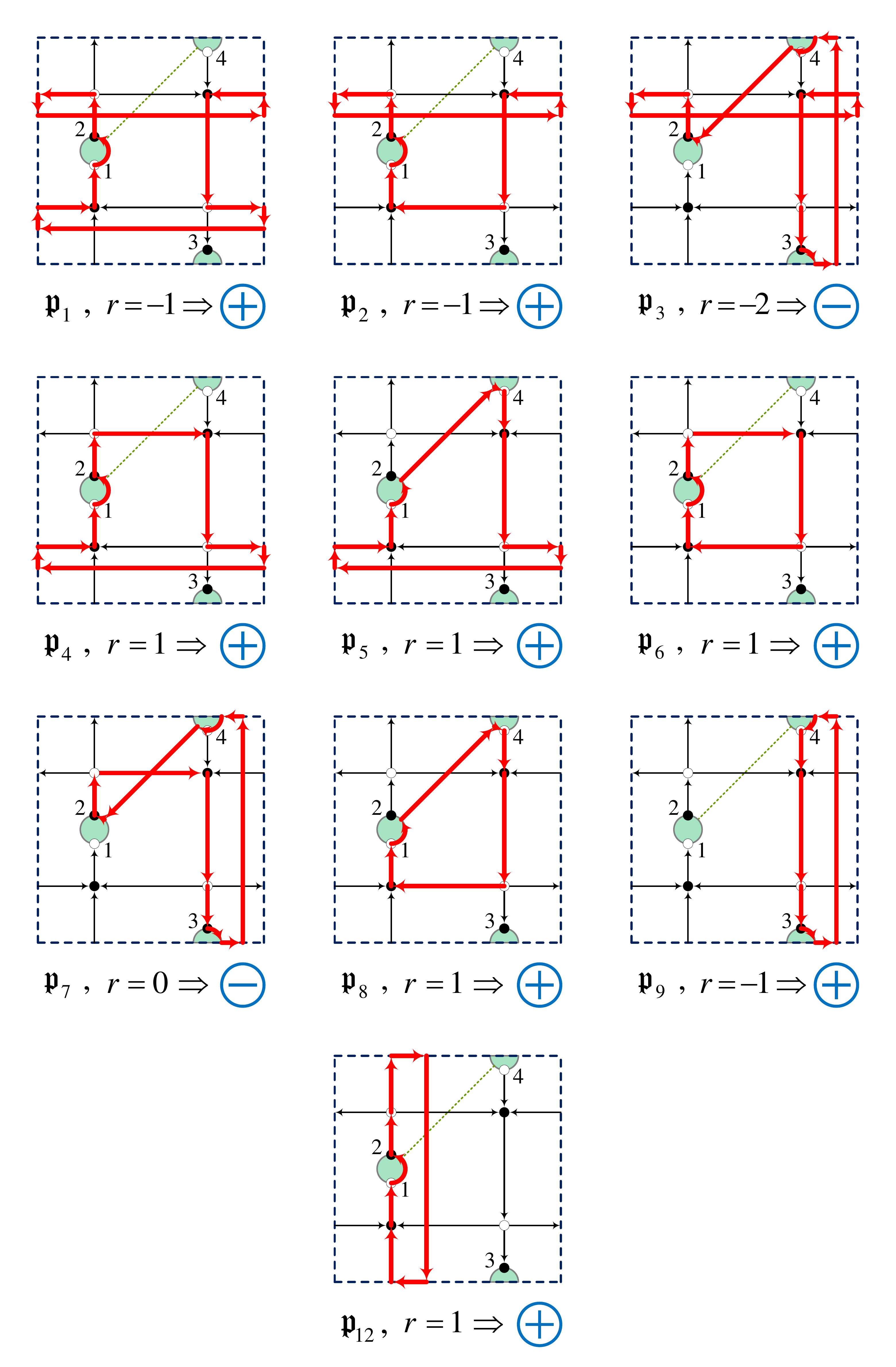}
\vspace{-0.2cm}\caption{Completion of flows into loops inside the unit cell for the example in \fref{fig:torus1} , their rotation numbers and the resulting signs.}
\label{fig:torus1good}
\end{center}
\end{figure}

\fref{fig:torus1good} shows all the flows for the diagram in \fref{fig:torus1} and their associated rotation numbers (and hence signs) as determined by the rules in \sref{sec:general-boundary-measurement}. It is easy to verify that these are the correct signs postulated above, which yield \eref{eq:correctedpluckers}.
%
%

\bigskip

\section{On-Shell Form for a Genus-One NMHV Diagram} \label{app:highergenus}

\begin{figure}[h]
\begin{center}
\includegraphics[scale=0.45]{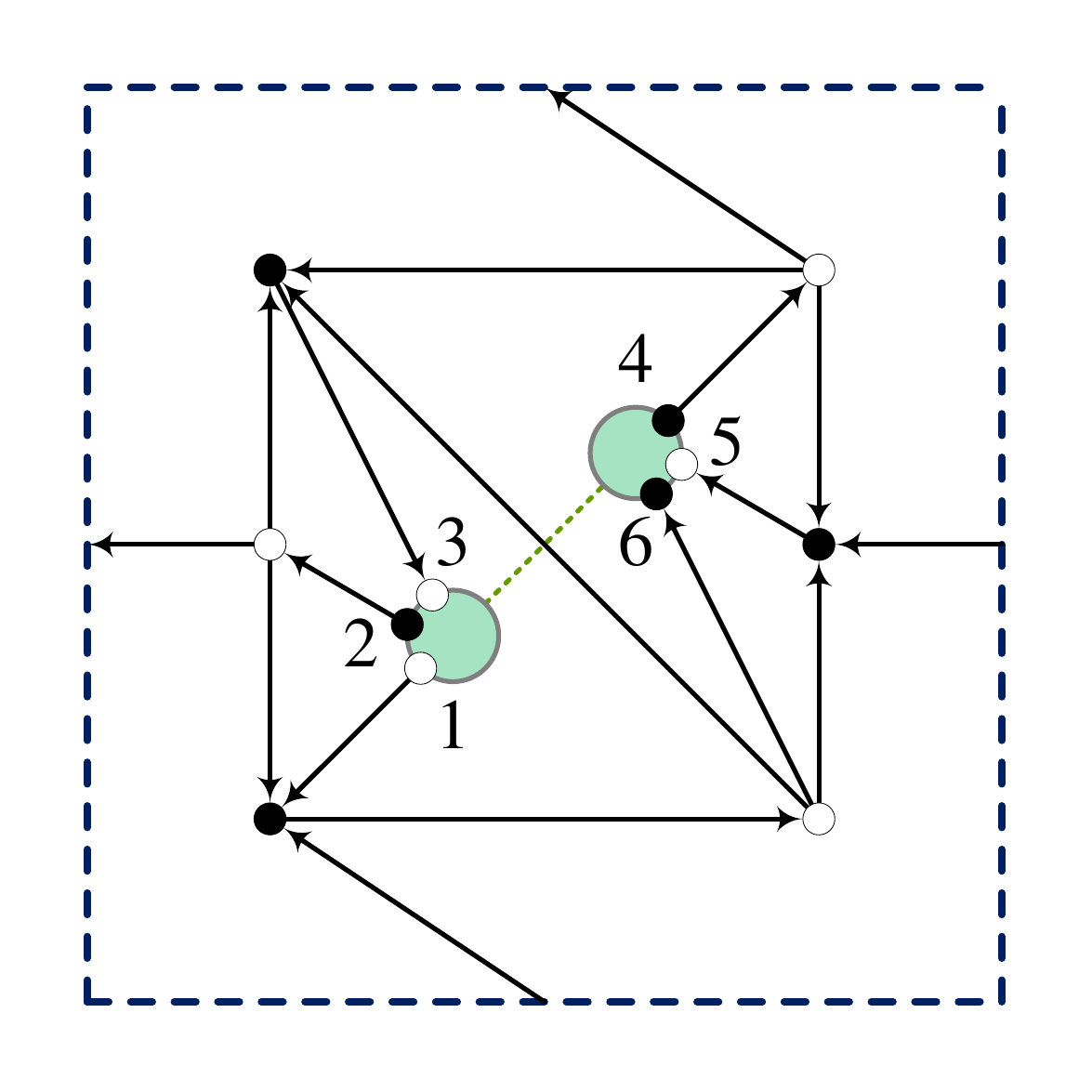}
\vspace{-0.3cm}\caption{An on-shell diagram embedded on a torus with two boundaries.}
\label{fig:nonplanarizableNolabel}
\end{center}
\end{figure}

To show that the method prescribed in \sref{sec:rules} works just as well for graphs with higher genus, we now consider the non-planarizable genus-1 example studied in \sref{sec:nonplanarizable} shown in \fref{fig:nonplanarizableNolabel}.

\noindent Following the prescription in \sref{sec:rules}, we find the matrices $T$ and $M$ to be
\begin{align}
T\,=\,\begin{pmatrix}
1 & 6 & 4 & 2 \\
3 & 2 & 4 & 6 \\
5 & 4 & 2 & 6
\end{pmatrix}\, ,\quad M\,=\,\begin{pmatrix}
(642) & -(164) & 0 & (162) & 0 & -(142) \\
0 & -(346) & (246) & (326) & 0 & -(324) \\
0 & (546) & 0 & -(526) & (426) & -(542)
\end{pmatrix}\ .
\end{align}

%
\noindent 
It is easy to see that the simplest way to obtain the on-shell form is by deleting columns \{2,4,6\}, 
\begin{align}
\widehat{M}_{2,4,6}\,&=\,\begin{pmatrix}
(642) &  0 & 0 \\
0 &  (246) & 0 \\
0 & 0 & (426)
\end{pmatrix},\quad \frac{\det \widehat{M}_{2,4,6}}{(246)}=(246)^2\, ,
\end{align}
which gives the on-shell form
\begin{align}
\label{eq:int-genus1}
\Omega\,=\,\frac{d^{3\times 6}C}{\text{Vol(GL}(3))}\frac{(246)^3}{(164)(421)(216)(324)(463)(632)(542)(265)(654)}\ .
\end{align}
We have checked that this result coincides with the result obtained by using the boundary measurement as described in \sref{sec:integrandfromface}, giving further evidence to both methods as well as to the validity of the boundary measurement in \sref{sec:general-boundary-measurement}.

\bigskip

\section{N$^2$MHV Example with Two Auxiliary Edges}
\label{app:NNMHV}

\begin{figure}[h]
\begin{center}
\includegraphics[scale=0.45]{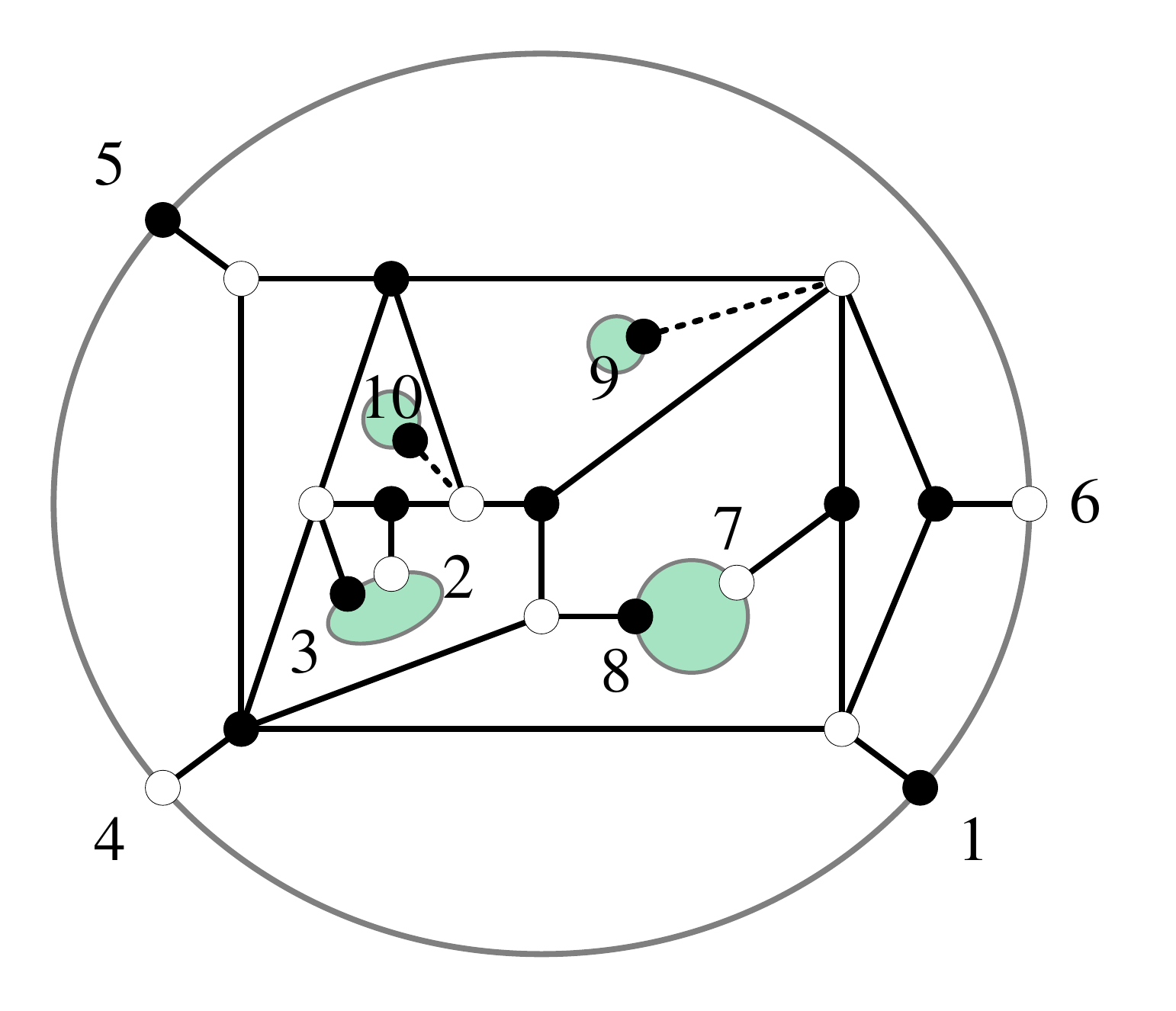}
\vspace{-0.3cm}\caption{An N$^2$MHV on-shell diagram for which $n_B=n-k+2$. In this case it is necessary to add two auxiliary external nodes, 9 and 10, for determining the on-shell form.}
\label{fig:NNMHV8}
\end{center}
\end{figure}

Let us consider the N$^2$MHV example in \fref{fig:NNMHV8}. The $T$ matrix is given by
{\small
\begin{align}
\label{eq:TNNMHV}
T\,=\,\begin{pmatrix}
\ 6 \ \ & \ 1 \ \ & \ 9 \ \ & \ * \ \ & \ * \ \ \\
1 & 7 & 9 & * & * \\
8 & 10 & 9 & * & *\\
10 & 3 & 5 & 9 & * \\
5 & 3 & 8 & 1 & 4 \\
2 & 3 & 10 & * & *
\end{pmatrix}\quad\xrightarrow{\text{Choice of }*}\quad T=\begin{pmatrix}
\ 6 \ \ & \ 1 \ \ & \ 9 \ \ & \ 3 \ \ & \ 8 \ \ \\
1 & 7 & 9 & 3 & 8 \\
8 & 10 & 9 & 1 & 3\\
10 & 3 & 5 & 9 & 1 \\
5 & 3 & 8 & 1 & 4 \\
2 & 3 & 10 & 1 & 8
\end{pmatrix} .
\end{align}
}

\noindent
This leads to the following matrix $M$
{\small
\begin{align}
M\,=\,\begin{pmatrix}
\label{eq:M82}
(9386) & 0 & (8619) & 0 & 0 & (1938) & 0 & (6193) & (3861) & 0\\
(7938) & 0 & (8179) & 0 & 0 & 0 & (9381) & (1793) & (3817) & 0 \\
(38109) & 0 & (81091) & 0 & 0 & 0 & 0 & (10913) & (13810) & (9138)\\
(10359) & 0 & (59110) & 0 & (91103) & 0 & 0 & 0 & (11035) & (3591) \\
(4538) & 0 & (8145) & (5381) & (3814) & 0 & 0 & (1453) & 0 & 0 \\
(82310) & (31018) & (10182) & 0 & 0 & 0 & 0 & (23101) & 0 & (1823) \\
\end{pmatrix} .\nonumber\\[5pt]
\end{align}} 
\noindent where we eliminated the minus signs on the entries of $M$ by using the fact that an equivalent way to write \eqref{eq:Cramer} for even $k$ is $ \vec{c}_{i_1}(i_2\cdots i_{k+1})+\text{cyclic}(i_1,i_2,\dots, i_{k+1})=0$. The result of the procedure in \sref{sec:rules} gives
{\small
\begin{align}
\begin{split}
\Omega\,&=\,\frac{d^{4\times 10}C}{\text{Vol(GL}(4))}\frac{(1358)^3(1389)^5(13810)^2(13910)^2}{(1238)(12310)(12810)(1345)(1348)(1359)(13510)(1368)(1369)(1378)(1379)}\\
&\times\frac{1}{(1458)(15910)(1689)(1789)(18910)(23810)(3458)(35910)(3689)(3789)(38910)}
.
\end{split}\nonumber
\end{align}
}
This can be simplified using the fact that the points $\{1,6,7,9\}$ are collinear,  $\{8,9,10\}$ are collinear,  $\{2,3,10\}$ are collinear and  $\{3,5,9,10\}$ are coplanar, as can be read off from \eqref{eq:TNNMHV}. After these simplifications, the dependence on nodes $9$ and $10$ is encoded in the ratio
\begin{align}
\label{eq:ratioNNMHV}
\left.I\right|_{9,10}\,&=\,\frac{1}{(38910)(12310)(1369)(1689)(18910)(23810)} ,
\end{align}
which after the residues around $C_{i9}=C_{i10}=0$ for $i=1,\dots 4$ gives
\begin{align}
\left.I\right|_{9,10}\,&=\,\frac{1}{(1368)^2(1238)^2}\, .
\end{align}
Putting everything together, we obtain the following on-shell form
{\footnotesize
\begin{align}
\Omega\,=\,\frac{d^{4\times 8}C}{\text{Vol(GL}(4))}\frac{(1358)^3(1386)}{(7812)(1345)(1348)(1356)(1458)(1568)(1376)(6781)(2345)(3528)(3568)(3782)} .
\end{align}
}
This differential form has been independently confirmed using the boundary measurement procedure from \sref{sec:integrandfromface}.

\bigskip

\section{Six-Point Leading Singularity} \label{appendix:ls}

Here we compute the six-point NMHV non-planar leading singularity considered in \sref{section:complicatedcase}. For convenience, we quote the Grassmannian formula
\begin{align}
\label{eq:36GrassFormula}
\mathcal{L}_{3,6} = \oint_{S=0} {d^{3 \times 6} C \over {\rm GL}(3)} {(346)^2(356)   \over (234)(345)(456)(561)(136)(236)\, S \, }  \prod^3_{\alpha=1}\delta^{4|4}\left( C_{\alpha a} \mathcal{W}^{a} \right) ,
\end{align}
where the contour of integration is taken around the pole $S=(124)(346)(365)-(456)(234)(136)$. After taking into account the $\delta$-functions, the six-point NMHV leading singularity is a one-dimensional contour integral of a parameter, which we shall denote $\tau$. It is straightforward to check that $S$ is a degree-three polynomial in $\tau$, rendering the solutions to $S=0$ rather complicated. Instead we apply the residue theorem, namely that the residue at $S=0$ is equal to minus the sum of the residues at $(234)=0, (345)=0, (456)=0, (561)=0, (136)=0$ and $(236)=0$.\footnote{We would like to remark the similarity between this leading singularity and the twistor string formula in the Grassmannian form, where one also uses residue theorems to change a polynomial pole into a sum of linear poles \cite{Spradlin:2009qr,Dolan:2009wf,Nandan:2009cc,ArkaniHamed:2009dg,Bourjaily:2010kw}.} In what follows we compute each residue separately. 
\begin{itemize}
\item For $(234)=0$, we have\footnote{We use standard spinor-helicity formalism: $ p_{\alpha \dot{\alpha}} = \lambda_{\alpha} \widetilde{\lambda}_{\dot{\alpha}} $, and scalar products $\lambda_i^\alpha\lambda_j^\beta\epsilon_{\alpha\beta}=\langle ij\rangle$, $\widetilde\lambda_{i\dot\alpha}\widetilde\lambda_{j\dot\beta}\epsilon^{\dot\alpha\dot\beta}=[ij]$, $ s_{i \ldots j}=(p_i + \ldots + p_j)^2$ and $ \langle i | k+l |m] =\langle ik\rangle [km]+ \langle il\rangle [lm]$.}   
\begin{align}
{\delta^{(8)}(\sum_i \lambda_i \eta_i) \delta^{(4)}([56]\eta_1 +[61] \eta_5 + [15] \eta_6 ) \over \langle 23\rangle \langle 24\rangle [56][61] \langle 4|5+6|1]  \langle 3|6+1|5]
s_{234}} .
\end{align}
\item For $(345)=0$, we have    
\begin{align}
{\langle 35\rangle [12] \delta^{(8)}(\sum_i \lambda_i \eta_i) \delta^{(4)}([61]\eta_2 +[12] \eta_6 + [26] \eta_1 ) \over \langle 45\rangle [61] \langle 5|4+3|2] \langle 3|4+5|2] \langle 3|4+5|1] (\langle 45\rangle [16] \langle 3|4+5|2]- \langle 35\rangle [12] \langle 4|5+3|6] )} .
\end{align}
\item For $(456)=0$, we have    
\begin{align}
{\langle 46\rangle \delta^{(8)}(\sum_i \lambda_i \eta_i) \delta^{(4)}([12]\eta_3 +[23] \eta_1 + [31] \eta_2 ) \over \langle 45\rangle \langle 56\rangle [23] \langle 4|5+6|1]  \langle 4|5+6|3]
\langle 6|4+5|1] \langle 6|4+5|2]} .
\end{align}
\item For $(561)=0$, we have    
\begin{align}
{ \langle 6|1+5|2]^2 \delta^{(8)}(\sum_i \lambda_i \eta_i) \delta^{(4)}([23]\eta_4 +[34] \eta_2 + [42] \eta_3 ) \over \langle 56\rangle \langle 61\rangle[23][24] \langle 5|1+6|2] \langle 1|5+6|2]  \langle 6|1+5|3] \langle 6|1+5|4] s_{561} } .
\end{align}
\item For $(136)=0$, we have    
\begin{align}
{[52] \delta^{(8)}(\sum_i \lambda_i \eta_i) \delta^{(4)}([24]\eta_5 +[45] \eta_2 + [52] \eta_4 ) \over  \langle 61\rangle[24][45] \langle 3|1+6|2]\langle 6|1+3|2] \langle 3|1+6|5] \langle 1|3+6|5]   } .
\end{align}
\item For $(236)=0$, we have    
\begin{align}
{\langle 36\rangle [14] \delta^{(8)}(\sum_i \lambda_i \eta_i) \delta^{(4)}([14]\eta_5 +[45] \eta_1 + [51] \eta_4 ) \over \langle 23\rangle [45] \langle 6|2+3|1] \langle 6|2+3|4] \langle 3|2+6|1] (\langle 23\rangle [45] \langle 6|2+3|1]- \langle 36\rangle [14] \langle 2|6+3|5] )} .
\end{align}
\end{itemize}
The three-loop non-planar leading singularity in \eref{eq:36GrassFormula} is then given by minus the sum of the residues above.

\bigskip

\bibliographystyle{JHEP}
\bibliography{references}
\end{document}